\documentclass[12pt,preprint]{aastex}
\pdfoutput=1

\newcommand\msol{{\cal M_{\odot}}}

\newcommand\teff{{T_{\rm eff}}}
\newcommand\amlt{{\alpha_{\rm MLT}}}
\newcommand\delv{\Delta V^{\rm HB}_{\rm TO}}
\newcommand\delc{\Delta H_{\rm TO,RGB}}

\newcommand\lta{\mathrel{\hbox{\raise 0.6 ex \hbox{$<$}\kern
                   -1.8 ex\lower .5 ex\hbox{$\sim$}}}}
\newcommand\gta{\mathrel{\hbox{\raise 0.6 ex \hbox{$>$}\kern
                   -1.7 ex\lower .5 ex\hbox{$\sim$}}}}
 
\shortauthors{VandenBerg et al.}
\shorttitle{Globular Cluster Ages}
 
\begin{document}
 
\title{THE AGES OF 55 GLOBULAR CLUSTERS AS DETERMINED USING AN IMPROVED
 $\delv$\ METHOD ALONG WITH COLOR-MAGNITUDE DIAGRAM CONSTRAINTS, AND
 THEIR IMPLICATIONS FOR BROADER ISSUES}

\author{Don A.~VandenBerg, K.~Brogaard\altaffilmark{1},
        R.~Leaman\altaffilmark{2,3,4}}
\affil{Department of Physics \& Astronomy, University of Victoria,
       P.O.~Box 3055, Victoria, B.C., V8W~3P6, Canada}
\email{vandenbe@uvic.ca, kfb@phys.au.dk, rleaman@uvic.ca}

\author{L.~Casagrande}
\affil{Research School of Astronomy \& Astrophysics, Mt.~Stromlo Observatory,
 The Australian National University, ACT 2611, Australia}
\email{luca@mso.anu.edu.au}

\altaffiltext{1}{Current Address: Stellar Astrophysics Centre, Department
 of Physics \& Astronomy, Aarhus University, 120 Ny Munkegade, Building 1520,
 8000 Aarhus C, Denmark}
\altaffiltext{2}{Instituto de Astrof\'isica de Canarias, Spain}
\altaffiltext{3}{Dept.~Astrof\'isica, Universidad de La Laguna, Spain}
\altaffiltext{4}{Current Address: Instituto de Astrofisica de Canarias,
 V\'ia L\'actea s/n, E-38200 La Laguna, Spain}

\begin{abstract}
Ages have been derived for 55 globular clusters (GCs) for which {\it Hubble
Space Telescope} ACS photometry is publicly available.  For most of them, the
assumed distances are based on fits of theoretical zero-age horizontal branch
(ZAHB) loci to the lower bound
of the observed distributions of HB stars, assuming reddenings from empirical
dust maps and metallicities from the latest spectroscopic analyses.  The age of
the isochrone that provides the best fit to the stars in the vicinity of the
turnoff (TO) is taken to be the best estimate of the cluster age.  The
morphology of isochrones between the TO and the beginning part of the subgiant
branch (SGB) is shown to be nearly independent of age and chemical abundances.
For well-defined CMDs, the error bar arising just from the ``fitting" of ZAHBs
and isochrones is $\approx \pm 0.25$ Gyr, while that associated with distance
and chemical abundance uncertainties is $\sim \pm 1.5$--2 Gyr.  The oldest GCs
in our sample are predicted to have ages of $\approx 13.0$ Gyr (subject to the
aforementioned uncertainties).  However, the main focus of this investigation
is on relative GC ages.  In conflict with recent findings based on the
{\it relative main-sequence fitting} (rMSF) method, which have been studied
in some detail and reconciled with our results, ages are found to vary from
mean values of $\approx 12.5$ Gyr at [Fe/H] $\lta -1.7$ to $\approx 11$ Gyr at
[Fe/H] $\gta -1$.  At intermediate metallicities, the age-metallicity relation
(AMR) appears to be bifurcated: one branch apparently contains clusters with
disk-like kinematics, whereas the other branch, which is displaced to lower
[Fe/H] values by $\approx 0.6$ dex at a fixed age, is populated by clusters
with halo-type orbits.  The dispersion in age about each component of the AMR
is $\sim \pm 0.5$ Gyr.  There is no apparent dependence of age on Galactocentric
distance (R$_{\rm G}$) nor is there a clear correlation of HB type with age.
As previously discovered in the case of M$\,$3 and M$\,$13, subtle variations
have been found in the slope of the SGB in the color-magnitude diagrams (CMDs)
of other metal-poor ([Fe/H] $\lta -1.5$) GCs.  They have been tentatively
attributed to cluster-to-cluster differences in the abundance of helium.
Curiously, GCs that have relatively steep ``M$\,$13-like" SGBs tend to be
massive systems, located at small R$_{\rm G}$, that show the strongest evidence
of {\it in situ} formation of multiple stellar populations.  The clusters in
the other group are typically low-mass systems (with 2--3 exceptions, including
M$\,$3) that, at the present time, should not be able to retain the matter lost
by mass-losing stars due either to the development of GC winds or to
ram-pressure stripping by the halo interstellar medium.  The apparent separation
of the two groups in terms of their {\it present-day} gas retention properties
is difficult to understand if all GCs were initially $\sim 20$ times their
current masses.  The lowest mass systems, in particular,
may have never been massive enough to retain enough gas to produce a 
significant population of second-generation stars.  In this case, the observed
light element abundance variations, which are characteristic of all GCs, were
presumably present in the gas out of which the observed cluster stars formed.
\end{abstract}
 
\section{Introduction}
\label{sec:intro}

Until about a decade ago, the age of the oldest globular cluster (GC) was of
widespread interest because it provided one of the best available constraints on
the age of the universe and thereby on the cosmological model used to describe
it (see, e.g., \citealt{vrm02}).  As it has turned out, observations of the
cosmic microwave background (CMB), taken with the WMAP and Planck satellites
have since yielded what appears to be a robust, and very precise, estimate of
the age of the universe (13.8 Gyr, with an uncertainty of $\lta \pm 0.1$ Gyr,
see \citealt{ksd11}, \citealt{aaa13}).  (This is the age predicted
by the flat $\Lambda\,$CDM model that accurately reproduces the observed
temperature power spectrum of the CMB.)  However, absolute (and relative) GC
ages are no less important today than they were prior to the CMB results.  In
particular, they are needed to test and constrain models for the origin of GCs
over the entire metallicity range spanned by them and to provide some insights
into the formation and early evolution of galaxies.

For instance, in their extensive review, \citet{bs06} argued that the most
metal-poor GCs formed in low-mass dark matter halos in the early universe (at
redshifts $z > 10$), whereas metal-rich systems were created during the
subsequent mergers of gas-rich structures that built up the parent galaxies.
In a later study, \citet{byn08} used $n$-body simulations combined with
semi-analytic treatments of the main processes that govern galaxy and GC
formation (e.g., merging, star formation, supernova feedback, radiative gas
cooling) to predict that $\sim 90$\% of GCs formed in low-mass galaxies at
$z > 3$ and that the mean ages of clusters that have [Fe/H] $\lta -1$ are $\sim
0.5$ Gyr older than more metal-rich systems.  More recently, \citet{emr12}
pointed out that dwarf star-forming (Lyman-$\alpha$ emitting) galaxies at
intermediate to high redshift have the right size, metallicity, luminosity, and
star formation rate to be a natural site for the formation of metal-poor GCs.
They suggest that ``low metallicities are not the exclusive result of an earlier
birth time compared to metal-rich disk and bulge GCs, but rather the result of
a lower mass host, considering the (observed) mass-metallicity relation in
galaxies".  In other words, metal-poor globular clusters could simply be those
systems that happened to form in dwarf galaxies of low metallicity at whatever
time the birth event occurred.

Even from these few examples, it is clear that the proposed formation scenarios
make different predictions for the age of the oldest GC, the dispersion in age
at a fixed iron abundance (perhaps especially at the lowest metallicities), and
the variation in age with [Fe/H] and galactocentric distance.  Unfortunately,
despite the efforts made by many researchers over the past $\sim 35$ years, it
has not yet been possible to firmly establish the absolute, or the relative,
ages of the Galactic GCs (see, e.g., \citealt{vbs90}; \citealt{cds96};
\citealt{bcp98}; \citealt{ros99}; \citealt{cgc00}; \citealt{van00};
\citealt{sw02}; and \citealt{dea05}).  This is apparent even if one considers
only those papers that were published after 2008.  For example, ages as young
as 11 Gyr (\citealt{dic10}) or as old as 13.5 Gyr (\citealt{vcs10}) have been
derived for M$\,$92, which has [Fe/H] $\approx -2.35$ (\citealt[][hereafter
CBG09]{cbg09a}).  (These seemingly discrepant results are, in fact, due mostly
to differences in the adopted distance.)  In addition, from an analysis of
homogeneous {\it HST} photometry for 64 GCs, \citet[][hereafter MF09]{map09}
found that the majority of the clusters with [$m$/H] $\lta -0.6$ are coeval (to
within the uncertainties) with the most metal-deficient systems.  However, there
are reasons to be concerned with this finding, especially after considering
recent estimates of the age of 47 Tucanae.

On the one hand, MF09 obtained a normalized age ($=$ absolute cluster age
divided by the mean age of GCs that have [Fe/H] $< -1.4$) of $0.96 \pm 0.07$
for 47 Tuc (which has [Fe/H] $= -0.71$ on the metallicity scale by
\citealt{zw84}), as compared with values of $1.00 \pm 0.04$ for M$\,$15
($-2.15$) and $1.02 \pm 0.04$ for M$\,$92 ($-2.24$), when \citet[hereafter
DSEP]{dcj07} isochrones are used in the analysis.  In a follow-up study of the
same photometry, \citet{dsa10} derived ages of $12.75 \pm 0.50$, $13.25 \pm
1.00$, and $13.25 \pm 1.00$ Gyr for 47 Tuc, M$\,$15, and M$\,$92, respectively,
using an isochrone-fitting procedure that permitted small adjustments to the
distances, reddenings, and metallicities from initial estimates of these
quantities given in the 2003 revision of the \citet{har96} catalog.  In fact,
the cluster parameters adopted by MF09 and Dotter et al.~are sometimes quite
different; e.g., the latter assumed [Fe/H] $= -2.40$ for both M$\,$15 and
M$\,$92, as well as lower values of [$\alpha$/Fe].  Moreover, the observed
color-magnitude diagrams (CMDs) are not reproduced very well by DSEP isochrones
at low metallicities; see Figs.~4 and 5 in the Dotter et al.~paper, which
show, in turn, that the fiducial sequences of NGC$\,$3201 and NGC$\,$7099
(M$\,$30), from just below the turnoff (TO) to the lower red-giant branch (RGB),
cross over isochrones spanning about a 3 Gyr range in age. Of particular concern
is the fact that the observed TOs are significantly redder than those of the 
isochrones in the adopted fits, as this has the effect of making the inferred
age somewhat too large (see the discussion in \S~\ref{subsec:m5} below).

If the same difficulties were found for M$\,$15 and M$\,$92, which is expected
to be the case since their CMDs in the vicinity of the turnoff are
morphologically indistinguishable from that of M$\,$30 (see VandenBerg 2000,
his Fig.~2), then the ages derived by Dotter et al.~for all three clusters
should be reduced to $\approx 12.75$ Gyr, which is identical to their estimate
of the age of 47 Tuc.  [Although Dotter et al.~(2010) do not actually show how
well their computations match the CMD of 47 Tuc, their Fig.~3 demonstrates that
DSEP isochrones for [Fe/H] $= -1.1$ faithfully reproduce the CMD of NGC$\,$6362,
which suggests that these models are not problematic in the metal-rich regime.
It is regrettable that Dotter et al.~provided such plots for only three GCs, as
it would have been helpful to see how well their models fare for more of the
most metal-deficient systems as well as those at the highest metallicities.]

On the other hand, using the same DSEP isochrones, \citet{tkr10} derived an
age of $11.25 \pm 0.21$ (random) $\pm\ 0.85$ (systematic) Gyr for 47 Tuc on the
assumption of [Fe/H] $= -0.70$, [$\alpha$/Fe] $= 0.4$, $Y = 0.255$, and an
apparent distance modulus, $(m-M)_V = 13.35 \pm 0.08$, which is based on the
properties of the eclipsing binary member known as V69.  This compares quite
favorably with the determination of 11.0 Gyr by VandenBerg et al.~(2010), if
$(m-M)_V = 13.40$ and similar chemical abundances are assumed, in an
investigation that also reported an age of 13.5 Gyr for M$\,$92.  Because the
distance moduli adopted by VandenBerg et al.~are within 0.05 mag of those
implied by fits of computed zero-age horizontal-branch (ZAHB) loci to the lower
bound of the distributions of cluster HB stars performed by VandenBerg (2000),
it is not a surprise that both studies found 47 Tuc to be about 2.5 Gyr younger
than M$\,$92.  In fact, notwithstanding small differences in the adopted cluster
properties, the binary- and ZAHB-based distance moduli for 47 Tuc agree very
well with the value of $(m-M)_V = 13.375$ that was obtained by \citet{bs09}
from fits of the cluster main-sequence (MS) to nearby subdwarfs having similar
metal abundances and well-determined trigonometric parallaxes from {\it
Hipparcos} (\citealt{vl07}).  Indeed, these results, coupled with the fact that
high ages ($\gta 13$ Gyr) are obtained for M$\,$92 when its distance is
similarly derived either from fits to local subdwarfs (\citealt[][see their
Table 2]{apm09}) or from the luminosity of its HB stars (\citealt[also see
\citealt{bmf11}]{van00}), appear to rule out a common age for 47 Tuc and the
most metal-deficient GCs.  At the very least, they call into question the
findings of MF09 and Dotter et al.~(2010).

The present investigation has been undertaken mainly to study and to try to
resolve the discrepancies discussed above concerning relative GC ages. 
However, we have also taken this opportunity to derive the absolute ages of
most of the clusters considered by MF09 on the assumption of ZAHB-based
distances.  \S~\ref{sec:input} briefly discusses the photometric data and the
stellar evolutionary models that have been used, while \S~\ref{sec:methods}
describes what we consider to be the most robust and least model-dependent way
of deriving the ages of globular clusters from their CMDs.  The reliability of
the adopted distance scale is examined in \S~\ref{sec:zahbdis}.  Ages
are determined in \S~\ref{sec:ages}: the results are supported by
plots for the majority of the GCs that we have analyzed to show how well the
isochrones match the observations.  Particular attention is paid to the
so-called ``second-parameter" clusters M$\,$3 and M$\,$13, as well as M$\,$5,
M$\,$12 (NGC$\,$6218), NGC$\,$288, and NGC$\,$362, and to those clusters in the
metallicity range $-2.0 \lta$ [Fe/H] $\lta -1.6$ with extended blue HBs.  Some
evidence is provided to support the possibility that enhanced helium abundances
are responsible for most (but not all) of these phenomena.  The main results of
this study are reported in \S~\ref{sec:summary}, which also explains (see
\S~\ref{subsubsec:mf09}) why MF09 obtained ages for metal-rich GCs that are too
high and, therefore, why they failed to find a significant variation of age with
[Fe/H], as obtained here.  Concluding remarks are given in \S~\ref{sec:final}.

\section{The Input Observational and Theoretical Data Bases}
\label{sec:input}

The same photometric data that were analyzed by MF09 are studied here; namely,
the $F606W, F814W$ observations that were obtained by \citet{sbc07} for $> 60$
globular clusters using the Advanced Camera for Surveys (ACS) on the {\it
Hubble Space Telescope}.\footnote{See http://archive.stsci.edu/prepds/acsggct}
The adopted zero points in the publicly available catalog are those given by
\citet{sjb05}, with the small adjustments to them subsequently determined by
\citet{bo07}: the latter have the effect of making the $F606W$ and $F814W$
magnitudes fainter by 22 and 25 mmag, respectively.\footnote{See
http://www.stsci.edu/hst/acs/analysis/zeropoints/old\_page/localZeropoints.}
In order to minimize the photometric scatter and to ensure that the principal
cluster sequences are well-defined, the CMDs have been limited to only those
stars for which the tabulated uncertainties in the magnitudes are $< 0.030$ mag.
(The ``readme" file associated with the catalog explains how these errors were
calculated.)  If the total number of stars that satisfies this criterion
exceeded $\sim 25,000$, this cutoff was reduced to 0.015 mag.  With relatively
few exceptions, each of the resultant GC data sets contained at least 10,000
stars.  To determine the intrinsic colors of the stars, we have assumed the
$E(B-V)$ values given by the \citet{sfd98} dust maps, except in a few cases
(which are flagged) where they were obviously problematic, together with
$E(m_{F606W}-m_{F814W}) = 0.984\,E(B-V)$ (\citealt{sjb05}).  (It will become
evident in the next section that, because the models are forced to match the
turnoff color in order to derive the correct estimate of the cluster age for
the assumed distance and metallicity, uncertainties associated with the adopted
value of $E(B-V)$ or with the relation between $E(m_{F606W}-m_{F814W})$ and
$E(B-V)$ are of no consequence for the age that is obtained.)

As far as the stellar models are concerned, we opted to use the grids of
evolutionary tracks computed by \citet[hereafter VBD12]{vbd12} for the
so-called ``GSCX" metals mixture.  This assumes the \citet{gs98} solar
abundances, with the enhancements in the abundances of the individual
$\alpha$-elements (at low metallicities) given by \citet{cds04}, appropriately
scaled to [Fe/H] values from $-3.0$ to 0.0, in 0.2 dex intervals.  (Because
[O/Fe] $= 0.5$, while most of the other $\alpha$-elements have [$m$/Fe]
$= 0.25$--0.4 dex enhancements, the overall value of [$\alpha$/Fe] is $+0.46$
in the very metal-deficient stars observed by Cayrel et al.)  As shown by
VandenBerg et al.~(2010), these models appear to provide good fits to optical
CMDs for both metal-poor and metal-rich GCs and they reproduce quite
satisfactorily the properties of local subdwarfs that have $-2.0 \lta$ [Fe/H]
$\lta -0.6$ and well-determined $M_V$ values, as derived from {\it Hipparcos}
parallaxes (\citealt{vl07}).  Since, in this investigation, the adopted GC
distances are based on the predicted luminosities of ZAHB models, we have
generated fully consistent ZAHB loci in the canonical way (see \citealt{vsr00})
using exactly the same stellar evolutionary code that is described in detail by
VBD12.

To transpose the models from the theoretical to the observed plane, bolometric
corrections (BCs) for several ACS bandpasses, including $F606W$ and $F814W$, have
been derived from synthetic spectra based on the latest MARCS model atmospheres
(\citealt{gee08}).\footnote{These transformations, along with those which are
applicable to many other widely used photometric systems, will be the subject 
of a forthcoming paper by L.~Casagrande et al.~(in preparation).  These have
been computed following the formalism described by \citet[][also see VandenBerg
et al.~2010, their \S 2]{cpf06}.}  As in the case of the Sarajedini et
al.~(2007) {\it HST} photometry, the \citet{bo07} spectrum of Vega was used
to set the zero points of the predicted BCs.  Because MARCS model atmospheres
were not computed for $\teff$\ values $> 8000\,$K, the aforementioned
transformations were supplemented by the BCs for hotter stars that were kindly
provided to us by S.~Cassisi (private communication, also see \citealt{bcc05}).
The latter, which apply to Castelli-Kurucz model atmospheres, were corrected by
amounts ranging from $-0.018$ to $+0.012$ mag, depending on the metallicity, in
order to ensure that there is good continuity of the BC values over the entire
temperature range that encompasses the stellar models.  (These numbers are the
average differences between the bolometric corrections in the two data sets at
8000 K, as calculated from the BC entries at $\log\,g$ values from 2.0 to 5.0.)
The only models in this study that have $\teff > 8000$ K are ZAHB models and,
as shown in \S~\ref{subsec:m5} and \S~\ref{sec:ages}, they generally provide
rather good fits to the HB populations of GCs that are nearly unreddened, when
the Schlegel et al.~(1998) $E(B-V)$ values are assumed; i.e., there is no
compelling evidence that the predicted colors should be further corrected in
any way.

When comparing isochrones and ZAHB loci with observed CMDs, we have assumed the
[Fe/H] values given by CBG09 for the GCs in our sample, along with
[$\alpha$/Fe] $= 0.46$ at [Fe/H] $\le -0.76$ (approximately the metallicity
where the [O/Fe] and [$\alpha$/Fe] versus [Fe/H] relations for field halo stars
change slope; e.g., \citealt{rmc12}).   At higher iron abundances,
[$\alpha$/Fe] has been assumed to decline linearly with [Fe/H]
in order to reach a value of 0.0 at [Fe/H] $= 0.0$.  We have also
adopted $Y = 0.2500$ at [Fe/H] $\le -1.0$, in good agreement
with the primordial helium abundance ($0.2485 \pm 0.0016$, \citealt[][see their
\S 4.8]{ksd11}), and slightly larger values at [Fe/H] $> -1.0$, to be consistent
with $\Delta\,Y/\Delta\,Z = 1.4$.  Approximately this value of the helium
enrichment parameter is obtained if $Y$ varies from 0.2485 at $Z = 0.0$ to the
initial $Y,\,Z$ values which are typically assumed in Standard Solar Models
(SSMs) that take diffusive processes into account (see, e.g., \citealt{bp95},
\citealt{trm98}).\footnote{We decided to use a slightly larger helium enrichment
factor than that implied by our SSM for the \citet{gs98} metals mixture because
the Victoria models do not take the diffusion of the metals into account ---
just the gravitational settling of helium (which is primarily responsible for 
the reduction in age at a given turnoff luminosity due to diffusive processes).
To satisfy the solar constraint, our SSM requires $Y_i = 0.2661$ and $Z_i =
0.0162$ for the initial helium and metals mass-fraction abundances, as well as
$\amlt = 2.007$ for the usual mixing-length parameter.  This model implies
$\Delta\,Y/\Delta\,Z \approx 1.1$.}  In this way, the values of [Fe/H],
[$\alpha$/Fe], and $Y$  specific to each GC have been set.  Isochrones for the
same values of these chemical abundance parameters were then obtained by
interpolating within the many grids of evolutionary tracks provided by
VBD12, using the code described by them.

\section{Ages Determined from $\delv$\ Observations}
\label{sec:methods}

\subsection{General Considerations}
\label{subsec:general}

The parameter $\delv$\ was originally (\citealt{san82}, \citealt{ir84}) used to
represent the difference in magnitude between the MS turnoff and the HB,
measured at the color of the TO.  However, it was appreciated early on that,
since a cluster fiducial is vertical at the TO (by definition) as well as a
slowly varying function of color over a fairly large range in magnitude above
and below this point, the turnoff luminosity and its variation with age cannot
be determined to very high precision.  To circumvent this difficulty,
\citet{cdk96} suggested that $\delv$\ be measured at a point that is brighter
and slightly redder than the turnoff (by 0.05 mag) as this would involve a
much smaller uncertainty.  The other well-known difficulty with the $\delv$
technique for measuring cluster ages is that, in many GCs, the horizontal part
of the HB is not populated; i.e., the core He-burning stars are located either
well to the blue of the instability strip or close to the RGB in a ``red clump".
However, in such cases, theoretical ZAHB loci may be used to extrapolate from
the observed distributions of HB stars to the color where $\delv$\ is evaluated,
especially if the same models perform well when applied to clusters of similar
metallicity in which their HB populations have colors that overlap those of
their TO and subgiant (SGB) stars.  Alternatively, a different method should be
used to determine the relative ages of GCs, such as a calibration of the
difference in color between the MSTO and the RGB in terms of age according to
the prescription described in the pioneering study by VandenBerg et al.~(1990).

Our implementation of the $\delv$\ method is more implicit than those versions
described above.  No attempt is made to determine the luminosity of the HB at
a particular color.  Rather, a theoretical ZAHB for the appropriate chemical
abundances is fitted to the lower bound of the observed distribution of HB
stars (since core He-burning stars are predicted to evolve to higher
luminosities as they age), after the observed colors have been dereddened, to
determine the apparent distance modulus.  Once the absolute magnitude scale has
been set in this way, there is only one isochrone for the assumed metallicity 
that will provide a simultaneous match to the observed turnoff color and the
beginning of the cluster SGB.  To identify it, isochrones for different ages
must each be shifted horizontally in color by whatever amount, if any, is
required to reproduce the observed turnoff color when they are overlaid onto the
photometric data.  If a given isochrone is too young or too old, its SGB segment
will be too bright or too faint, respectively, relative to the observed stars.
By iterating on the age, the best-fit isochrone can be readily determined.

The same thing can be accomplished by first shifting all of the isochrones for
an appropriate range in age to a common TO color (specifically, to the observed
turnoff color), and then overlaying the resultant grid onto the cluster CMD.
As before, the age of the cluster is equated to the age of the isochrone that
provides the best superposition of the subgiant stars just past the TO.   It
should be appreciated that the color adjustments which are applied to the models
do not affect the predicted age-luminosity relations for any point along the
isochrones (e.g., at the turnoff or at the location on the SGB that is 0.05 mag
redder than the TO).  These offsets are needed because, in general, the 
predicted and observed color scales will not be in perfect agreement due to,
for instance, errors in the color--$\teff$\ relations, uncertainties in the
treatment of some of the stellar physics ingredients (e.g., superadiabatic
convection, surface boundary conditions), or incorrect assumptions regarding
the cluster properties (reddening, distance, metallicity).  Only by matching the
predicted and observed TOs are we are able to reliably ascertain which isochrone
provides the best fit to an observed CMD in the vicinity of the TO, including
the turnoff luminosity.

The main difference between our approach and that advocated by Chaboyer et
al.~(1996) is that we fit isochrones directly to the observed distributions of
stars that have colors within approximately 0.05 mag of the turnoff ---
especially brighter than the TO, but also below it --- instead of relying on
just two points (i.e., the TO color and the selected SGB fiducial point) to 
infer an age.  In the case of the Chaboyer et al.~technique, one has no idea
how well the isochrones are able to match the morphology of the observed CMD,
though it is implicitly assumed that they do.  In fact, as shown below, modern
stellar models that take diffusive processes into account do reproduce the
shapes of GC CMDs in the vicinity of the turnoff very well.  As a consequence,
essentially identical ages would be obtained using either the method that we
have employed or that described by Chaboyer et al.

\begin{figure}[t]
\centering
\includegraphics[width=0.8\textwidth]{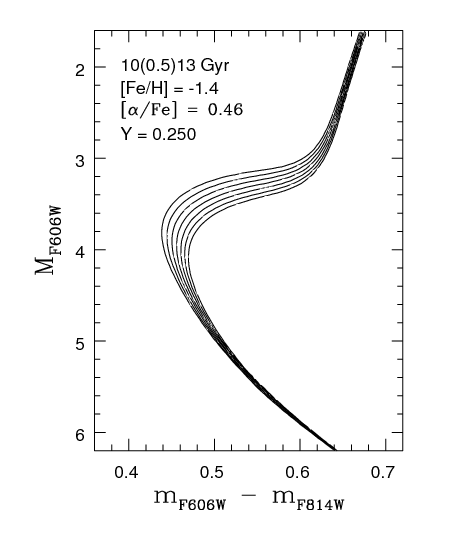}
\caption{Isochrones from VandenBerg et al.~(2012, GSCX model series) for the 
indicated ages and chemical abundances.  They have been transposed to the
observed plane using color transformations based on the latest MARCS model
atmospheres.} 
\label{fig:fig1}
\end{figure}

To better appreciate the reasons for, and the advantages of, comparing
isochrones with observations in the particular way that we have described, it
is helpful to consider several plots, beginning with Figure~\ref{fig:fig1}.
This illustrates the appearance on the [$(M_{F606W} -
M_{F814W}),\,M_{F606W}$]-plane of a representative set of isochrones; in this
case, for [Fe/H] $= -1.4$ (roughly the mean metallicity of the Galactic GCs),
[$\alpha$/Fe] $= 0.46$, $Y = 0.250$, and ages of 10 to 13 Gyr, in 0.5 Gyr
increments.  Due to the use of the state-of-the-art interpolation code developed
by P.~Bergbusch (see \citealt[][and references therein, as well as
the latest updates reported by VBD12]{bv01}), and to the well-behaved
derivatives of $\log\,L$ and $\log\,T_{\rm eff}$ with respect to time along the
computed evolutionary tracks, the isochrones have especially smooth
morphologies.

\begin{figure}[t]
\centering
\includegraphics[width=0.9\textwidth]{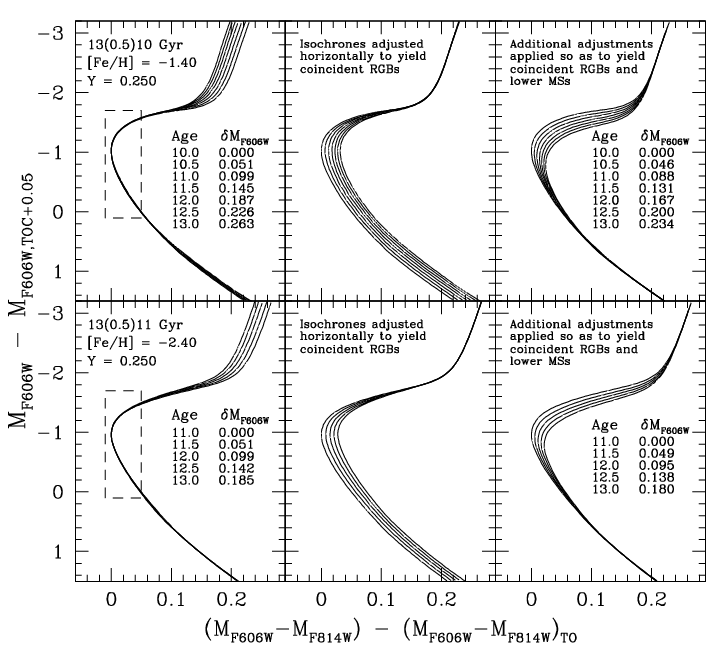}
\caption{{\it Top row}: isochrones from the previous figure have been shifted in
color and magnitude to achieve coincident turnoffs ({\it left-hand panel}), then
adjusted horizontally in color by the amounts that are needed in order for the
RGBs of all of the older isochrones to coincide with the location of the giant
branch of the 10 Gyr isochrone ({\it middle panel}), and finally corrected in
both magnitude and color (see the text) in order to obtain a simultaneous
coincidence of both their giant branches and their lower main sequences ({\it
right-hand panel}).  The legends in the {\it left-} and {\it right-hand panels}
list the differences in the turnoff magnitudes of the isochrones with ages $\ge
10.5$ Gyr relative to that of the 10.0 Gyr isochrone.  Note that the isochrones
superimpose each other nearly exactly in the vicinity of the turnoff (notably
the region inside the dashed box that has been plotted in the {\it left-hand
panel}).  {\it Bottom row:} as in the top row, except that 11--13 Gyr 
isochrones for [Fe/H] $= -2.40$ have been intercompared.}
\label{fig:fig2}
\end{figure}

In the upper left-hand panel of Figure~\ref{fig:fig2}, the 10 Gyr isochrone from
the previous figure has been replotted so that the abscissa gives the color
relative to the turnoff color and the ordinate specifies the magnitude with
respect to a point on the upper MS that is 0.05 mag redder than the TO. 
Isochrones for all of the other ages were then superimposed in such a way that,
after registering them to the same abscissa and ordinate zero points, additional
(small) vertical shifts were applied, if needed, in order to minimize the
differences in their magnitudes at the aforementioned color offset both above
{\it and} below the TO.  It is clear that, as the result of this centering
procedure, which yielded the $\delta M_{F606W}$ information provided in the
legend, all of the isochrones have essentially identical turnoff luminosities.
Thus, we have determined, for instance, that the turnoff luminosities of 10.5
and 13.0 Gyr isochrones are, respectively, 0.051 mag and 0.263 mag fainter than
that of the 10.0 Gyr isochrone.  These numbers are very precise, with
uncertainties at the level of $\lta \pm 0.005$ mag.  However, we do not need,
or use, these numerical results in our age-dating technique.  What is important
is that we have demonstrated that the shapes of isochrones within 0.05 mag of
the TO are essentially independent of age; consequently, there is no basis for
preferring one isochrone over another (in a given grid) from a morphological
perspective when they are restricted to comparisons with observed CMDs in the
vicinity of the turnoff.  The bottom left-hand panel shows that this conclusion
holds for other metallicities.

We realized, after creating the upper left-hand panel of Fig.~2, that we could
easily evaluate the errors associated with the MF09 ``relative main-sequence
fitting" (rMSF) method, at a fixed metallicity, through further manipulations of
the isochrones.  Beginning with the grid as plotted in that panel, all of the
older isochrones were shifted horizontally to the red until their RGB segments
overlaid that of the 10 Gyr isochrone --- resulting in the middle panel (top
row).  From this starting point, it is easy to achieve a simultaneous
coincidence of the RGB and lower MS segments of all of the isochrones, which is
the essence of the MF09 approach, by simply shifting them along a line that is
parallel to the giant branch.  This can be easily accomplished if a horizontal
line is drawn in the middle panel at, say, an ordinate value of 1.0 and the
difference in color between any two isochrones is evaluated from their
intersection with that line.  A triangle can then be defined in which this color
difference is the length of one of its sides and the angles at each vertex may
be determined from fact that the other two sides have slopes equal to those of
the MS and the RGB.  This is sufficient information that, with the aid of simple
trigonometry, the small horizontal and vertical shifts that are needed to obtain
the results shown in the upper right-hand panel are easily calculated.  (The
plots in the bottom row present similar results for isochrones computed on the
assumption of [Fe/H] $= -2.40$.)

The legends in the right-hand panels for both metallicities list the adjustments
in $M_{F606W}$ so derived.  These are the differences in the turnoff
luminosities, relative to that of the youngest isochrone in each grid, that are
obtained when isochrones are aligned according to the MF09 formalism.  (Note
that these results would be more uncertain if we had attempted to determine the
TO luminosities explicitly, instead of employing our indirect, but more accurate
and precise method.)  Interestingly, they agree quite well with the results
reported in the legends of the left-hand panels (particularly at the lowest
[Fe/H] value); and, as expected, the former are somewhat less than the latter.
Some compression of the range of TO luminosities for the same age range can be
expected because, as readily appreciated by considering Fig.~\ref{fig:fig1},
it is necesssary to move the older isochrones upwards and to the left, in a
direction parallel to the lower MS, in order to force their giant branches to
match the RGB of the youngest isochrone without causing any separation of their
respective lower main sequences.  The errors, which are small compared with
other sources of uncertainty, arise simply because the location of the RGB is
not independent of age, though the dependence is quite weak (as noted by MF09).
Indeed, Fig.~\ref{fig:fig2} provides encouraging support for the rMSF method,
at least when applied to GCs having similar metal abundances, which implies that
any problems with the MF09 study must arise in the connection of their results
for the different metallicity groups.  This is fully discussed in
\S~\ref{subsubsec:mf09}.

\begin{figure}[t]
\centering
\includegraphics[width=0.8\textwidth]{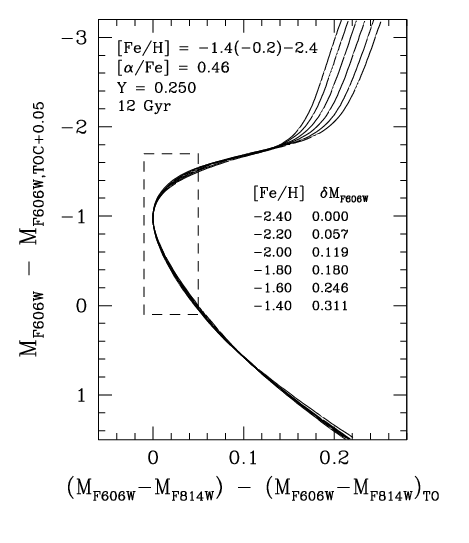}
\caption{As in the left-hand panels of the previous plot, except that
isochrones for a range in [Fe/H], instead of age, are intercompared.  The table
lists the differences in the turnoff magnitudes of the isochrones for [Fe/H]
values $\ge -2.20$ relative to that of the isochrone for [Fe/H] $= -2.40$.  
Note that the portions of the isochrones that are contained within the dashed
box are morphologically nearly identical, though there is some dependence of the
slope of the subgiant branch on [Fe/H], with the lowest metallicity isochrone
having the steepest slope.}
\label{fig:fig3}
\end{figure}

Figure~\ref{fig:fig3} is similar to the left-hand panels of Fig.~\ref{fig:fig2}
except that isochrones for the same age (12 Gyr), helium content, and
$\alpha$-element abundances, but different [Fe/H] values ($-2.4 \le$ [Fe/H]
$\le -1.4$, in 0.2 dex increments) are considered.  The thinness of the line at
the turnoff, as well as slightly above and below it, indicate that all of the
isochrones have exactly the same TO luminosities as a result of the color and
magnitude adjustments that have been applied.  (The table contained within the
plot lists the differences in the TO magnitudes between the more metal-rich
isochrones and that for [Fe/H] $= -2.4$.  We see from a comparison of these
results with those given in the left-hand panels of Fig.~\ref{fig:fig2} that the
effect on the turnoff luminosity of varying the metallicity by 0.2 dex is
slightly larger than the impact of varying the age by 0.5 Gyr.)  Although the
remaining parts of the isochrones that lay within the region enclosed by the
dashed rectangle are not quite as ``perfectly" coincident as they are at the TO,
the models span a much larger range in metallicity than one would normally
consider when comparing isochrones with the CMD of a given cluster.  Most would
agree that the [Fe/H] values of the majority of GCs are known to within $\pm
0.1$--0.2 dex (compare, e.g., the [Fe/H] determinations for clusters in common
to the studies by \citealt{zw84}, \citealt{ki03}, and CBG09), and had we
restricted Fig.~\ref{fig:fig3} to a plot of isochrones that span a range in
[Fe/H] of only 0.2--0.3 dex, a noticeable reduction in the morphological
variations near the TO would have been apparent.  Be that as it may, the main
conclusion to be drawn from Fig.~\ref{fig:fig3} is that the shapes of isochrones
(for a given heavy-element mixture) within $\sim 0.05$ mag of the turnoff are
predicted to be {\it nearly} independent of [Fe/H], at least at values of $-1.4$
and less.  

\begin{figure}[t]
\centering
\includegraphics[width=0.8\textwidth]{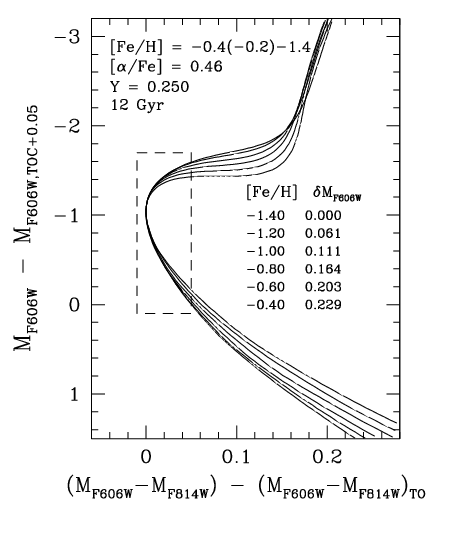}
\caption{As in the previous figure, except that isochrones for $-1.40 \le$
[Fe/H] $\le -0.40$ have been intercompared.  The table lists the differences in
the turnoff magnitudes of the isochrones for [Fe/H] values $\ge -1.20$ relative
to that of the isochrone for [Fe/H] $= -1.40$.}
\label{fig:fig4}
\end{figure}

At higher metallicities, the opacities in stellar interiors become larger at an
increasingly rapid rate, which impacts the predicted mass-luminosity relations
(i.e., the turnoff mass at a fixed age) to an ever greater extent.  As a result
of the latter, the subgiant branch of, say, a 12 Gyr isochrone becomes
noticeably flatter as the [Fe/H] value increases above $\approx -1.2$.  This is
illustrated in Figure~\ref{fig:fig4}.  (The drop in SGB luminosities occurs
because it takes more energy to expand the envelope of a more massive star as it
evolves towards the RGB.)  Interestingly, when measured at the same luminosity
offset from the turnoff, the MSTO-to-RGB color difference at a fixed age is
predicted to have no more than a slight dependence on metal abundance at [Fe/H]
$\gta -1.4$.

\begin{figure}[t]
\centering
\includegraphics[width=0.8\textwidth]{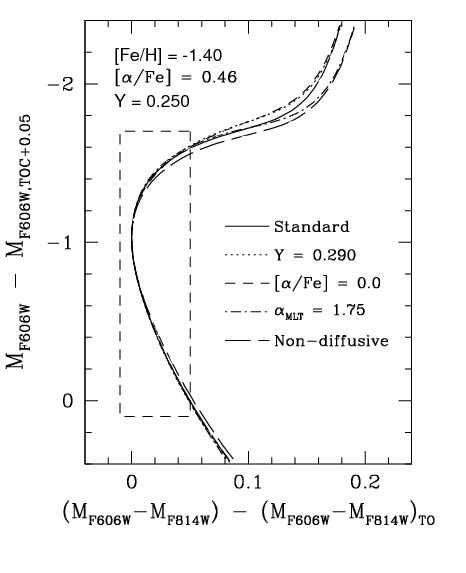}
\caption{As in the previous three figures; in this case, a 12 Gyr isochrone for
the chemical abundances specified in the top left-hand corner (the {\it solid
curve}) are compared with those for higher $Y$ (11.3 Gyr, {\it dotted curve}),
lower [$\alpha$/Fe] (13.5 Gyr, {\it short-dashed locus}), a smaller value of
the mixing-length parameter by 0.255 (12 Gyr, {\it dot-dashed curve}), and one
in which diffusive processes have been neglected (13.3 Gyr, {\it long-dashed
locus}), as indicated.  Different ages have been assumed in order that all of
the isochrones have approximately the same turnoff luminosity.}
\label{fig:fig5}
\end{figure}

As shown in Figure~\ref{fig:fig5}, the near invariance of the morphology of the
turnoff portions of isochrones for high ages is also found when $Y$,
[$\alpha$/Fe], and even the value of the mixing-length parameter, $\amlt$,
are varied.  Except for the beginning of the SGB (the part inside the dashed
rectangle), the shapes of isochrones for post-TO phases, as well as the
location of the RGB relative to the turnoff, clearly depend quite sensitively
on these parameters (as well as on age and [Fe/H]).  As a result, one should
not be concerned if the isochrone that is fitted to the turnoff observations
fails to reproduce the location of the giant branch: there are too many 
uncertainties that affect the predicted $\teff$\ and color scales to
expect perfect consistency.  The only models that we have found which are
appreciably offset from the others within the region enclosed by the dashed
rectangle are those that neglect diffusive processes (as represented by the
{\it long-dashed curve}), though the differences are still quite small.
However, helioseismic studies established the importance of this physics many
years ago (see, e.g., \citealt{cpt93}; \citealt{bpb97}; \citealt{trm98}).
The main point of
Figs.~\ref{fig:fig2},~\ref{fig:fig3}, and~\ref{fig:fig5} is that 
parameter variations appear to have little or no impact on the shapes of
isochrones near the turnoff (at least for ages $\gta 10$--11 Gyr), and hence
that our isochrone-fitting procedure has a solid footing.  Small differences
are, in any case, neither observationally detectable nor of significant
importance for the inferred ages.

\begin{figure}[t]
\centering
\includegraphics[width=0.85\textwidth]{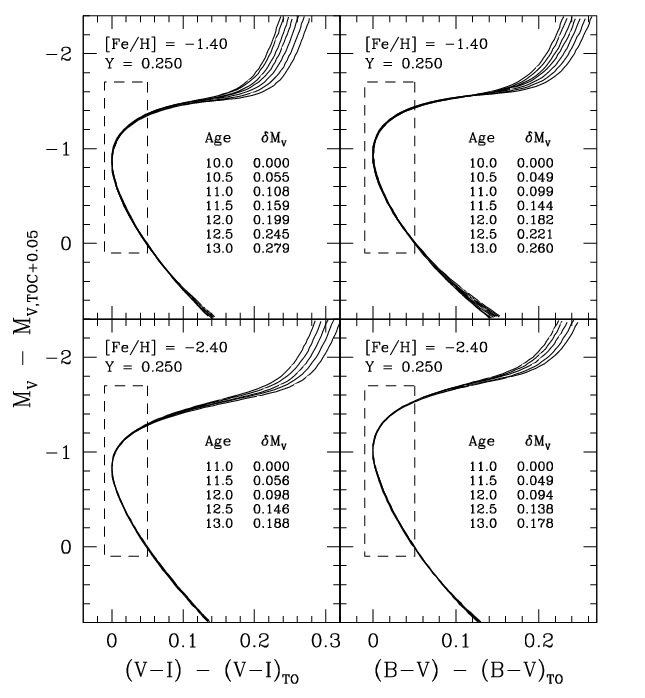}
\caption{As in the left-hand panels of Fig.~\ref{fig:fig2}, except that the
isochrones are compared on the $V-I,\,V$ and $B-V,\,V$ diagrams.}
\label{fig:fig6}
\end{figure}

It is worth pointing out that the same 
conclusions would have been reached had the isochrones been plotted on the
$V-I,\,V$ or $B-V,\,V$ diagrams.  For instance, the superpositions of the
various loci in Figure~\ref{fig:fig6} look no different from those shown in the
left-hand panels of Fig.~\ref{fig:fig2}.  The same thing will likely be found
when other photometric systems and filter bandpasses are used because the total
range in $\teff$\ spanned by $\sim 10$--13 Gyr between the TO and the beginning
of the SGB is quite small.  However, this is something that will need to be
checked on a case-by-case basis: it is beyond the scope of this study to do so
here.  Similarly, our results should not be extrapolated to ages outside the
ranges that we have considered.  Had we included isochrones for younger ages in
the aforementioned plots, some deviations between them and the older isochrones
in the vicinity of the turnoff would have been evident.  Our results have been
restricted to the age ranges that are relevant to the GCs in our sample.

\subsection{Application to M$\,$5}
\label{subsec:m5}

\begin{figure}[t]
\centering
\includegraphics[width=0.8\textwidth]{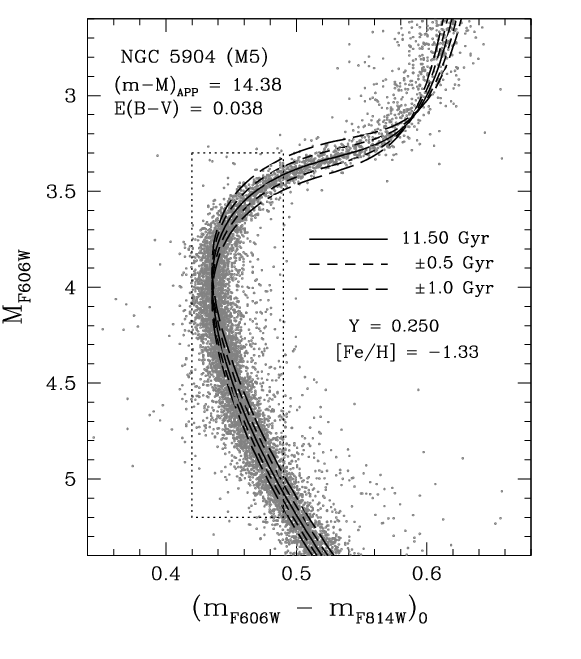}
\caption{A comparison of the CMD of M$\,$5 with isochrones for the indicated
chemical abundances when an apparent distance modulus of 14.38 and a reddening
corresponding to $E(B-V) = 0.038$ are assumed.  The {\it solid curve}
represents an 11.5 Gyr isochrone, which provides the best fit to the cluster
subgiants (notably those within the dotted rectangle), once it has been
corrected by $\delta (color) = -0.025$ mag, while {\it dashed loci} represent
isochrones that differ in age by $\pm 0.5$ and $\pm 1.0$ Gyr.  The latter have
also been arbitrarily shifted to the observed TO color: they will provide
equally good fits to the turnoff region of the observed CMD if the adopted
distance modulus is appropriately adjusted.}
\label{fig:fig7}
\end{figure}

In the case of a well-defined CMD, our implementation of the $\delv$\ method 
will provide a highly precise age, whose accuracy depends almost entirely on
the adopted distance and chemical abundances.  So-called 
``fitting errors" are not of significant importance.  Consider, for example,
the CMD of M$\,$5 (NGC$\,$5904) from the \citet{sbc07} database that is shown
in Figure~\ref{fig:fig7} for the region within $\sim \pm1.5$ mag of the
turnoff.  From least-squares fits to the median points that are derived after
the photometry has been sorted into 0.10 mag bins, the observed turnoff color
is found to be 0.4738.  (As long as this number has been determined to better
than $\sim 0.005$ mag, which is appreciably larger than its uncertainty, the age
that is found from an overlay of isochrones onto the turnoff photometry is not
significantly affected; see below.)  To place the data on the [$(m_{F606W} -
m_{F814W})_0,\,M_{F606W}$]-plane, we have assumed the indicated reddening and
an apparent distance modulus of 14.38, which is based on the fit of a ZAHB to
the cluster counterpart (to be discussed shortly).

To determine the age of M$\,$5, isochrones for $Y = 0.250$, [Fe/H] $= -1.33$
(CBG09), and enhanced $\alpha$-element abundances (see \S~\ref{sec:input}) were
generated for ages from 10.5 to 12.5 Gyr, in steps of 0.25 Gyr, once preliminary
fits had narrowed the age range.  Each isochrone was then shifted horizontally,
in turn, by whatever amount was needed in order for the predicted and observed
turnoff colors to match, until the one was found that provided the best fit to
the beginning part of the SGB (the stars within the dotted rectangle brighter
than $M_{F606W} \sim 4$).  Just by inspection, it is obvious that an isochrone
for 11.5 Gyr (the solid curve) reproduces the mean stellar distribution in the
vicinity of the TO very well and that isochrones for ages which differ by $\pm
0.5$ Gyr or $\pm 1.0$ Gyr (the dashed curves) do not.  The ``fitting" error,
which appears to be at the level of $\lta \pm 0.2$ Gyr, is a small fraction of
the total uncertainty associated with the derived age given that, in particular,
a change of $\pm 0.10$ mag in the assumed distance modulus would result in an
age that differs by about $\pm 1.0$ Gyr (see the tabular results in the
right-hand panels of Fig.~\ref{fig:fig2}).  Moreover, varying the adopted
abundances of helium or oxygen would affect the age at a given turnoff
luminosity by a small or a large amount, depending on the size of the variation;
see VBD12.  For instance, a change in the oxygen abundance by a factor of two
would alter the inferred age by $\sim 1.0$ Gyr.

Fortunately, the other heavy elements are much less important for predicted TO
luminosity versus age relations (see VBD12), though they do have some
impact on ZAHB-based distance determinations.  In the left-hand panel of
Figure~\ref{fig:fig8}, a ZAHB for the same chemical abundances that were assumed
in the previous figure has been fitted to the lower bound of the distribution of
HB stars in M$\,$5, yielding $(m-M)_{APP} = 14.38$.  As shown in the right-hand
panel, an equally good fit to the cluster HB is obtained on the assumption of 
an apparent modulus of 14.34 and [Fe/H] $= -1.18$ (instead of $-1.33$), with
the values of $Y$ and [$\alpha$/Fe] left unchanged.  The ages that are obtained
by matching the corresponding isochrones onto the TO observations differ by only
0.25 Gyr.  Thus, a 0.15 dex change in the adopted [Fe/H] value does not have
major consequences for the derived age, though (of course) the {\it actual}
distance and reddening of M$\,$5 must be known in order to determine the true
age of this system.  However, this is not a concern for {\it relative} GC ages.
In fact, it is a further advantage of using the $\delv$\ method to determine
cluster-to-cluster variations in age that the inferred ages have a reduced
dependence on the adopted metal abundance --- because both the HB and the TO,
at a fixed age, become fainter as the [Fe/H] value increases, albeit not at
exactly the same rate.  On the other hand, as shown in our analysis of the
M$\,$13 CMD in \S~\ref{subsubsec:m3m13}, $\delv$\ ages are highly sensitive to
the assumed helium content given that the luminosity of the HB is predicted to
be a strong function of $Y$.

As in the case of M$\,$5, we have found that the best-fit isochrones generally
need to be offset to the blue by 0.01--0.025 mag (see \S~\ref{sec:ages}) in
order to reproduce the observed turnoff colors, even though our ZAHBs appear to
be able to match the cluster HBs without such color adjustments.  Indeed,
throughout this investigation, the $\delta$(color) values that are specified in
plots similar to Fig.~\ref{fig:fig8} have been applied only to the isochrones,
not to the ZAHB loci.  (This practice will have little, or no, impact on our
results because our distance determinations are based primarily on the nearly
horizontal part of a ZAHB.  If small color offsets were applied to our ZAHB
models, they would have noticeable consequences only at the blue end.)  The
cause of such discrepancies is not known, though one can speculate that, among
other possible explanations, the assumed [Fe/H] or [O/Fe] values are slightly
too high, the gravity and/or temperature dependencies of the adopted color
transformations are not quite right, or there are minor problems with, say, the
treatment of the diffusion and extra-mixing physics or of the atmospheric
boundary condition.\footnote{In fact, preliminary work has revealed that 
evolutionary tracks and isochrones for low metallicities are hotter/bluer by
approximately the requisite amount if MARCS model atmospheres (\citealt{gee08})
are attached to the interior structures at the photosphere or at an optical
depth $\tau = 100$.  Thus, we have good reason to suspect that the treatment
of the atmosphere is mainly responsible for this difficulty.}  However, a
resolution of this matter is not needed here because discrepancies between the
predicted and observed turnoff colors do not affect the derived $\delv$\ ages,
{\it provided} that the isochrones are corrected for them when the age is
evaluated. 

This is demonstrated in Figure~\ref{fig:fig9}.  The middle panel is equivalent
to Fig.~7, except that isochrones for the same chemical abundances have been
plotted for ages of 10 to 13 Gyr, in steps of 1 Gyr, have not been shifted in
the horizontal direction to a common TO color.  The models clearly provide
a very good match to the observed morphologies of the MS and SGB and, as deduced
from the earlier plot, the implied age of M$\,$5 is very close to 11.5 Gyr.  In
the left- and right-hand panels, the same isochrones have arbitarily been 
shifted by 0.015 mag to the red and to the blue, respectively, to force a 
mismatch between theory and observations.  Indeed, if we had not shifted our
isochrones by 0.025 mag to the blue (middle panel), their superposition onto 
the cluster CMD would have looked similar to that shown in the left-hand panel,
but appearing even more discrepant.  On the other hand, the case shown in the
right-hand panel bears considerable similarity to the fits of isochrones to the
NGC$\,$3201 and M$\,$30 CMDs that were reported by Dotter et al.~(2010).
Regardless of how well the models reproduce the colors of real stars, these
sorts of discrepancies could easily arise if, for instance, the assumed
reddenings are wrong.

It is clearly much more difficult to derive the age of M$\,$5 if the isochrones
do not fit the MS and TO observations.  Whereas an age near 11 Gyr seems to be
indicated by the overlay of the models onto the observed SGB in the left-hand
panel, a significantly higher age (at least 12 Gyr) would be favored by the
comparison shown in the right-hand panel.  These estimates differ from the
correct age (for the assumed distance and chemical abundances) by about 0.5 Gyr.
The reason why isochrones are compared with observed CMDs is to determine which
one provides the optimum match to the locus of stars from just below the
turnoff to the beginning of the SGB, as this isochrone is presumably the one
that best reproduces the observed turnoff luminosity.  This can be evaluated
only if each isochrone that is considered is first adjusted in color, if
necessary, so that the predicted and observed TO colors agree.  The fact that
Dotter et al.~(2010) failed to do this in the case of NGC$\,$3201 and
M$\,$30 is the reason why the ages that they determined for these GCs
(and possibly other metal-deficient clusters) are too high for the assumed
distances by $\sim 0.5$ Gyr, judging from the results presented in
Fig.~\ref{fig:fig9}.  [In comparing isochrones with observed CMDs, Dotter et
al.~(see their \S 4.2) applied ``minor adjustments" to initial estimates of
[Fe/H], distance modulus, and reddening from the Harris (1996; 2003 revision)
catalog ``to improve the fit to the unevolved main sequence first and the RGB
second".  As discussed in \S~\ref{subsubsec:caution}, there is no reason to
expect that such fine-tuning will lead to improved interpretations of observed
CMDs, especially if the resultant overlays of the models onto the TO and SGB
observations make it difficult to identify the best-fit isochrone.]  

\section{The ZAHB-based Distance Scale}
\label{sec:zahbdis}

It is relatively straightforward to apply the $\delv$\ method described above
to at least 30 of the GCs in the Sarajedini et al.~(2007) sample, less so for
most of the rest.  However, before considering those clusters with the most 
easily fitted HB populations, it is important to show that the luminosities of
our ZAHBs are in satisfactory agreement with observational constraints, as this
implies that the derived distance moduli (and hence ages) are reasonably
accurate in an absolute sense.  The open circles in the top panel of
Figure~\ref{fig:fig10}, which are connected by the solid curve, give the
predicted $M_V$ values for the ZAHB models at each of nine [Fe/H] values from
$-2.40$ to $-0.80$, assuming $Y = 0.25$.  They were evaluated at $\log\,\teff =
3.85$, which is characteristic of variable stars near the center of the
instability strip (see, e.g., \citealt{san90a}, \citealt{cac13}).  Since the
mean magnitude of the RR Lyrae stars in GCs that have [Fe/H] $\approx -1.5$ is
about 0.10 mag brighter than the ZAHB (\citealt{san93}), the dotted curve,
which is obtained if the solid curve is shifted upwards by 0.10 mag, should
be compared with empirical determinations of RR Lyrae luminosities at similar
metal abundances.

From an analysis of $> 100$ RR Lyraes in the Large Magellanic Cloud (LMC),
\citet{cgb03} obtained $<$$V_0$$>\ = 19.064 \pm 0.064$ at [Fe/H] $= -1.5$,
together with $\Delta\,M_V/\Delta\,$[Fe/H] $= 0.214 \pm 0.047$.  These findings,
which are based on stars that have $-1.8 \lta$ [Fe/H] $\lta -1.2$ (with only a
few outliers), are represented by the dashed line and the attached errorbar if
it is assumed that the true distance modulus of the LMC is 18.50 mag
(\citealt{fm10}, \citealt{ljp12}, \citealt{mrm12}).  Both the
slope and the zero-point clearly agree quite well with the predictions of our
models.  However, based on new astrometric data for five field RR Lyrae
variables from the Fine Guidance Sensors on the {\it HST},
\citet[see their table 10]{bmf11} derived $M_V =
0.45 \pm 0.05$ at the reference metallicity.  This is just barely consistent at
the $1\,\sigma$ level with the Clementini et al.~results (see
Fig.~\ref{fig:fig10}), athough the differences are negligible in the case of RR
Lyr, which is the star with the smallest errorbar in the Benedict et al.~sample.
For this variable, the latter obtained $M_V = 0.54 \pm 0.07$, which places it
on the dashed line if it has the metallicity that they adopted ([Fe/H] $=
-1.41$).  On the other hand, UV$\,$Oct ($M_V = 0.35 \pm 0.13$, [Fe/H] $= -1.47$)
is brighter than RR Lyr by nearly 0.2 mag, which leads one to wonder if it is
in an advanced evolutionary stage that is much more luminous than the ZAHB for
its metal abundance.  (This is quite possible since, for instance, about 14\%
of the RR Lyrae pulsators in M$\,$3 are overluminous according to
\citealt{ccc05}.)  Just one such star would have an impact on the Benedict et
al.~determination of the mean value of $M_V$ and its uncertainty because their
sample contains only a few stars.  On the other hand, improved consistency
between the Benedict et al.~and Clementini et al.~results would be obtained if
the LMC has $(m-M)_0 > 18.50$. 

Statistical parallax and Baade-Wesselink studies have favored considerably
fainter absolute magnitudes for RR Lyrae stars.  Using the first of these two
methods, \citet{gp98} obtained $M_V = 0.77 \pm 0.13$ at [Fe/H] $= -1.60$, which
may be compared with $M_V = 0.74 \pm 0.14$ at [Fe/H] $= -1.53$ (\citealt{fbs98})
from the application of the second technique.  These results, which are
represented, in turn, by the open and filled triangles in Fig.~\ref{fig:fig10},
would imply rather short distance scales and high ages (greater than the age of
the universe) if they are accurate.  However, the errorbars are large enough
that they do not present a problem at a $\gta 1$--$2\,\sigma$ level.  From our
perspective, it is encouraging that the dotted curve provides a reasonable
compromise of all of the above results, though its uncertainty in the vertical
direction is quite large ($\sim \pm 0.10$ mag).

Insofar as the slope of the variation of $M_V$(RR) with [Fe/H] is concerned,
we (and Benedict et al.~2011) are inclined to favor the value derived by
Clementini et al.~(2003) because it is based on so many variable stars.  Worth
mentioning is the fact that BASTI models (\citealt{pcs06}) and those reported
by \citet{cps04} predict almost the same slopes as our models.  Although not
shown here, the former are $\sim 0.07$ mag brigher than ours at [Fe/H] $= -1.5$,
while the latter are fainter by $\sim 0.03$ mag, but both run roughly parallel
to the dotted and dashed loci in the upper panel of Fig.~\ref{fig:fig10}.  It
should be kept in mind, however, that there is also significant support for 
$\Delta\,M_V/\Delta\,$[Fe/H] $> 0.214$ (see the recent review by
\citealt{cac13}).  

Further support for our models is provided in the bottom panel of
Fig.~\ref{fig:fig10}.  This compares the predicted luminosities (at $\log\,\teff
= 3.85$) from the same ZAHB models that were plotted in the upper panel with the
values derived by \citet{dsc99} for seven GCs from an analysis of the
pulsational properties of member $ab$-type variables.  Although De Santis \&
Cassisi adopted a different metallicity scale in their investigation, they state 
that the uncertainties in this scale do not significantly affect their 
evaluation of $\log\,L_{3.85}^{ZAHB}$.  Consequently, we have adopted the latest
[Fe/H] values from CBG09, as assumed throughout this study, in preparing this
plot.  Except in the case of NGC$\,$1851, which is more luminous than the
theoretical locus at its metallicity by $> 3\,\sigma$, the solid curve provides
a satisfactory fit to the data.  (A possible explanation of the relatively high
luminosity of the NGC$\,$1851 HB is that this system has a somewhat higher
helium abundance than the other GCs considered by De Santis \& Cassisi.)  On
the other hand, it could be argued from these data that the dependence of
$\log\,L_{3.85}^{ZAHB}$ on [Fe/H] is somewhat shallower than that predicted by
our ZAHB models.

To conclude this section, a few remarks concerning subdwarf-based distances are
warranted.  As already mentioned in \S~\ref{sec:input}, our isochrones reproduce
the properties of local subdwarfs with accurate distances from {\it Hipparcos}
quite well (see VandenBerg et al.~2010).  In fact, they do so without requiring
any $\teff$\ or color shifts whatsoever, if the temperatures derived for them by
Casagrande et al.~(2010) and the color-$\teff$\ relations based on the latest
MARCS model atmospheres (Gustafsson et al.~2008) are assumed.  Yet, as shown in
the next section, the same isochrones must be adjusted to the blue by $\sim
0.02$ mag, on average, in order for them to match the observed turnoffs of GCs
when our ZAHBs are used to determine the cluster distance moduli.  The
direction of these offsets is such that, were we to use the field subdwarfs to
evaluate the cluster distances through the MS-fitting technique, we would
obtain larger values of $(m-M)_{APP}$ by $\approx 0.10$ mag (on average).
However, the local field subgiant HD$\,$140283 suggests that, if anything, our
ZAHB-based distances are already too high (see VandenBerg et al.~2002,
\citealt{bnv13}).  This apparent inconsistency has yet to be resolved, though
it may indicate, for instance, that GC metallicities, which are based primarily
on bright giants, are not on precisely the same scale as those of the nearby
field halo dwarfs, that there are some important differences in the chemical
abundances of cluster and field stars of similar [Fe/H], or $\ldots$

This problem is not new, as the same difficulty was reported by \citet{bv01}
in their analysis of field subdwarfs using temperatures that were derived for
them primarily by \citet{gcc96}.  The Gratton et al.~and Casagrande et
al.~$\teff$\ scales are quite similar: both are considerably hotter than, e.g.,
the \citet{aam96} scale, and the assumption of either one enables our models to
satisfy the subdwarf constraint.  In principle, we {\it should} adopt [Fe/H]
values for the GCs which are close to those reported by \citet{cg97} (instead
of those given by CBG09) since they are based on the hot $\teff$\ scale of
Gratton et al, but doing so would increase the difficulty of matching the 
turnoff colors.  That is, the assumption of more metal-rich isochrones in fits
to a given CMD would require even larger blueward color offsets in order to
reproduce the observed photometry.  This is obviously a complex and important
problem that needs to be resolved as soon as possible.  In any case, because
distances based on local subdwarfs are much more dependent on the assumed
$\teff$\ and [Fe/H] scales than ZAHB-based distances (see, e.g., the caption of
Fig.~\ref{fig:fig8}), we have chosen to rely on the latter.  However, the
associated distance moduli uncertainties are at least $\sim \pm 0.10$ mag.

\section{Globular Cluster Age Determinations}
\label{sec:ages}

Although the number of globular clusters that contain multiple, chemically
distinct, discrete stellar populations appears to be increasing with time,
they are {\it obviously} present in the optical CMDs of only a few of them,
including $\omega\,$Cen (e.g., \citealt{bbp10}), NGC$\,$2808 (\citealt{pba07}),
NGC$\,$1851 (\citealt{mbp08}), and M$\,$22 (\citealt{mmp09}).  They have been
found in several others, but the use of ultra-violet (e.g., broad-band $U$) or
specialized filters (e.g., see the investigations of M$\,$4, NGC$\,$288, and
M$\,$2 by \citealt{mvp08}, \citealt{rlj11}, and \citealt{lpm12}, respectively)
and/or very careful and precise photometry near the cluster centers has been
required to discover them; see, the studies of 47 Tuc by \citet{apk09} and
\citet{mpb12}, of NGC$\,$6752 by \citet{mpk10}, and of NGC$\,$6397 by
\citet{mmp12}.  The pioneering Str\"omgren work that F.~Grundahl carried out
more than 10 years ago should also be acknowledged since the measured $c_1$
indices in every GC that he observed revealed the existence of large
star-to-star variations in the abundance of nitrogen at sufficiently faint
magnitudes on the giant branch (see \citealt{gvb00}, \citealt{gb01}) that
they must have been present in the gas out of which the current MS and lower
RGB stars formed.  In fact, C--N and O--Na anticorrelations appear to be a
universal property of all globular clusters (\citealt{cbg10b}).

For the vast majority of the GCs, such chemical abundance variations do not 
have a significant impact on the optical CMD analyses presented here.  With few
exceptions, the total [(C+N+O)/Fe] abundances in the cluster stars appear to be
constant to within the observational errors (e.g., \citealt{ssb96}), despite the
omnipresence of CN-weak and CN-strong stars, and the [Fe/H] variations are known
to be very small (\citealt{kra94}, CBG09).  In fact, it is to be expected that
the {\it HST} CMDs considered here generally will exhibit relatively little
scatter because the adopted color index, ($m_{F606W} - m_{F814W}$) is similar
to Johnson-Cousins $V-I_C$ in not being particularly sensitive to metallicity.
Such CMDs can discriminate between stellar populations that have significantly
different helium and C$+$N$+$O abundances because these elements affect the
H-R diagram locations and morphologies of isochrones, but $uv$ photometry
(preferably narrow-band, perhaps centered on a suitable spectral feature)
appears to be needed to detect CMD splittings due to O--Na or Mg--Al
anticorrelations (see \citealt{ssw11}, \citealt{cmp13}).  It is still of some
interest to compare isochrones and ZAHB loci for normal He and CNO abundances
with the CMDs of GCs that are known to have enhanced abundances of these
elements in order to obtain a visual impression of how well such models
reproduce the observations.  Consequently, such comparisons are presented below
for, e.g., NGC$\,$1851 and NGC$\,$2808, but not for $\omega\,$Cen, which is
dropped from further consideration because of the wide range in [Fe/H] that is
found in this system.  (We note, however, that a detailed investigation of
{\it HST} photometry of $\omega\,$Cen, using the same Victoria-Regina isochrones
as those employed here, and of its chemical evolution, has been recently carried
out by \citealt{hvn12}.)

Section~\ref{subsec:std} presents and discusses our results for 24 GCs that are more
metal-poor than [Fe/H] $= -1.0$, according to CBG09, and that have HB
morphologies which can be fitted most reliably by our models.  Because this
procedure does involve some level of subjectivity, the University of Victoria
participants in this project (DAV, RL, and KB) independently determined the
cluster distance moduli and ages following the methods described in
\S~\ref{sec:methods}.  In general, the three
estimates of the ages so obtained agreed to within $\pm
0.25$ Gyr: no cases were found where the derived ages differed by
$> 0.5$ Gyr.  (These age differences, which provide a measure of the internal
uncertainty of our age-dating procedure, are listed in the summary table in
\S~\ref{sec:summary}.)  Similar good agreement was found for 10 clusters
that have [Fe/H] $> -1.0$ (see \S~\ref{subsec:mrich}), once it was decided how
best to fit a ZAHB with an appreciable vertical component at the red end to the
observed HB ``clump" that is characteristic of metal-rich systems.  Most of the
remaining GCs in our sample are considered in \S~\ref{subsec:msrg}, where the
difference is color between the turnoff and the lower giant branch is used as
the primary age constraint.  The CMDs of a few clusters, including those
associated with the Sagittarius dwarf galaxy, are analyzed in 
\S~\ref{subsec:other}.  Section~\ref{sec:summary}, tabulates the ages that have
been derived for all 55 GCs that have been considered in this study, describes
the resultant age--metallicity and age versus Galactocentric distance relations,
reconciles our findings with those  reported by MF09, and compares various
properties of the globular clusters in an attempt to understand why the most
metal-deficient clusters appear to be split into two groups.

\subsection{Metal-poor (${\rm [Fe/H]} < -1.0$) Globular Clusters with Easily
Fitted HB Populations}
\label{subsec:std}

Figure~\ref{fig:fig11} illustrates that our ZAHB models provide a good match to
the luminosities and colors of the HB populations that belong to six of the most
metal-deficient GCs (those with [Fe/H] $< -2.2$) on the assumption of the
distance moduli and reddenings that are specified in the upper left-hand corner
of each panel.  Except for the fact that the observed RGBs tend to be somewhat
bluer than the model predictions, the isochrones also reproduce the cluster CMDs
quite well, yielding ages that range from 12.0 Gyr, in the case of M$\,$68
(NGC$\,$4590), to 12.75 Gyr, in the case of M$\,$92 (NGC$\,$6341), M$\,$15
(NGC$\,$7078), and M$\,$30 (NGC$\,$7099).  The other two clusters are predicted
to have intermediate ages, resulting in a mean age of 12.5 Gyr for the entire
group.  The adopted $E(B-V)$ values are identical to those found from the
Schlegel et al.~(1998) dust maps, though a somewhat smaller value may be
applicable to M$\,$15 given that it may otherwise be hard to explain why the
$\delta$(color) value, which is the discrepancy between the predicted and
observed turnoff colors, is significantly larger for this cluster than for
M$\,$92 or M$\,$30, which have very similar metallicities (unless, perhaps,
M$\,$15 has a higher helium abundance).  If the $E(B-V)$ value of M$\,$15 were
reduced by about 0.02 mag, improved overall consistency would be obtained
insofar as all of the best-fit isochrones would have to be adjusted to the blue
by 0.015--0.020 mag in order to match the observed TO colors.

The difficulty of obtaining accurate {\it absolute} ages is highlighted by the
fact that recent determinations of the age of M$\,$92 do not agree particularly
well.  For instance, \citet{btg12}, who were primarily interested in the ages of
three ultra-faint dwarf (UFD) galaxies {\it relative} to that of M$\,$92,
derived an age of 13.7 Gyr for the latter.  The higher age, by nearly 1 Gyr
compared with our estimate, is due to their adoption of a shorter distance
modulus by about 0.03 mag as well as a lower (absolute) oxygen abundance by
0.24 dex.  Similarly, differences in the assumed distances are the main reason 
why VandenBerg et al.~(2010) and \citet{dic10} obtained ages of 13.5 Gyr
and 11.0 Gyr, respectively, for M$\,$92.  However, the adopted distance
scale in the present study is consistent with that implied by the RR Lyrae
standard candle (though the uncertainties are large; see the previous section).
In addition, we favor a moderately high [O/Fe] value because the latest non-LTE
analyses of the spectra of extremely metal-deficient field halo stars have
yielded [O/Fe] $\sim 0.6$ at [Fe/H] $\lta -2$ (Ram\'irez et al.~2012; see their
Fig.~3), and because a relatively high oxygen abundance appears to be needed for
ZAHB models to match the reddest HB stars in M$\,$15. 

In fact, it is apparent in Fig.~\ref{fig:fig11} that our ZAHB models do not
extend far enough to the red to match those stars, which suggests that M$\,$15
may have an even higher O abundance than we have assumed --- see, e.g., 
\citealt{vb01}, who have investigated the impact of varying the O abundance on
computed ZAHB loci for extreme Population II stars.  (The reddest HB stars in
the other five GCs considered in this figure are likely evolved stars given
their small numbers and somewhat elevated luminosities.)  Whether or not M$\,$15
and M$\,$92 have significantly different oxygen abundances is not known, but
just as the measured O abundances in field stars at a given [Fe/H] show a
scatter of $\sim 0.2$ dex (e.g., \citealt{fna09}), there could well be similar 
cluster-to-cluster variations.  

Similar analyses are presented in Figure~\ref{fig:fig12} for GCs that have
$-1.50 >$ [Fe/H] $\ge -2.20$.  It is not necessary to say very much about these
results, given that each panel is self-explanatory.  The most robust ages are
those which have been determined for clusters with low reddenings and/or that
have substantial red HB populations.  For the seven clusters with [Fe/H] values
between $-1.50$ and $-1.80$, the derived mean age is 11.7 Gyr.  Higher ages are
predicted for the two more metal-poor systems in this sample, but most of the
clusters in the $-1.80 >$ [Fe/H] $\ge -2.20$ metallicity bin have extremely blue
HBs.  Their ages are determined in \S~\ref{subsec:other}.  As in
the previous figure, Schlegel et al.~(1998) $E(B-V)$ values have been adopted,
except for NGC$\,$3201, where a higher reddening by 0.04 mag has been assumed
in order to accommodate the location of the bluest HB stars while preserving a
good fit to the lower bound of the distribution of red HB stars.  It is a
concern that the $\delta$(color) offsets, which are needed for the best-fit
isochrones to match the observed turnoffs, vary as much as they do (from
$-0.011$ to $-0.040$), but what this means is not clear.  It is possible that 
the variations in $\delta$(color) arise from errors in the adopted [Fe/H]
values.  Alternatively, it may be the assumed reddenings that need to be
revised.

Suppose, for instance, that the actual reddening of NGC$\,$5286 is lower than
the Schlegel et al.~value by 0.010 mag.  In order to achieve a comparable fit to
the blue HB as that shown in Fig.~\ref{fig:fig12}, the apparent distance modulus
would have to be increased by 0.10 mag, resulting in a significantly reduced age
(to 11.0 Gyr).  Although not shown here, that isochrone provides an equally good
fit to the turnoff observations, assuming $\delta$(color) $= -0.025$ mag, as
that obtained by the 12.0 Gyr isochrone if $(m-M)_{APP} = 15.94$ and
$\delta$(color) $= -0.040$ mag. However, if the higher distance modulus is more
accurate, then all of the stars near the red end of the densest part of the HB
population in NGC$\,$5286 would lie well above the ZAHB; i.e., they would all
have to be evolved stars.  This seems less likely than in the adopted fit,
wherein the faintest of these stars are adjacent to the ZAHB, simply because
there are so many stars in this group: the densest concentration of HB stars
should be near the ZAHB where the evolutionary timescales are the longest.
Regardless of which variation of the $\delv$\ method is implemented, the ages
that are inferred for some GCs will depend on the assumption that is made
concerning the evolutionary state of the HB stars that are
observed.\footnote{Our ZAHB models were sufficiently relaxed/evolved over many
short timesteps that they should mark the beginning of the slowest part of the
core He-burning phase.  Still, it would be much more instructive to compute
synthetic HB distributions for comparisons with the observed HB populations
(see, e.g., \citealt{ccc04}, Catelan et al.~2004) as they have the potential to
discriminate between different possible interpretations of the data.  Such work
will be carried out once a subroutine to deal with semi-convection has been
implemented in the Victoria code.}

There are no such difficulties when globular clusters contain prominent red HB
populations, such as those that are the subject of Figure~\ref{fig:fig13}.
Indeed, the $\delv$\ ages for most of the clusters in this group should be
especially robust.  The exceptions are NGC$\,$2808 which, as already noted,
contains discrete stellar populations, and M$\,$107 (NGC$\,$6171), which suffers
from significant differential reddening.  (The difficulty of selecting the
best-fit isochrone in the latter case is most apparent when a much larger plot
of its CMD is examined.)  The ZAHB does match the morphology of the HB
populations of NGC$\,$2808 quite well on the assumption of the Schlegel et
al.~(1998) reddening, despite the known chemical abundance variations in it.
Furthermore, its ZAHB-based distance should be of quite high precision
given that the lower bound to the distribution of its reddest HB stars is
so well-defined.  In any case, if the two problematic clusters are ignored,
the mean age of the remaining GCs, which have $-1.07 \ge$ [Fe/H] $\ge -1.44$,
is 11.6 Gyr.  Interestingly, the ages of NGC$\,$6362, NGC$\,$6717, and
NGC$\,$6723 (12.5, 12.5, and 12.25 Gyr, respectively) are similar to those
found for the most metal-deficient GCs, while most of other [Fe/H] $\approx
-1.3$ clusters have ages of $\lta 11.0$ Gyr, which are less than any of the
ages that have been determined at lower metallicities.  

Given the compelling evidence that NGC$\,$2808 contains stars with different
helium abundances (perhaps up to $Y = 0.40$, see \citealt{pba07}), it is perhaps
not surprising that isochrones for $Y = 0.25$ require an especially large
blueward color correction in order to match the turnoff color of this cluster.
(On the other hand, most of the spread in its CMD occurs below the turnoff;
consequently, the median TO color, to which the best-fit isochrones are matched,
could well be representative of the dominant, normal-$Y$ sub-population.)  In
fact, even larger $\delta$(color) values would have been obtained for
M$\,$4 (NGC$\,$6121) and M$\,$107 had we adopted Schlegel et al.~(1998)
$E(B-V)$ values for them.   However, in these two cases, it seems more likely
that such large color offsets are due to an over-estimation of the dust-map
reddenings, which are very high ($\gta 0.45$ mag), than to some other
explanation.  Hence, we decided to adopt smaller $E(B-V)$ values for M$\,$4 (see
\citealt{hsv12}) and M$\,$107 by 0.03--0.05 mag in order that the resultant
$\delta$(color) values are much more consistent with those obtained for nearly
unreddened GCs of similar metallicity.  NGC$\,$6723 is the only other cluster in
this group where the Schlegel et al.~reddening, $E(B-V) = 0.177$, was not
adopted.  Indeed, such a high value is unrealistic since, if the distance
modulus is determined from the horizontal part of the distribution of HB stars,
the location of the blue tail can be matched by the same ZAHB {\it only} if
$E(B-V) \approx 0.07$.  (Since the dust maps provide ``line of sight"
reddenings, it may be possible to reconcile these, or similar, discrepancies if
a large fraction of the absorbing gas/dust is behind the GC in question, rather
than being entirely in front of it.)

No attempt has been made here (but see \citealt{pba07}, \citealt{mpb12c})
to try to fit isochrones for higher $Y$ to the CMD of
NGC$\,$2808, in part because the photometry that we have used does not separate
the observations into discrete main sequences.  However, as shown by Piotto
et al., the latter merge into a relatively narrow, single locus of stars
near the turnoff, resulting in a well-defined (single) SGB.  This morphology is
fully consistent with the predicted behavior of isochrones for a large range in
helium abundance, but essentially the same age (see VBD12).  In view of this,
we expect that our fit of isochrones to the median turnoff color and to the
distribution of SGB stars just redder than the TO will have yielded quite a 
good estimate of the cluster age (11.0 Gyr, see Fig.~\ref{fig:fig13}).  In the
case of NGC$\,$1851, the double subgiant branch discovered by \citet{mbp08} is
clearly present in our data and we have fitted isochrones to the dominant,
brighter population (although this is not apparent at the scale of the plot in
the top right-hand corner).  The reduced luminosity of the
fainter SGB population is most easily explained if the stars in that sequence
have higher C$+$N$+$O abundances than in the majority of the clusters stars
(see \citealt{csp08}).

\subsection{Globular Clusters that have ${\rm [Fe/H]} \ge -1.0$}
\label{subsec:mrich}

The most well-studied GC in the metal-rich group is 47 Tucanae.  In addition to
being nearly unreddened, its distance modulus has been determined to relatively
high precision by \citet{tkr10}, who obtained $(m-M)_V = 13.35 \pm 0.08$ from
their analysis of the eclipsing binary member V69.  As this estimate depends on
the assumed reddening and adopted color--$\teff$\ calibration, we used their
derived radii, masses, colors, and magnitudes for V69, along with $E(B-V) =
0.032$ (Schlegel et al.~1998) and [Fe/H] $= -0.76$ (CBG09), to obtain an
apparent distance modulus for 47 Tuc of $(m-M)_V = 13.35$ if the
color--temperature relations of \citet{crm10} are employed, or 13.33 if the
MARCS transformations, which are also been used to transpose
our isochrones from the theoretical to the observed plane, are adopted instead.
The difference between these estimates is much smaller than the uncertainty,
and arguments can be presented in support of both transformations.  However, we
are inclined to favor the larger value of $(m-M)_V$ because the Casagrande et
al.~transformations are based on direct measurements of real stars.

Accordingly, we adopted $(m-M)_V = 13.35$, which, for $E(B-V) = 0.032$, is
equivalent to $(m-M)_{F606W} = 13.34$, since $A_{F606W} \approx 2.80\,E(B-V)$
(\citealt{sjb05}) as compared with $A_V \approx 3.07\,E(B-V)$ (\citealt{mcc04}).
Assuming this distance modulus, together with $E(B-V) = 0.032$, we produced the
fit of a ZAHB to the HB population of 47 Tuc that is shown in
Figure~\ref{fig:fig14}.  Although there is considerable evidence for helium
abundance variations ($\delta\,Y \lta 0.03$) in this system (e.g.,
\citealt{ngp11}, \citealt{gls13}), the faintest and reddest HB stars
presumably have have normal helium contents, and it is to these stars that the
ZAHB has been fitted.  As indicated, the models assume [Fe/H] $= -0.76$ and
an initial helium abundance corresponding to $Y = 0.257$.  Moreover, with these
choices for the cluster properties, the age of 47 Tuc is predicted to be 11.75
Gyr.  We decided to adopt an independent estimate of the 47 Tuc distance because
there is some ambiguity in how a ZAHB should be compared with the type of HB
morphology that is characteristic of metal-rich systems.  Previous work (e.g.,
\citealt{dvl89}) suggested that the slightly sloped line which defines the
lower boundary of the faintest HB stars in 47 Tuc does not coincide with the
observed ZAHB, but rather that the latter is located along the upturn that is
predicted to occur near the red end.

The same thing is indicated by the particular comparison that is shown in
Fig.~\ref{fig:fig14}.  According to these results, stars arriving on the HB
after undergoing the helium flash populate
only that part of a ZAHB between its red end and the
short horizontal line just below it, which has been drawn at the minimum
luminosity of the observed HB clump.  Because masses tend to ``pile up" when
a ZAHB approaches its minimum $\teff$\ (maximum color), before bending back to
the blue, the mass range in this small region of the ZAHB is considerable.  To
be specific, the reddest ZAHB model that has been plotted has a mass which is
consistent with the neglect of mass loss over the preceding evolution, while the
point that is defined by the intersection of the short horizontal line with the
ZAHB has a lower mass by $0.20 {{\cal M}_\odot}$.  (The ranges in luminosity and
color encompassed by the entire HB population must therefore be a reflection of
the tracks that are followed during post-ZAHB evolution; see the paper by
Dorman et al.~1989.)  Thus, differences in the adopted distance have implications
for the inferred mass loss that occurred in the stars prior to reaching the
ZAHB.

It is clearly fortuitous that the ZAHB for the adopted abundances provides such
a good fit to the observations, given all of the parameters at play. In fact,
the models that have been used to fit the CMD of 47 Tuc do not reproduce the
properties of the binary (V69) as well as one would like. Figure~\ref{fig:fig15}
plots the components of V69 on the mass--radius plane, together with the
predicted variations of radius with mass along isochrones for ages that vary in
0.5 Gyr increments from 10.0 to 12.0 Gyr in some panels, or from 10.5 Gyr to
12.5 Gyr in others (as specified just above the abscissa of each panel).  The
upper left-hand panel contains the isochrones from the previous figure: they
favor an age of $11.0 \pm 0.5$ Gyr for V69 (and hence 47 Tucanae), which is
0.75 Gyr less than the age that was derived from the turnoff photometry.  The
other panels illustrate the impact of varying each of the chemical abundance
parameters in turn.  Higher ages by about 0.5 Gyr are obtained if $Y$ is reduced
by 0.005 or if the assumed [Fe/H] value is increased by 0.06 dex.  Although not
shown, the implication of the lower left-hand panel is that a higher age (by
about the same amount) would also be the result of increasing [$\alpha$/Fe] by
0.11 dex.

Whether or not 47 Tuc stars have such large enhancements of the
$\alpha$-elements is presently uncertain.  Whereas \citet{cgb13} obtained
[Mg/Fe] $= 0.53$ and [Si/Fe] $= 0.44$ from their spectroscopic survey of about
100 cluster giants, \citet{gls13} derived [Mg/Fe] $= 0.39$ and [Si/Fe] $=0.26$
from a similar investigation of 110 HB stars.  For the initial oxygen abundance,
the situation is further complicated by the O--Na anticorrelation, though we
suspect that the best estimate is close to the upper end of the observed range.
In the investigation by Carretta et al., the mean and upper envelope [O/Fe]
values for their sample of stars are 0.26 and $\approx 0.40$, respectively.  A 
much higher average value of [O/Fe] ($= 0.60$) was derived \citet{km08} from
8 RGB stars that have $0.48 \le$ [O/Fe] $\le 0.66$.  These determinations may be
compared with [O/Fe] $\approx 0.50$ in thick-disk field stars that have [Fe/H]
$\sim -0.8$ (Ram\'irez et al.~2012).  Regardless, just as an increase in
[$\alpha$/Fe] would tend to reduce the difference between the ages inferred
from the CMD and the binary, the age discrepancy would be exacerbated if the
models assumed lower $\alpha$-element abundances (see the bottom, left-hand
panel of Fig.~\ref{fig:fig15}).  (Note that different choices for the chemical
abundance parameters would affect the corresponding ZAHB-based distances, but
only slightly given that the adopted variations are small.  Nevertheless, this
effect on the CMD age has not been taken into account.) 

Another way of reconciling the binary mass-radius age with the CMD age is
to increase the model temperatures.  It is quite possible that the predicted
$\teff$\ scale is too cool because of deficiencies in, for instance, our
treatment of the atmospheric boundary condition, convection, or diffusive
processes.  The impact of such a problem on the M-R diagram is illustrated by
the short- and long-dashed curves in the upper left-hand panel.  These show, in
turn, how the location of the 11.0 Gyr isochrone that is represented by the
solid curve would be affected if the predicted temperatures were increased, or
decreased, by 75 K.  The effect on the inferred age is apparently $\pm 0.25$
Gyr, respectively.  Thus, it would be possible to obtain ages from both
the M-R diagram and the CMD that agree to within $1\,\sigma$ if the models were
hotter by about 75 K.  It turns out that, at the masses of the binary
components, the temperatures along 11.0--11.75 Gyr isochrones are cooler than
those obtained from the radii and colors of V69, but within $\approx 60$ K.
Since an upward revision in the predicted temperatures would improve the
agreement, it seems likely that this is the correct way to obtain the best
consistency between Figs.~\ref{fig:fig14} and~\ref{fig:fig15}.  Note that
improved consistency would not be obtained if we simply adopted an increased
distance modulus by a few hundredths of a magnitude.   Because the components
of V69 have measured magnitudes, models that match the observed radii, in
combination with the predicted temperatures, also imply a particular value of
the apparent distance modulus.  Since our adopted modulus is already at the
high end of the range indicated by the uncertainties in the properties of the
binary, any arbitary increase in $(m-M)_V$ will cause even larger discrepancies
with the distance based on the binary.

As already mentioned, other potential solutions include lowering the helium
abundance and/or increasing the [Fe/H] value, and we are considering such
adjustments in a much more detailed study of 47 Tuc and its binary V69
(K.~Brogaard et al., in preparation).  However, for the present study we have
chosen to adopt the CBG09 metallicity scale and a particular variation of $Y$
with [Fe/H] for the entire GC sample.  Even if it is possible to argue for
adjustments of these choices for a given cluster, it is not known whether they
are unique to that cluster or whether they should be applied to the entire
sample.  Therefore, no such variations have been explored here, especially as
our main goal is to obtain the best estimates of {\it relative} GC ages.

With the adopted composition and apparent distance modulus, the implied mass
loss agrees rather well with that determined by \citet{mbs12} in the case of
the super-metal-rich open cluster NGC$\,$6791 using asteroseismology.  These
investigators found a mean mass difference of $0.09 {{\cal M}_\odot}$ between
the lower RGB and the red HB clump, which suggests that the red HB stars in
NGC$\,$6791 have masses that do not differ by more than $\approx 0.18
{{\cal M}_\odot}$.  Such a dispersion in the mass loss is also supported by
the analysis of the cluster CMD carried out by \citealt{bvb12}.)  To within the
uncertainties, this is the same as the value of $\approx 0.20 {{\cal M}_\odot}$
found here from our fit of a ZAHB to the HB population of
47 Tuc\footnote{Even in low metallicity clusters, the range in mass spanned
by HB stars may be quite similar.  For instance, our fit of ZAHB models to
the horizontal branch of M$\,$92 (see Fig.~\ref{fig:fig11}) suggests that
its lowest mass HB star is only $\sim 0.16 {{\cal M}_\odot}$ less massive than
a 12.75 Gyr star that has evolved to the HB without undergoing any mass loss
whatsoever during its previous evolution.  However, such estimates will be
quite uncertain given the strong sensitivity of HB models to the assumed
chemistry.  For an excellent review of mass loss and its implications for the
HB phase, see \citet{cat09}.}

Given the apparent near constancy of the difference in the mass of the most and
least massive stars that populate the red HBs of NGC$\,$6791 and 47 Tuc, whose
metallicities bracket those of the GCs in the metal-rich group under
consideration, we believe that it is quite reasonable to assume that the same
result applies to the latter.  Thus, to determine the distance modulus of each
metal-rich GC in our sample, we have determined which ZAHB model has a mass
that is $0.20 {{\cal M}_\odot}$ less than the RGB tip mass for the adopted age
(ignoring mass loss along the giant branch): that model is identified by the
intersection of a short horizontal line with the ZAHB.  Then the dereddened
cluster photometry is shifted in the vertical direction until that horizontal
line coincides with the faintest HB star in the cluster HB population, resulting
in the adopted value of $(m-M)_{APP}$.  If the maximum amount of mass loss that
occurs prior to the ZAHB were, for instance, as small as $0.10 \msol$\ or as
large as $0.30 \msol$, the derived distance modulus of 47 Tuc (and the other
metal-rich GCs in our sample) would be larger, or smaller, than our adopted
modulus by only $\approx 0.05$ mag.  This is not insignificant, but neither
does it completely overwhelm other uncertainties.

When distance moduli are derived in this way for the GCs that have [Fe/H] $\ge
-1.0$, the ZAHBs and isochrones generally provide satisfactory matches to the
observed photometry, as shown in Figure~\ref{fig:fig16}.  The least reliably
determined ages are those for NGC$\,$5927, NGC$\,$6304, NGC$\,$6624, and M$\,$69
(NGC$\,$6637), as their CMDs appear to be affected by significant differential
reddening and/or substantial field-star contamination.  Note that, in none of
the nine GCs, is the $E(B-V)$ value less than 0.10 mag.  Except in the case of
NGC$\,$6366, where the HB stars are offset to the blue of their expected 
location relative to the ZAHB (possibly due to a chemical abundance effect),
the HB populations are quite well matched by the ZAHBs --- though
this is partly by design.  Schlegel et al.~(1998) reddenings have been assumed
for all of the clusters except those considered in the top row as well as
M$\,$71 (NGC$\,$6838).   In order to obtain a similar fit to their HB
populations as those found for the other five clusters, it was necessary to
adopt reduced reddenings by as little as 0.02 mag (NGC$\,$6304) to as much as
0.08 mag (NGC$\,$5927).  An important consequence of making such revisions is
that the resultant $\delta$(color) values become much more similar for all of
the GCs in the sample.  The mean age of this group, including 47 Tuc, is 11.1
Gyr, implying that metal-rich GCs are about 1.5 Gyr younger than the most
metal-deficient systems.  This finding is at odds with the conclusion reached
by MF09, who found little or no difference in their ages.  A resolution of
these conflicting results is provided in \S~\ref{subsubsec:mf09}.

\subsection{MSTO-to-RGB Color Constraints on Relative GC Ages}
\label{subsec:msrg}

It is readily appreciated that the $\delv$\ method cannot be used to determine
very precise ages for globular clusters that have extremely blue HBs.  In those
systems in which the horizontal-branch populations are nearly vertical on the
CMD, even small offsets in color (due, e.g., to reddening uncertainties) would
have a big effect on the distances, and therefore ages, that are derived from
fits of computed ZAHBs to the observations.  In such cases, the relative age
technique described by VandenBerg et al.~(1990; hereafter VBS90) can be used to
provide much tighter constraints on their absolute ages if, in particular, it
is used to compare the CMDs of two clusters that have similar chemical
abundances but one of them has a red HB that is amenable to a $\delv$\ analysis.
The essence of the VBS90 procedure is to superimpose the principal photometric
sequences of two GCs, simply by applying the necessary vertical and horizontal
shifts to one of them, so that both have the same turnoff color and the same
magnitude at a point on the upper MS that is 0.05 mag redder than the TO.  Then
whatever separation of their RGBs is found at, say, 2.8 mag above the latter
fiducial point will be a direct measure of the difference in the age of the two
clusters.  In effect, relative GC ages are found from the difference in color
between the MSTO and the RGB at a pre-selected, arbitrary point on the giant
branch.  This relative age diagnostic, which will henceforth be referred to as
$\delc$, where the ``H" represents ``horizontal" just as the ``V" in
$\delv$\ can be thought of as indicating ``vertical", is independent of
distance, reddening, and the zero-point of color calibrations.

\citet[][also see MF09]{mca10} have stated that this so-called ``horizontal
method" of determining relative ages is sensitive to the assumed value of the
mixing-length parameter ($\amlt$), or more generally, the treatment
of super-adiabatic convection, as well as the adopted color--$\teff$\ relations.
They say that, as a consequence, this approach is ``strongly model dependent".
However, these are valid concerns only if the $\delc$\ method is used to
determine {\it absolute} ages; indeed, VBS90 advised against such a practice in
the paper that introduced this new {\it relative}-age-dating technique.  The
latter relies only on how the color of the RGB, at a fixed magnitude, is
affected by variations in age.  Because it involves such a narrow range in RGB
colors and because it has no dependence on the zero-points of the color
transformations, errors in the color--$\teff$\ relations will be
inconsequential.

To support these assertions we have used the $\delc$\ method to evaluate
the relative ages of two hypothetical GCs whose RGBs differ in color by 0.015
mag at 2.8 mag above the ordinate zero-point when their CMDs have been 
registered to one another according to the directions given by VBS90 (see 
above).  Suppose both clusters have [Fe/H] $= -1.4$ and normal helium and
$\alpha$-element abundances for that metallicity, and let us consider two sets
of isochrones, for the same chemical abundances, which are otherwise identical
except that one assumes $\amlt = 1.50$ while the other assumes $\amlt = 2.007$.
When similarly registered to one another, the isochrones based on the smaller
value of the mixing-length parameter predict a {\it much} larger difference in
color at 2.8 mag above the ordinate zero-point than those which assume $\amlt
= 2.007$.  In fact, depending on which of the two grids of isochrones is used
to evaluate the age corresponding to a given (observed) MSTO-to-RGB color
difference, absolute ages that differ by $\gta 3$ Gyr would be obtained.  
(Clearly it is very risky to derive the absolute ages of star clusters in this
way.)

However, for relative ages, it is only the difference in color between the RGB
segments that matters.  At the same magnitude offset from the ordinate
zero-point, the difference in color between, say, 11 and 13 Gyr isochrones is
0.0221 mag if derived from the models for $\amlt = 1.50$, versus 0.0195 mag in
the case of those which assume $\amlt = 2.007$.  That is, the former predict a
variation in color with age of 0.0109 mag/Gyr as opposed to 0.0097 mag/Gyr for
the latter.  (Such slopes are not strictly constant as the difference in color,
at a fixed luminosity between, e.g., 11 and 12 Gyr isochrones will generally be
slightly larger that that found for 12 and 13 Gyr isochrones --- but this is a
second-order effect.)  Returning to the example of two clusters whose RGBs
differ in color by 0.015 mag, the isochrones for $\amlt = 1.50$ predict an age
difference of 1.55 Gyr, while those for $\amlt = 2.007$ yield 1.38 Gyr.  Thus,
despite using isochrones with very different color offsets between the TO and
the RGB, the derived relative ages are nearly the same.

Even though this analysis considered isochrones that were computed for very
different values of $\amlt$, we do not believe that there is
compelling evidence that this parameter varies {\it significantly} with mass,
metallicity, or evolutionary state.  If it did, we would not obtain such good
agreement between isochrones that assume a constant value of $\amlt$\ and
the observed CMDs of GCs that span at least 2 dex in [Fe/H] (see, e.g.,
Figs.~\ref{fig:fig11}--\ref{fig:fig14}, \ref{fig:fig16}).  In fact, models based
on the same physics have been just as successful in explaining the observations
of the old, super-metal-rich open cluster NGC$\,$6791 (\citealt{bvb12}),
whose properties have been especially tightly constrained through analyses of a
few of the detached, eclipsing binary stars that it contains.  The VandenBerg
et al paper also showed that Victoria-Regina isochrones reproduce to within 
$\sim 10$ K (in the mean) the temperatures derived by \citet{crm10} using the
infrared-flux method for field subdwarfs that have $-2.2 \lta$ [Fe/H] $\lta
-0.5$ and accurate {\it Hipparcos} parallaxes.  Similar results have been
obtained in the past (\citealt{vbd06}), and importantly, no significant
dependence of the mixing-length parameter on metallicity has been found in
empirical calibrations of infrared photometry for giants in 28 GCs that have
$-2.2 \lta$ [Fe/H] $\lta -0.2$ (\citealt{fvs06}, also see \citealt{pps02}).

These results are completely at odds with the recent findings of
\citet[also see \citealt{bvc12}]{btb12}, who concluded that $\amlt$\ increases
by $\approx 0.5$ per dex in [$m$/H] from their analysis of
asteroseismic data for a sample of dwarfs and and subgiants in the Kepler field.
To resolve this dilemma, work needs to be undertaken to compare the photometric
and spectroscopic $\teff$\ scales, to study the extent to which star-to-star
differences in the detailed heavy element abundances affect the results, to
evaluate the impact of stellar models that neglect (in the Bonaca et al.~study)
convective core overshooting and diffusive processes, and to carefully examine
other sources of systematic error.  In this paper, we can only point out that
comparisons of modern isochrones with photometric data for star clusters do not
support a steep dependence of $\amlt$\ on metallicity.

For example, since isochrones for $\amlt = 2.005$ provide a superb
fit to the CMD of NGC$\,$6791 (see \citealt{bvb12}), the Bonaca et al.~results
would suggest that models relevant to 47 Tuc, which is $\approx 1$ dex more
metal-poor than NGC$\,$6791, should assume $\amlt \approx 1.5$.  To
examine the implications of this choice, this (low) value of the mixing-length
parameter has been adopted in the computation of a grid of models for $Y =
0.257$ and [Fe/H] $= -0.76$ (i.e., the same chemical abundances that were
assumed in the models plotted in Fig.~\ref{fig:fig14}).  Although not shown,
these calculations predict temperatures that are 320 K and 390 K cooler near the
base and the tip of the RGB, respectively, than those shown in that figure.
Furthermore, because of the steep dependence of the
bolometric corrections and colors on $\teff$\ in the case of cool giants, both
the location and the slope of the predicted RGB are very discrepant relative to
the observed giant branch.  To be sure, the uncertainties are such that a small
variation of $\amlt$\ with mass, metallicity, and/or evolutionary
state cannot be ruled out, but even if this parameter varied by $\sim 10$--15\%
across the H-R diagram, which would require some compensating adjustments to the
atmospheric boundary conditions and/or to the color--$\teff$\ relations in order
to obtain comparable fits to observed CMDs, our isochrone-fitting procedure
would yield essentially identical ages (as shown in Fig.~\ref{fig:fig5}).

While, in principle, it should be possible to obtain rather precise relative
ages using the VBS90 technique, the results do depend quite critically on
whether or not the clusters whose CMDs are intercompared have the same chemical
abundances.  For instance, \citet{vs91} pointed out early on that the same value
of $\Delta(B-V)_{\rm MS,RGB}$ would be obtained for two GCs if they had
identical metal abundances but differed in age by $\approx 10$\%, or if they
were coeval but had different [O/Fe] values by $\sim 0.3$ dex (assuming no other
chemical differences).  Indeed, it turns out that cluster-to-cluster variations
in $Y$ or in the abundances of several of the other metals (notably Ne, Mg, or
Si; see VBD12), would complicate the interpretation of $\delc$\ measurements,
especially at higher metallicities.  This is shown in Figure~\ref{fig:fig17},
where 12 Gyr isochrones for [Fe/H] $= -2.0$ and 11 Gyr isochrones for [Fe/H] $=
-1.0$ have been plotted with, and without, the assumed chemical abundance
variations that are indicated in the top left-hand corner of each of the six
panels that comprise this plot.  The isochrones, which have been taken from
VBD12, have been registered to one another using the VBS90
prescription: the zero-point of the abscissa coincides with the TO color, while
the magnitude at the point on the upper MS that is 0.05 mag redder than the
turnoff defines the zero-point of the ordinate (note the location of the filled
circle on the 12 Gyr isochrones).

The insert in the upper left-hand panel plots the relative locations of the RGB
segments of isochrones for the two values of [Fe/H], and ages from 10.5 to 13.5
Gyr in 1.0 Gyr steps.  They were arbitarily moved to the region enclosed by the
dotted rectangle (simply for display purposes) from original ordinate values
ranging from $\approx -2.4$ to $-3.2$.  By comparing these results with the
predicted color offsets of the RGBs (at an ordinate value of $\sim -2.8$) that
arise from chemical composition variations, one finds, for instance, that a 0.4
dex increase in the value of [O/Fe] at [Fe/H] $= -2.0$ (right-hand plot, top
row) mimicks the effect on the RGB location that is comparable to the shift
produced by about a 1 Gyr increase in age.  Similarly, a higher value of
[Mg/Fe] by 0.4 dex at [Fe/H] $= -1.0$ (middle panel, bottom row) apparently
causes a redward offset of the RGB (relative to the TO) that is approximately
equivalent to that predicted for a $\sim 1.5$ Gyr reduction in age.  These are
just rough estimates: Fig.~\ref{fig:fig17} has been included here mainly to
illustrate qualitatively how cluster-to-cluster differences in age and chemical
abundances affect the $\delc$\ parameter.

In fact, the observed variations in the abundances of individual metals in
clusters of similar [Fe/H] are closer to $\delta$[$m$/Fe] $= 0.1$--0.2 dex
(\citealt{cbg09b}) than to 0.4 dex, so the concerns that are raised by
Fig.~\ref{fig:fig17} are not nearly as serious as this plot suggests.  Moreover,
since the majority of the GCs that have extremely blue HBs have [Fe/H] $\lta
-1.6$, the only abundance variations that appear to have detectable consequences
for $\delc$-based estimates of their relative ages are those of helium and/or
oxygen.  Fortunately, differences in $Y$ can be distinguished from differences
in [O/Fe] because the former have bigger effects on the slope of the subgiant
branch than the latter.  This is most easily seen when isochrones are
superimposed according to the VBS90 procedure (compare the plots in the upper
left- and right-hand panels).  It is worth reiterating that, as discussed by
VBD12, the age-sensitive parts of isochrones {\it for low
metallicities} depend almost entirely on the absolute C$+$N$+$O abundances (or
[CNO/H] coupled with the CNO abundances in the adopted solar mixture).  The
[Fe/H] value is a useful way of ``tagging" GCs because, with few exceptions, it
does not vary significantly from star-to-star within a given cluster (as already
mentioned).  It also provides a good ``label" for isochrones given that a given
metals mixture, with or without some adjustments to the solar $m$/Fe number
abundance ratios, is normally scaled to any desired [Fe/H] value.  However, the
iron abundance is far less important than the [O/H] ($=$ [Fe/H] $+$ [O/Fe])
value at low $Z$.

Although Fig.~\ref{fig:fig2} showed that the morphology of isochrones (for
[Fe/H] $\le -1.4$) in the vicinity of the turnoff is nearly independent of age,
there is some age dependence of the luminosity, and slope, of the SGB at the
lowest metallicities.  As illustrated in Figure~\ref{fig:fig18}, 11--13 Gyr
isochrones for [Fe/H] $= -2.40$ begin to deviate from one another just past the
turnoff, while those for higher metallicities do so at colors, relative to the
TO, that become steadily\ redder as the [Fe/H] value increases.   (Note that
the intercomparisons of isochrones presented in
Figs.~\ref{fig:fig2}--\ref{fig:fig5} minimized the
differences between them at a color that is 0.05 mag redder than the TO, both
above and below the turnoff.  As a consequence, the differences between their
SGB segments are considerably smaller in those plots than in 
Fig.~\ref{fig:fig18}.)  While these differences are still rather small, they
nevertheless resemble the effects of enhanced $Y$ on the SGB at a fixed age
(see the top left-hand panel of Fig.~\ref{fig:fig17}).  This should be kept in
mind when interpreting any cluster-to-cluster differences in the morphologies
of their CMDs that are found when the latter are registered to one another
according to the directions given by VBS90.

It is obviously important to apply the $\delc$\ method to the GCs that have
already been considered, in order to check that the relative ages of clusters
with similar metallicities are consistent with the differences in their
$\delv$\ ages.  However, we will first turn our attention to a few
``second-parameter" clusters and then consider several systems with extremely
blue HBs.  The advantage of proceeding in this way is that the latter appear to
provide some valuable insights (or at least ``food for thought") concerning the
second-generation stars in GCs that may be relevant to a subset of the [Fe/H]
$\lta -1.6$ clusters that were the subject of \S~\ref{subsec:std}.

\subsubsection{M$\,$3 and M$\,$13}
\label{subsubsec:m3m13}

Figure~\ref{fig:fig19} illustrates the fiducial sequences that have been
derived for M$\,$3 and M$\,$13.  Why these clusters have such different HB
morphologies, despite having similar metallicities, is still an open question
even though extensive efforts have been made during the past 40 years to try to
solve this puzzle (see, e.g., the review by \citealt{cat09}).  Since age and/or
chemical abundance differences, notably of He or the CNO elements, are the most
likely second parameters (after [Fe/H], which is the first parameter), it is
possible that variations in the properties of the multiple stellar populations
that all GCs appear to contain are responsible for this phenomenon. Furthermore,
given that age, $Y$, and [CNO/Fe] affect isochrones in different ways (see
Fig.~\ref{fig:fig17}), the differences between the CMDs of M$\,$3 and M$\,$13
for turnoff to lower-RGB stars should provide vital clues concerning the
identification of the second parameter, or of their relative importance if
there is more than one.

Accordingly, considerable care has been taken to accurately determine the median
photometric sequence through the MS, SGB, and lower RGB of each cluster.  To do
this, we sorted the photometry into 0.10 mag bins, determined the median
magnitude and color in each bin, and then (for the lower RGB) performed a
least-squares cubic fit to the 18-20 points so obtained.  To determine the
turnoff color, the median points for three bins both above and below the TO
were fitted by a least-squares quadratic, which was then evaluated over small
intervals of $M_{\rm F606W}$ to determine the bluest color.  A similar process
was used to derive the magnitude at the color on the upper MS which is 0.05 mag
redder than the TO color, using the median points from 8 neighboring bins.
(Approximate distance moduli and the adopted cluster reddenings have been
assumed in constructing Fig.~\ref{fig:fig19} so that the same regions of both
CMDs are displayed, thereby facilitating visual comparisons.)

Unfortunately, the age-dependent $\delc$\ parameter is also predicted to be a
fairly sensitive function of [Fe/H] at a fixed age (see Fig.~\ref{fig:fig3}).
Although the VBS90 registration procedure was not used to produce the latter,
it is quite obvious that nearly the same figure would have been obtained had it
been employed: the isochrones plotted therein would require only slight vertical
shifts in order for them to pass through the usual registration zero-points for
the $x$- and $y$-axes.  From suitable plots, we find that a 0.1 dex difference
in the relative metal abundances of two GCs whose CMDs are intercompared would
affect the inferred age difference by 0.25--0.5 Gyr.  This can be taken as the
uncertainty of relative age estimates based on the horizontal method since most
[Fe/H] determinations are uncertain by $\pm 0.05$--0.10 dex in a relative sense
(versus $\gta \pm 0.10$--0.20 dex in an absolute sense).  To obtain the best
consistency with $\delv$\ ages, it is, in fact, necessary (see
\S~\ref{subsubsec:blue}) to take metallicity differences into account when
using MSTO-to-RGB colors to evaluate relative GC ages.

How we have done this is illustrated in the upper left-hand panel of
Figure~\ref{fig:fig20}, which plots the the fiducial sequences of M$\,$3 and
M$\,$13 after they have been registered to one another.  (To minimize space
requirements, only the region from just below the TO to the lower RGB has been
plotted.)  The dashed lines plot the RGB segments of 10.5 to 13.5 Gyr 
isochrones, in 0.5 Gyr steps, for $Y = 0.25$ and [Fe/H] $= -1.50$ (i.e., for
chemical abundance parameters that we have adopted for M$\,$3).  However, we
could have opted to plot the giant-branch loci for a somewhat higher or lower
metallicity without any misgivings because the age-dependent separations of the
RGBs vary only weakly with [Fe/H] (see Fig.~\ref{fig:fig18}).  Given the
uncertainties that were discussed in the previous paragraph, it does not matter
that the horizontal displacement between the giant branches of, say, 10.5 and
11.0 Gyr isochrones is slightly larger than that of 12.5 and 13.0 Gyr
isochrones; consequently, it is not necessary to assume a specific age (in Gyr)
for one of the GCs and then deduce the absolute age of the other cluster from
the overlap of its giant branch onto the isochrone RGBs.  In fact, it is our
usual practice to adjust the location of the grid of dashed lines by whatever
color offset is needed so that the giant branch of one of the clusters coincides
with the third or fourth isochrone.  Then, just by inspection of the observed
separation of the GC giant branches relative to the dashed lines, which differ
in age by 0.5 Gyr (in the direction from ``younger" to ``older", as indicated),
the difference in the cluster ages is readily evaluated.

The length of the horizontal line which connects the small filled circle at an
ordinate value of $-2.8$ to the short solid line that has the same slope as the
dashed loci represents the correction to the color of the M$\,$3 RGB that should
be applied to it to remove the effect on the $\delc$\ parameter of the
difference in the cluster metallicities.  Thus, whereas the separation of the
solid and dashed curves suggests that the two GCs differ in age by about 0.5
Gyr, the models actually predict an age difference closer to 1.0 Gyr if the
nearly 0.1 dex difference in their [Fe/H] values is taken into account.
However, according to the right-hand panel of Fig.~\ref{fig:fig18}, the cluster
SGBs should overlay one another if age is the main distinguishing property of
M$\,$3 and M$\,$13.  This is clearly not the case.  Indeed, the top right-hand
panel of Fig.~\ref{fig:fig20} indicates that a difference in $Y$ is a much more
viable way of explaining the offset in the cluster subgiant branches.

The lower left-hand panel shows that, if M$\,$3 has $Y = 0.25$ and an age of
11.75 Gyr (which will be justified shortly), it is then possible to reproduce
the observations in the panel just above it if M$\,$13 has $Y = 0.29$ and an
age of 12.0--12.25 Gyr.  We initially thought that similar consistency could be
obtained if M$\,$13 had an age of $\gta 11.75$ Gyr, $Y = 0.29$, and an enhanced
O abundance by $\delta$[O/Fe] $\approx 0.3$ dex, since the effect of the latter
is to displace the RGB for a fixed age to somewhat bluer colors {\it relative to
the turnoff} (see the lower right-hand panel).  [It is, of course, the C$+$N$+$O
abundance that is important; i.e., the same results would be found if N, instead
of O, were enhanced by a sufficient amount that $\delta$[CNO/Fe] $\approx 0.3$.
In fact, this alternative is arguably the favored one given that helium-rich
stars in $\omega\,$Cen (see, e.g., \citealt{jp10}, \citealt{mmp11}) and in
NGC$\,$2808 (\citealt{bcg10}) have low oxygen, and rather high nitrogen,
abundances.]

However, the main effect of increasing the C$+$N$+$O abundance on isochrones
is to reduce the age at a given turnoff luminosity (or equivalently, to displace
an isochrone for a fixed age to a lower TO luminosity).   This tends to increase
the separation between the MSTO and the giant branch; compare the dashed and
dotted curves in the lower right-hand panel.  To obtain a reduced separation,
either the helium enhancement must be large enough to overwhelm the effects of
increased CNO, or alternatively, the abundances of the CNO elements should be
reduced.  In the latter case, fits of the corresponding models to the observed
SGB on an absolute magnitude versus color plane would imply a higher age.  The
main conclusion to be drawn from Fig.~\ref{fig:fig20} is that M$\,$13 appears
to have a higher helium abundance than M$\,$3, as there does not appear to be
any other way of explaining the differences in the morphologies of their CMDs
between the turnoff and the RGB.  (The same solution of this problem was first
proposed by \citealt{jb98}.)  It is certainly very fortunate that variations in
the abundance of helium can be detected in this way --- if the model predictions
can be relied upon --- in view of the fact that $Y$ is normally very difficult
to determine in cool stars, due to the lack of detectable He lines in their
spectra.

This does not necessarily imply, however, that there are no significant
star-to-star variations of $Y$ in either cluster.  The more general conclusion
to be drawn from Fig.~\ref{fig:fig20} is that the {\it average} helium content
of M$\,$13 appears to be higher than that of M$\,$3 (by, say, $\delta\,Y \sim
0.04$).  In fact, the possibility that all of the stars in M$\,$13 had the same
high value of $Y$ when they formed presents some difficulties for the
interpretation of the cluster CMD.  Suppose, for instance, that M$\,$13 has an
age of 12.25 Gyr and $Y = 0.29$, versus 11.75 Gyr and $Y = 0.25$ for M$\,$3;
i.e., that their properties differ in ways that are compatible with the results
shown in Fig.~\ref{fig:fig20}.  As illustrated in Figure~\ref{fig:fig21}, the
isochrone for the adopted parameters is able to reproduce the turnoff of M$\,$13
only if the apparent distance modulus is 14.40, assuming $E(B-V) = 0.017$
(Schlegel et al.~1998).  In fact, that isochrone provides an especially
agreeable fit to the MS-to-RGB populations, including, in particular, the slope
of the subgiant branch.  However, the fully consistent ZAHB locus is clearly too
bright relative to the observed HB stars with $M_{F606W} \lta 1.5$.

One way of achieving a satifactory simultaneous fit of the models to the entire
CMD of M$\,$13 is to assume that this cluster has two or more chemically
distinct stellar populations characterized by different helium (and possibly
CNO) abundances.  The left-hand panel of Figure~\ref{fig:fig22} shows that the
lower bound to the distribution of HB stars can be matched quite well by a ZAHB
for $Y = 0.25$ and [Fe/H] $= -1.58$ if M$\,$13 has $(m-M)_{APP} = 14.45$.
Since, on this color plane, the location of the blue HB tail is insensitive to
the value of $Y$, it is not possible to distinguish between the high- and
low-$Y$ HB populations if the former evolve to ZAHB structures that are well to
the blue of the instability strip.  The brightest/reddest HB stars may be either
evolved stars from zero-age locations along the blue tail or they may be 
near-ZAHB stars with $Y \sim 0.29$ (given their proximity to the dashed curve),
especially if they also have high CNO abundances, which tends to drive HB stars
of a given mass to redder colors (see, e.g., \citealt{pcs09}).
[\citet{sgl10} also suggested that the reddest HB stars in M$\,$13 are probably
evolved stars, in part because this is implied by the distance modulus that is
obtained using the RGB-tip standard candle (\citealt{bfs04}).]

The observed widths of the M$\,$13 MS and RGB certainly appear to be large
enough to accommodate a large spread in helium abundance, as indicated by the
12.0 Gyr isochrones for $Y = 0.25$ (solid curve) and $Y = 0.33$ (dashed curve).
In fact, the widths of the principal photometric sequences (also in the case of
M$\,$3, see the right-hand panel) are artificially broadened due to photometric
errors arising from crowding and other effects, which can be expected to be
especially severe in the most populous clusters.  Although we generally selected
stars with tabulated errors of $< 0.01$ mag, a large fraction of the stars,
especially those on the giant branch, appear to have been measured only once
(based on our limited examination of the impact of culling the data in various
ways).  If we further restrict the number of stars to those with 4 or more
observations in both $m_{F606W}$ and $m_{F814W}$, the resultant CMDs for the
main sequences of M$\,$3 and M$\,$13 are tightened considerably, though the MS
widths still remain somewhat larger than the difference between the solid and
dashed loci in the left-hand panel of Fig.~\ref{fig:fig22}.  We are unsure as
to whether or not this is the true intrinsic width, but we have assumed that it
is.  Moreover, it is our impression that the tightest CMDs that we are able to
obtain for the most massive clusters are broader, with thicker subgiant
branches ($\sim 0.10$--0.15 mag, at a given color), than those of lower mass,
but further work is needed to do a proper evaluation.  In any case, this is an
important caveat that should be kept in mind.

M$\,$3 could well have appreciable star-to-star helium abundance variations as
well, though comparisons of HB models with observations of M$\,$3 on various
Str\"omgren CMDs and on the $(\log\teff,\,\log\,g)$-plane suggest that there is
little or no variation in $Y$ along its entire horizontal branch
(\citealt{cgs09}).  Of course, for a difference in $Y$ to be the main cause of
the different SGB morphologies (see Fig.~\ref{fig:fig20}), it is only necessary
that M$\,$13 have a higher $Y$, {\it in the mean}, than M$\,$3 by about 0.04.  A
very similar suggestion, based in part on a consideration of the location of the
RGB luminosity function bump in the two clusters, was put forward by 
\citet{cd05}, who noted that a large variation in $Y$ might be expected in
M$\,$13 if it has a much higher proportion of second-generation stars than
M$\,$3.  Further support for $Y({\rm M}\,13) - Y({\rm M}\,3) \sim 0.02$--0.04 is
provided in the very recent analysis of far--$uv$/optical CMDs by \citet{dsf13}.
Clearly such a difference would also provide a natural way of explaining why the
HB in M$\,$13 is so much bluer than that of M$\,$3.\footnote{Interestingly,
calibrations of the helium-sensitive $R$ parameter ($= N_{\rm HB}/N_{\rm RGB}$,
where $N_{\rm HB}$ is the number of horizontal-branch stars and $N_{\rm RGB}$
is the number of red giants that are brighter than the luminosity level of the
HB) do not indicate that the helium abundances of these two clusters are
appreciably different (e.g., \citealt[also see \citealt{sgl10}]{sq00}).  In
fact, they suggest that $Y$ is very nearly constant for {\it all} GCs, with no
statistically significant dependence on [Fe/H] (\citealt{src04}).  However, such
calibrations cannot be expected to apply to clusters that contain multiple
stellar populations.}

As shown in Fig.~\ref{fig:fig22}, our preferred fits of isochrones to the CMDs
of these two clusters suggest they are nearly coeval: the difference in their
ages is estimated to be $0.25 \pm 0.25$ Gyr.  We note, as well, that the
thickness of the SGB (in both GCs) suggests that something other than, or in
addition to, $Y$ is varying since the location of the subgiant branch at a fixed
age has very little dependence on the assumed helium abundance (see, e.g., the
isochrones that have been plotted in the left-hand panel).  Presumably this is
caused by a variation in the CNO abundances and/or age, although
\citet{cm05a} have concluded from their spectroscopic analyses that both
clusters have the same, constant value of [CNO/Fe], even at the tip of the RGB
of M$\,$13 where some stars have super-low oxygen abundances.  A final point 
worth mentioning is that, if our distance estimates are accurate, the RGB bump
in M$\,$13 is intrinsically $\approx 0.1$ mag brighter than that of M$\,$3,
(assuming the apparent V magnitudes of the bump given by \citealt{ngp13}),
which would be consistent with M$\,$13 having somewhat higher helium and/or
lower CNO abundances than M$\,$3 (for recent predictions of the RGB bump
luminosity, see \citealt{van13}). 

\subsubsection{M$\,$5, M$\,$12, NGC$\,$288, and NGC$\,$362}
\label{subsec:n288n362}

\begin{figure}[t]
\centering
\includegraphics[width=0.9\textwidth]{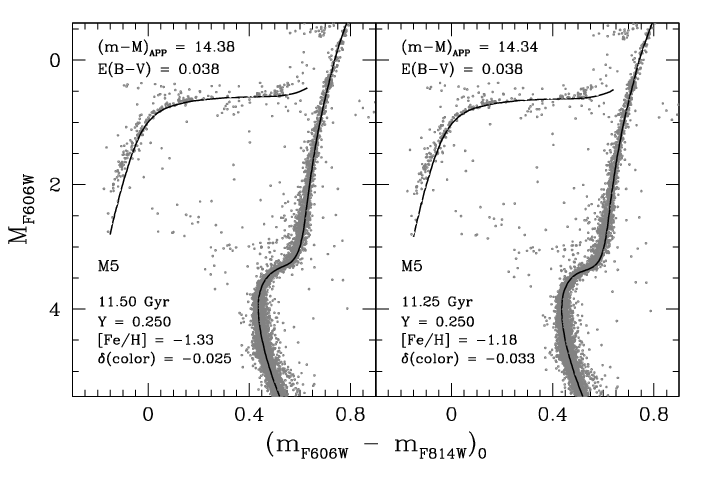}
\caption{Fits of ZAHB loci to the lower bound of the distribution of HB stars
in M$\,$5, assuming [Fe/H] $= -1.33$ ({\it left-hand panel}) and [Fe/H] $=
-1.18$ ({\it right-hand panel}), assuming the same values of $E(B-V)$ and $Y$
(as indicated).  The models provide an excellent match to the observed HB
morphology, except at $M_V > 1.5$ where some discrepancies become apparent.
The ZAHB-based apparent distance moduli are 14.38 and 14.34 for the two cases.
(The difference in the derived distance modulus would have
been nearly a factor of two larger had they been based on fits of the observed
CMD to the lower MS segments of the isochrones given that they differ by 0.069
mag at a fixed color.)  To match the observed turnoff color, it was necessary
to shift the best-fit isochrones for 11.50 and 11.25 Gyr horizontally by
$-0.025$ and $-0.033$ mag, respectively.  These color offsets were not applied
to the ZAHB models.}
\label{fig:fig8}
\end{figure}

\begin{figure}[t]
\includegraphics[width=1.0\textwidth]{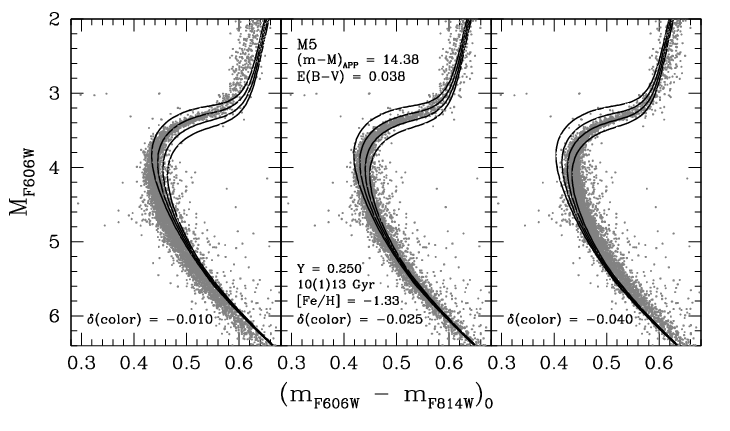}
\caption{The {\it middle panel} presents a similar fit of isochrones to the CMD
of M$\,$5 as that shown in Fig.~\ref{fig:fig7}, except that the various loci,
which represent ages from 10 to 13 Gyr, in 1.0 Gyr increments, have not been
shifted to a common turnoff color.  This more traditional comparison of theory
and observations indicates that an age of approximately 11.50 Gyr is required
to match the observed subgiant stars, as found in the previous figure.
In the {\it left-hand} and {\it right-hand panels}, different color offsets
have been applied to the set of four isochrones, as specified just above the
abscissa, to illustrate the difficulties of deriving an age when the best-fit
isochrone does not reproduce the observed turnoff color.}
\label{fig:fig9}
\end{figure}

\begin{figure}[t]
\centering
\includegraphics[width=0.8\textwidth]{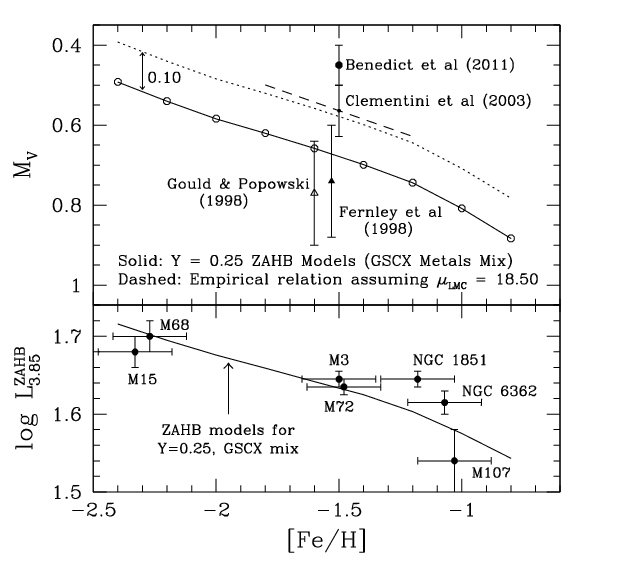}
\caption{{\it Upper panel}: RR Lyrae $M_V$ zero-points as determined from
trigonometric parallaxes (Benedict et al.~2011), the apparent magnitudes
of RR Lyraes in the LMC assuming $(m-M)_0 = 18.50$ (Clementini et al.~2003),
statistical parallaxes (Gould \& Popowski (1998), and the Baade-Wesselink
method (Fernley et al.~1998).  The slope of the {\it dashed line} is
$\Delta\,M_V/\Delta\,$[Fe/H] $= 0.214$, as reported by Clementini et al.  The
{\it solid curve}, which connects individual ZAHB models for specific [Fe/H]
values ({\it open circles}), gives the predicted dependence of $M_V$ on [Fe/H].
Since RR Lyraes in GCs that have [Fe/H] $\approx -1.5$ are observed to be
$\approx 0.10$ mag brighter than the faintest non-variable HB stars adjacent to
the instability strip (\citealt{san93}), the {\it dotted curve} ($=$ {\it solid
curve} minus 0.10 mag) represents the theoretical results that should be directly
compared with the observations.  {\it Lower panel}: The same ZAHB is compared
with with the $\log\,L$ values that have been determined for ZAHB stars at
$\log\,\teff = 3.85$ (see the text) from the pulsational properties of variables
in seven clusters, as indicated, by \citet{dsc99}.}
\label{fig:fig10}
\end{figure}

According to CBG09, M$\,$5 and M$\,$12 (NGC$\,$6218) have [Fe/H] $= -1.33$,
while NGC$\,$288 and NGC$\,$362 have [Fe/H] $= -1.32$ and $-1.30$, respectively.
Despite having virtually identical metallicities, NGC$\,$362 has a very red
horizontal branch, the HB populations of NGC$\,$288 and M$\,$12 are well
to the blue of the instability strip, and the HB of M$\,$5 spans a wide color
range that the other three GCs sample only partially.  These four GCs thus
provide an even better example of the second-parameter phenomenon than M$\,$3
and M$\,$13.  Since the $\delv$\ method has already been used to derive reliable
ages for M$\,$5 (11.5 Gyr, see Figs.~\ref{fig:fig8},~\ref{fig:fig9}) and
NGC$\,$362 (10.75 Gyr, see Fig.~\ref{fig:fig13}), a comparison of the CMD of
M$\,$5 (which is better defined than that of NGC$\,$362) with those of
NGC$\,$288 and M$\,$12 via a $\delc$\ analysis should help to explain why
the latter have much bluer HBs than the former.

As in the case of M$\,$3 and M$\,$13, we begin with a plot of the photometric
data.  Figure~\ref{fig:fig23} illustrates the CMDs of NGC$\,$288 and M$\,$12
from the upper MS to the lower RGB, along with the fiducial sequences that have
been derived to represent those observations (using the methods described in
connection with Fig.~\ref{fig:fig19}).  The densities of stars along the cluster
giant branches are less than ideal, but the least-squares fits (the solid
curves) appear to provide satisfactory representations of the RGB data.  The
result of registering these mean fiducial loci to that of M$\,$5 (in the usual
way) is shown in Figure~\ref{fig:fig24}.  Interestingly, the cluster subgiant
sequences are coincident to within their uncertainties, which suggests that all
three GCs have the same helium abundances (recall the upper left-hand panel of
Fig.~\ref{fig:fig17}).  It is also somewhat surprising to find that there is
no indication of an age difference between NGC$\,$288 and M$\,$5, while the 
latter seems to be $\sim 2$ Gyr younger than M$\,$12.\footnote{Because the
fiducial sequence that we derived for M$\,$5 appears to be especially well
determined, a plot to illustrate how well it reproduces the cluster photometry
has not been included here.  As a additional space-saving measure, a
$\delc$\ analysis of NGC$\,$362 has been omitted, as well, because it serves
only to confirm that NGC$\,$362 is $\approx 0.75$ Gyr younger than M$\,$5, as
found using the $\delv$\ technique.  This implies, of course, that NGC$\,$288
is older than NGC$\,$362 by only $\approx 0.75$ Gyr.}

The issue of whether or not an age difference is primarily responsible for the
very different HB types of NGC$\,$288 and NGC$\,$362 has been extensively
debated in the scientific literature.  Some studies (e.g., \citealt[][also see
\citealt{van00}]{svb96}) have argued that the two clusters are nearly the same
age, while others (\citealt[][\citealt{cat01}]{bff01}) have
concluded that they differ in age by $2 \pm 1$ Gyr.  While the result obtained
here lies approximately midway between these findings, it is based on the
assumption that both clusters, as well as M$\,$5, have the same abundances of
iron and the other heavy elements.  This may not be the case.
Figure~\ref{fig:fig25} compares isochrone and ZAHB loci with the CMD of
NGC$\,$288, on the assumption that its age is 11.5 Gyr, which requires that
$(m-M)_{APP} = 14.90$ if $E(B-V) = 0.012$ (Schlegel et al.~1998).  Comparing
this plot with its counterpart for M$\,$5 in the left-hand panel of
Fig.~\ref{fig:fig8} reveals that the RGB of NGC$\,$288 has a noticeably
shallower slope than that of M$\,$5 (or, incidently, that of NGC$\,$362, which
is morphologically very similar to the giant branch of M$\,$5).

The simplest explanation of this difference is that NGC$\,$288 has a higher
abundance of Fe, Mg, or Si given that these are the main elements that affect 
the predicted temperatures of giants (see VBD12).  Some support for this
possibility is provided by \citet{sk00}: according to their tabulated
spectroscopic results, NGC$\,$288 has $<$[Mg/Fe]$>$\ $= 0.43 \pm 0.03$ and
$<$[Si/Fe]$>$\ $= 0.43 \pm 0.03$ (where the uncertainty is the standard error
of the mean), as compared with $<$[Mg/Fe]$>$\ $= 0.36 \pm 0.03$ and
$<$[Si/Fe]$>$\ $= 0.35 \pm 0.04$ in the case of NGC$\,$362.   Whether the former
has a higher Mg$+$Si$+$Fe abundance than the latter depends on their relative
[Fe/H] values --- something that appears to be quite uncertain.  For instance,
although Shetrone \& Keane concluded from their work that NGC$\,$288 has a lower
iron abundance than NGC$\,$362 by 0.06 dex (in close agreement with the findings
of \citealt{ki03}), CGB09 reported a difference of only 0.02 dex (in the same
sense), while \citet{cg97} found that NGC$\,$288 had a {\it higher} [Fe/H] value
by 0.08 dex.

If NGC$\,$288 does have a higher Mg$+$Si$+$Fe abundance, then one would expect
that its SGB should be slightly flatter than those of NGC$\,$362 and M$\,$5
(see Figs.~\ref{fig:fig4} and~\ref{fig:fig24}, which indicate that this effect
is expected to become important at [Fe/H] $\approx -1.3$).  Since this effect
is opposite to that produced by a higher helium abundance, it is possible that
Fig.~\ref{fig:fig24} is giving us the misleading impression that M$\,$5 (and
NGC$\,$362) have the same helium abundances as NGC$\,$288 (and M$\,$12).
That is, NGC$\,$288 may actually have higher abundances of helium {\it and} Fe,
Mg, and/or Si.  Furthermore, isochrones for the same age and metal abundances,
but higher $Y$, have somewhat reduced RGB slopes (compare the solid and dashed
isochrones in the left-hand panel of Fig.~\ref{fig:fig22}).  If our estimated
relative ages are accurate, then a helium abundance difference may be the best
way of explaining most of the differences in the HB morphologies of NGC$\,$288,
NGC$\,$362, and M$\,$5.  [Curiously, \citet{sk00} found from their spectroscopic
analyses that NGC$\,$288 has a somewhat higher oxygen abundance than NGC$\,$362,
which goes in the wrong direction to help explain the differences in their HBs.
On the other hand, the earlier investigation by \citet{dcc91} had found nearly
idential Fe and C$+$N$+$O abundances in these two clusters.]

While the same concerns may apply to M$\,$12, the differences in the
MSTO-to-RGB colors of M$\,$5 and M$\,$12 are too large (see
Fig.~\ref{fig:fig24}) to be explained solely in terms of a helium abundance
variation, as there is no evidence from the superposition of the cluster SGBs
to support this possibility.  Some difference in the helium abundance or in the
mixture of the heavy elements cannot be excluded, but the simplest
interpretation of Fig.~\ref{fig:fig24} is that M$\,$12 is significantly
older than M$\,$5.  If its age is 13.0 Gyr, making it one of the oldest GCs in
our sample, the relevant isochrone and ZAHB fit to the observed CMD will be that
shown in the right-hand panel of Figure~\ref{fig:fig26}.  To obtain a
satisfactory fit of the ZAHB to the cluster HB stars, it is necessary to adopt
$(m-M)_{APP} = 14.05$ and $E(B-V) = 0.225$, which is higher than the reddening
derived from the Schlegel et al.~(1998) dust maps by $\approx 0.05$ mag.
Although a rather large color offset must be applied to the isochrone in order
to reproduce the turnoff color, the isochrone provides a better fit to
the photometry than one for, say, 11.5 Gyr (see the left-hand panel).  Both the
slope of the SGB and the location of the RGB are less problematic if the higher
age is assumed.  Perhaps the main argument in support of this possibility is
that the fit of stellar models to the MS and the RGB in the right-hand panel is
of comparable quality to that obtained for M$\,$5 (see Fig.~\ref{fig:fig8}).
In cases like M$\,$12, which has a steeply sloped, extremely blue HB, as
well as a relatively high (and possibly uncertain) reddening, the HB stars
clearly do not provide a very tight constraint on the cluster distance, and
hence its age.

\subsubsection{Low Metallicity (${\rm [Fe/H]} < -1.4$) Globular Clusters}
\label{subsubsec:blue}

The horizontal method, or the approach used by MF09, to determine the relative
ages of GCs that have similar metallicities is clearly preferable to the 
$\delv$\ technique when the clusters possess blue or extremely blue HBs.  It is
a concern, however, that the $\delc$\ diagnostic is sensitive to metal abundance
differences, as already pointed out.   Consequently, it is important to 
intercompare the CMDs of clusters that have as close to the same [Fe/H] value
as possible so that the effects of the observed [Fe/H] differences on the
measured cluster-to-cluster variations in the $\delc$\ parameter are small.
For this reason, we initially selected M$\,$5, M$\,$3, NGC$\,$4147, and M$\,$53
(NGC$\,$5024), which have well-determined $\delv$\ ages, to be the reference
clusters at their metallicities ([Fe/H] $= -1.33$, $-1.50$, $-1.78$, and
$-2.06$, respectively, according to CBG09).  However, during the course of our
investigation, we discovered that, whereas the aforementioned GCs appear to
have subgiant branches with similar slopes, despite spanning a wide range in
metallicity, approximately half of the target clusters had more steeply sloped
SGBs.  In fact, the differences in the MSTO-to-RGB morphologies of the two
groups of clusters appear to resemble those of M$\,$3 and M$\,$13; recall the
discussion in \S~\ref{subsubsec:m3m13}.

The upper left-hand panel of Figure~\ref{fig:fig27} reproduces the comparison
of the M$\,$3 and M$\,$13 fiducials from Fig.~\ref{fig:fig20}, together with
that derived for NGC$\,$4147 to show that the latter looks much more
``M$\,$3-like" than ``M$\,$13-like" insofar as the slope of its SGB is
concerned.  (We are interested only in the part of each CMD from just past the
turnoff to an abscissa value of $\approx 0.13$, as the location of the base of
the RGB in a cluster will depend on its age.)  The fact that the turnoffs and
the SGBs of NGC$\,$4147 and M$\,$3 are nearly coincident, despite having [Fe/H]
values that differ by 0.28 dex, indicates that the differences between the
solid and dashed loci are approximately independent of metallicity (at least in
the range spanned by these two clusters).  Moreover, since the slope of the SGB,
at a fixed [Fe/H] value, has no more than a slight dependence on age (see
Fig.~\ref{fig:fig18}), the near overlap of the NGC$\,$4147 and M$\,$3 fiducials
suggests that they have similar helium abundances.  The remaining panels in the
top two rows of Fig.~\ref{fig:fig27}, which consider GCs with [Fe/H] values
within 0.15 dex of those of M$\,$3 and M$\,$13 likewise suggest that NGC$\,$6584
and M$\,$70 (NGC$\,$6681) have approximately the same $Y$ as M$\,$3 while
NGC$\,$6752 probably has an enhanced helium abundance comparable to that
previously deduced for M$\,$13 (see \S~\ref{subsubsec:m3m13}).  Indeed, support
for this possibility is provided by the very recent discovery of three distinct
sequences in {\it HST}/WFC3 CMDs of NGC$\,$6752, which are characterized by
$\Delta\,Y$ values ranging from 0.01 to 0.03 (see \citealt{mmp13}).

The bottom row of Fig.~\ref{fig:fig27} presents similar analyses of the TO and
subgiant CMDs of two somewhat more metal-deficient GCs using NGC$\,$4147 and
NGC$\,$4833 as the adopted reference clusters.  The M$\,$13 fiducial has
been included in the left-hand panel to show that its SGB slope is quite 
similar to that of NGC$\,$4833.  Curiously, M$\,$53, which is about 0.3
dex more metal-poor than NGC$\,$4147, appears to belong to the same ``family"
as NGC$\,$4147 (and M$\,$3) given that their SGBs have nearly the same slope,
while the appreciably steeper subgiant branch of NGC$\,$5286 is a shared
property with NGC$\,$4833 (and M$\,$13).  In the first (submitted) version of
this paper, additional plots similar to those in Fig.~\ref{fig:fig27} were
included for all of the GCs that have [Fe/H] $\lta -1.5$.  They showed, among
other things, that M$\,$92 and M$\,$15 appear to have steep SGBs which resemble
those of NGC$\,$4833 and M$\,$13, and that NGC$\,$5466 has the connection of a
relatively flat SGB via M$\,$53 and NGC$\,$4147 to M$\,$3. 

This is especially interesting in view of the fact that \citet{smw10} have found
little evidence for star-to-star variations in the abundances of the light
elements in NGC$\,$5466, which suggests that the latter consists almost entirely
of first-generation stars.  This would be consistent with our working hypothesis
that those metal-poor clusters with the flattest SGBs have close to the
primordial helium abundance and presumably, therefore, a much smaller fraction
of second-generation stars than GCs that have steeper subgiant branches and
(possibly) higher $Y$.  An example of a cluster that is chemically very
different from NGC$\,$5466 is M$\,$15, which happens to be one of the most
massive clusters in the Milky Way.  \citet{cbs05} have found that the C and N
abundances in its subgiant and lower RGB stars vary by $\sim 1.5$ and $\sim 2.5$
dex, respectively.  Although they did not measure the abundance of oxygen, C$+$N
was found to increase as C decreases, which requires (as noted by them) the
incorporation of ON-processed material.  Because these stars are too faint for
deep mixing to alter surface abundances (see \citealt{dv03}), Cohen et
al.~concluded that there must have been an extended period of star formation in
M$\,$15 involving at least two generations of massive stars.  (Variations in
$Y$ {\it and} [CNO/H] could well be the reason why its horizontal branch extends
to both very blue and very red colors.)

Until the effects of differential reddening (when it is important) and chemical
abundance variations have been quantified and disentangled, it is not possible
to determine the relative ages of the different stellar populations in M$\,$15,
or in any other GC that has a ``thick" SGB.  At this stage, we are thus unable
to say anything about O (or CNO) variations when attempting to interpret the
photometric data.  However, because the slope of the subgiant branch appears to
be a fairly sensitive function of the helium abundance, and not much else, this
diagnostic may enable us to separate the GCs with [Fe/H] $\lta -1.5$ into low-
and higher-$Y$ groups.  However, a valid question to ask is how accurately can
the SGB slopes be determined, and indeed, one of the referees of this paper
urged us to find a way of attaching an errorbar to the derived slopes so as to
provide a measure of the significance of the results.  In response to this
recommendation, we decided to retain only the few comparisons of fiducial
sequences that are shown in Fig.~\ref{fig:fig27} and to evaluate the SGB slopes
using the linear least-squares subroutine ``FIT" from {\it Numerical Recipes}
(\citealt{pft86}), as it yields both the slope and its uncertainty for the
best-fit line through data.

Some care must be taken in selecting the part of the SGB that is fitted by a
straight line since there is obvious curvature near the turnoff and near the
lower RGB.  Based on an inspection of isochrones and the observed GC CMDs, we
opted to perform a least-squares fit to those SGB stars from 0.045 mag redder
than the TO to 0.13 mag redder than the TO, in the case of clusters with [Fe/H]
$< -1.9$.  Because more metal-rich GCs have shorter subgiant branches, the
adopted color range was slightly reduced at higher metallicities to include
only those stars from 0.05 mag redder than the TO to 0.125 mag redder than the
TO.  Except for this slight difference in the selection criteria, all clusters
were treated in the same way.  Figure~\ref{fig:fig28} shows some examples of
the observed CMDs in which dashed rectangles delineate the regions of the SGBs
that have been subjected to a linear least-squares analysis.  These CMDs are
not the best, nor the worst, ones in our data base insofar as the star-to-star
scatter is concerned.  The two clusters in the left-hand panels have nearly
the same [Fe/H], as do those in both right-hand panels, to within 0.03 dex
according to CBG09, and it is apparent even from a visual inspection that
M$\,$15 has a steeper SGB than NGC$\,$5466 and NGC$\,$6752 has a steeper
subgiant branch than NGC$\,$6934.  The derived slopes and their uncertainties
are indicated in each panel.

Note that the median fiducial sequences which have been determined for the MS
and TO stars (the filled circles) are very smooth, as in similar plots which
have already been discussed (see Figs.~\ref{fig:fig19} and~\ref{fig:fig23}).
We investigated whether these sequences depended on the photometric errors by
deriving them for CMDs that included only those stars with errors in the
tabulated magnitudes $< 0.01$ mag, or $< 0.02$ mag, or $< 0.03$ mag and,
although we made such tests for only a limited number of the clusters, the
resultant fiducial sequences were always nearly identical.  There were slight
shifts of the median points for just a few bins in some clusters, but regardless
of whether the CMD was well populated and tightly defined (e.g., M$\,$13), had
relatively few stars and considerable scatter (like NGC$\,$288), or was
especially broad due (likely) to differential reddening (as in the case of
NGC$\,$5986), the curves defined by the median points did not differ
significantly for the  different data sets.  Recall from Fig.~\ref{fig:fig3}
that the slope of the SGB at a fixed age tends to increase inversely with
[Fe/H].

Figure~\ref{fig:fig29} plots the slopes, and their uncertainties, for the GCs in
our sample with [Fe/H] $\lta -1.5$.  (Because the slope of the SGB becomes quite
a sensitive function of [Fe/H] at higher metallicities, as already discussed,
little is to be gained by extending the analysis to more metal-rich clusters.)
The dashed line represents the best-fit line through all of the points, as 
derived from the {\it Numerical Recipes} FIT subroutine.  This separates the
clusters into groups that have relatively steep or shallow SGBs (the filled and
open circles, respectively): the few GCs that straddle the line and hence have
intermediate SGB slopes are depicted as open circles with black dots at their
centers.  The clusters comprising each of the three groups are identified in
Table~\ref{tab:tab1}, which also lists the information that has been plotted.
For the most part, the derived slopes support the division of the GCs into
``M$\,$3-like" or ``M$\,$13-like" as found in our initial analysis, which was
based on comparisons of median SGB sequences like those shown in
Fig.~\ref{fig:fig27}.  Only in a few cases have the classifications changed from
our initial determinations, but this is to be expected given the size of the
$1\,\sigma$ errorbars for several cases.  Indeed, at the $2\,\sigma$ level,
most of the errorbars would intersect, which shows that the cluster-to-cluster
differences in their SGB slopes are really quite subtle.

Before examining whether these two groups can be differentiated in other ways,
it is necessary to determine the ages of those clusters which are not amenable
to a $\delv$\ analysis (since age may be one of the distinguishing
characteristics).  It is unfortunate that the MSTO-to-RGB color difference is
a function of the metallicity since a $\pm 0.1$ dex uncertainty in the relative
[Fe/H] values implies an error bar of $\sim \pm 0.25$--0.5 Gyr in the age
difference that is derived using the VBS90 approach.  For instance, we found
that M$\,$3 is $\lta 0.25$ Gyr younger than NGC$\,$4147, while M$\,$53 is
$\approx 1$ Gyr older than NGC$\,$4147, if the $\delc$\ technique is used to
estimate their relative ages.  However, these results are at odds with those
already obtained: according to Figs.~\ref{fig:fig12} and~\ref{fig:fig22},
M$\,$3, NGC$\,$4147, and M$\,$53 have ages of 11.75, 12.25, and 12.25 Gyr,
respectively.  It is possible to reconcile the relative ages of the three
clusters from the horizontal and vertical methods if the [Fe/H] value of
NGC$\,$4147 is close to $-1.9$ (instead of $-1.78$, as reported by CBG09) and
either M$\,$53 is slightly more metal-rich that our adopted value of [Fe/H]
and/or it is $\sim 0.25$ Gyr older the $\delv$\ estimate.  (It would be very
encouraging, indeed, if future spectroscopic studies confirm these predictions.)

Fortunately, such cases were found to be in the minority; in general, there was
reasonably good consistency of the relative $\delc$\ ages of a given GC when
determined with respect to both higher and lower metallicity reference clusters.
Nevertheless, in view of the M$\,$3--NGC$\,$4147--M$\,$53 example, the
$1\,\sigma$ uncertainties of ages that are based solely on the horizontal method
must be at least $\pm 0.5$ Gyr.  As a result of this example, and a few others
that showed similar discrepancies (see below), considerable time and effort was
spent examining the impact on the inferred ages of using different reference
clusters and of checking the degree of consistency with the $\delv$\ ages, even
if the latter had not been derived previously.  It is helpful to consider, in
particular, if isochrones for the adopted ages provide fits to the cluster
turnoffs and giant branches of comparable quality as those reported in 
\S~\ref{subsec:std} irrespective of how well the ZAHBs match up with the HB
populations.  Blindly accepting the $\delc$\ ages without such cross-checks
would have resulted in different ages for several of the GCs, though the net
effect on the mean age--[Fe/H] relation that is derived in this study would not
have been very large, given that the number of age determinations that are
based primarily on the horizontal method is $< 20$\% of the total.

Figure~\ref{fig:fig30} presents the application of the $\delc$\ method to those
clusters for which the horizontal approach is deemed to be more reliable than
the $\delv$\ technique, along with some counter-examples.  In each panel, the
solid and dashed curves represent, in turn, the lower giant-branch portions of
the fiducial sequences of the reference and target GCs, which are identified in
the upper left-hand corner.  These sequences were registered to the usual
abscissa and ordinate zero-points (as in, e.g., Fig.~\ref{fig:fig24} and
Figs.~\ref{fig:fig27}--\ref{fig:fig29}), but only the comparison of their RGB
segments are shown, together with a set of isochrones for the metallicity of the
reference cluster, since nothing more is needed to derive the difference in
their ages.  Thus, considering the top row, and noting the corrections that
should be applied to the solid curve to account for metallicity differences,
one finds that M$\,$3 is predicted to be 0.25--0.5 Gyr older than M$\,$5,
$\lta 0.25$ Gyr older than NGC$\,$3201 (though the CMD of the latter is very 
poorly defined), coeval with NGC$\,$6584, and $\sim 1.5$ Gyr younger than
M$\,$70.  Except for M$\,$70, which was not considered in
\S~\ref{subsec:std}, these results are consistent to within $\pm 0.25$ Gyr with
the differences in the absolute cluster ages previously derived using the
$\delv$\ method.  Indeed, the same conclusion is reached in the case of
NGC$\,$6934 (see the left-hand panel in the second row from the top) and
M$\,$72 (not shown, but which is clearly coeval with M$\,$3).

In the next five panels, M$\,$13 is the adopted reference cluster and the 
target clusters include several of those systems which, like M$\,$13, appear
to have steeper SGBs than M$\,$3.  For three of them (M$,$10, NGC$\,$6752,
and M$\,$2), the level of consistency with the results of $\delv$\ analyses
is quite satisfactory.  As a space-saving measure, plots showing fits of
isochrones and ZAHB loci to the CMDs of clusters with blue HBs have generally
not been included in this paper, though they have been produced and carefully
examined.  Indeed, in the case of NGC$\,$6752, which has a relatively low
and well-determined reddening ($E(B-V) = 0.056$; Schlegel et al.~1998), the fit
of a ZAHB for [Fe/H] $= -1.55$ (see CBG09) to the cluster HB stars yields
$(m-M)_{APP} = 13.23$, implying an age of 12.50 Gyr (i.e., 0.50 Gyr older than
M$\,$13, which is also favored by the comparison shown in Fig.~\ref{fig:fig30}).
However, the M$\,$13--NGC$\,$5286 and M$\,$13--NGC$\,$5986 intercomparisons
suggest that M$\,$13 is younger than either of the other clusters by $\gta
1.25$ Gyr.  The age of NGC$\,$5286 was previously found to be 12.25 Gyr (see
Fig.~\ref{fig:fig12}, which is unlikely to be wrong by more than
0.5 Gyr; consequently, we suspect that the much older age that is suggested by
Fig.~\ref{fig:fig30} is due, in part, to the assumption of a metallicity that
is too low.  If the difference in the adopted [Fe/H] values is reduced, the
inferred ages of M$\,$13 and NGC$\,$5286 would be more similar.  The same
explanation may apply to NGC$\,$5986, though the CMD of this system is clearly
affected by differential reddening, which my affect the determination of its
mean photometric sequence.  Stellar populations effects may also
complicate the interpretation of the CMDs of NGC$\,$5286 and NGC$\,$5986 if our
suspicion that age and chemical abundance variations are greater in the group of
GCs that has been classified as ``M$\,$13-like" is correct.

NGC$\,$4833 is another cluster that presented some difficulties when its CMD
is compared with those of other clusters that have similar [Fe/H] values.
Hence, for the next group of GCs in Fig.~\ref{fig:fig30}, NGC$\,$5286 was chosen
to be the reference cluster.  With this choice, the ages of NGC$\,$6144,
NGC$\,$6541, M$\,$22, and M$\,$56 relative to that of NGC$\,$5286 were
found to be in satisfactory agreement with the differences in their absolute
ages as derived from the $\delv$\ technique.  In the case of NGC$\,$4833, an age
of 12.50 Gyr has been adopted, which represents a compromise of the ages that
are found from the two methods.  The results shown in the remaining panels of
Fig.~\ref{fig:fig30} were all found to be in acceptable agreement with those
based on fits of isochrones and ZAHBs to the observed CMDs.  For none of them
were the derived age differences from the $\delc$\ and $\delv$\ analyses
$> 0.50$ Gyr.  For instance, whereas Fig.~\ref{fig:fig11} indicates that M$\,$92
and M$\,$30 are coeval, Fig.~\ref{fig:fig30} suggests that they differ in age by
0.50 Gyr.  Accordingly, we have adopted an age of 13.0 Gyr for M$\,$30 and
thereby achieved consistency with both determinations to within $\pm 0.25$ Gyr.
As noted in \S~\ref{sec:summary}, similarly averaged ages (sometimes taking a
subjective assessment of the reliability into account) have been taken to
be the ``best" estimates of the ages of a few other GCs.

Aside from a few problematic cases, our age determinations appear to be quite
well constrained --- though one must continue to be wary of the $\delc$-based
results given their very considerable sensitivity to metal abundances.  Not
surprisingly, the ages also depend on whether or not we have correctly
classified a given cluster into one of the two groups of GCs that are
distinguished by different SGB slopes, and on whatever is responsible for that
distinction.  (Helium abundance differences, which could arise as a consequence
of cluster-to-cluster variations in the chemical properties of the stellar
populations that they contain, have been tentatively identified as the cause of
such morphological differences, but there may be other, as yet unidentified,
factors at play.)  Figure~\ref{fig:fig31} shows, for example, that much older
ages ($\gta 13$ Gyr) would have been derived for NGC$\,$6752 and M$\,$56 had
we used M$\,$3 and M$\,$53 as the reference clusters, respectively (instead
of M$\,$13 and NGC$\,$5286, which appear to belong to the same ``family").
Indeed, the oldest clusters in the group in which M$\,$13 has been taken to be
the prototype would have predicted ages as old, or older, than the age of the
universe if the $\delc$\ method were used to obtain their ages relative to
that of M$\,$3.  By the same token, the oldest GCs in the M$\,$3 group would
have been found to have younger ages by $\sim 0.5$ Gyr than our adopted
estimates had their fiducial sequences for the MSTO-to-RGB stars been compared
with that of M$\,$13. 


\subsubsection{Higher Metallicity (${\rm [Fe/H]} \ge -1.4$) Globular Clusters} 
\label{subsubsec:red}

Because the effects of variations in the abundances of the heavy elements on
isochrones become larger as the metallicity increases (see VBD12), relative
age determinations based on the horizontal method will be especially uncertain
at higher [Fe/H] values.  Fortunately, the red HBs of metal-rich systems are
more easily fitted by computed ZAHBs than the blue horizontal branches that are
typical of metal-poor GCs, and since the $\delv$\ technique has a reduced
dependence on the metal abundance (see \S~\ref{subsec:m5} for an instructive
discussion of this point), the vertical method is clearly the ``method of
choice" when determining ages (particularly cluster-to-cluster differences in
in age) at higher values of [Fe/H].  For these reasons, no attempt has been made
to apply the $\delc$\ method to GCs that have [Fe/H] $\ge -1.4$.  We note, as
well, that we were unable to detect any SGB slope variations in the CMDs of
metal-rich clusters; see, e.g., the comparisons of the fiducial sequences of
M$\,$5, NGC$\,$288, and M$\,$12 that are shown in Fig.~\ref{fig:fig24}.
It may, in any case, be difficult to determine whether such variations exist
because the slope of the SGB is predicted to be an increasingly sensitive
function of the metal abundance at [Fe/H] values above $\sim -1.4$ (see
Fig.~\ref{fig:fig4}).  Consequently, the effects on the SGB due to a metal
abundance increase could be comparable to, and compensate for, those caused by
a moderate helium abundance enhancement, and vice versa.  

\subsection{Six Additional Globular Clusters, Including Four Associated with the
Sagittarius Dwarf Galaxy}
\label{subsec:other}

Although there are 64 GCs in the Sarajedini et al.~(2007) data set, the CMDs
for several of them are not sufficiently well defined to permit the derivation
of accurate fiducial sequences.  This is especially true in the case of the
[Fe/H] $\sim -0.45$ clusters NGC$\,$6388 and NGC$\,$6441, due to field-star
contamination and/or the effects of significant differential reddening.   (Since
they are among the most massive and tightly-bound clusters in the Milky Way,
stellar populations effects likely contribute to the breadth of their
photometric sequences as well.  In fact, distinct sub-populations of stars have
been discovered in them by \citealt{bpm13}.)
NGC$\,$2298 ([Fe/H] $= -1.96$) and NGC$\,$6093
([Fe/H] $= -1.75$) were found to be similarly problematic, in that their stellar
distributions in the vicinity of the turnoff are clearly asymmetric, which makes
the determination of the turnoff color, the slope of the SGB, and the value of
the $\delc$\ parameter in either cluster quite uncertain.  Furthermore, our
attempt to fit ZAHB loci to their blue HBs suggests that the reddenings given by
the Schlegel et al.~(1998) dust maps are too low by $\gta 0.04$ mag.  Among
those GCs without NGC numbers, the paucity of stars in the CMDs of E$\,$3
and Pal$\,$1, and the ambiguous location of the subgiant branch and the very
broad giant branch in the CMD of Lynga 7, preclude the possibility of precise
age determinations for these systems.  In view of these difficulties, we decided
to drop these seven clusters from our sample, along with Terzan 7, because our
model grids do not extend to sufficiently high metal abundances to produce
isochrones and ZAHBs for its metallicity ([Fe/H] $= -0.12$, according to CBG09).

Up to this point, ages have been derived for 50 of the remaining 55 globular
clusters.  For the rest, the observed CMDs and the derived $\delv$\ ages are
presented in Figure~\ref{fig:fig32}.  [In order to have two complete rows of
three panels, an isochrone fit to the CMD of M$\,$70 is included, even
though its age was previously determined using the horizontal method.  This
particular cluster was selected because, despite its very blue HB, the
ZAHB-based distance that is obtained on the assumption of the reddening given
by Schlegel et al.~(1998) implies an age which agrees well with that inferred
from the $\delc$\ approach; see Fig.~\ref{fig:fig30}.  By assessing the
consistency of the results from the two age-dating methods, we have obtained
an age for M$\,$70 that appears to be quite well constrained.  Similar 
assessments have been made for other clusters with blue HBs, but with varying
degrees of success.]  Because NGC$\,$6535 has such a sparsely populated CMD,
the best estimate of its age is obtained using the $\delv$\ method.  However,
unlike the example of M$\,$70, it is not possible to obtain a satisfactory
fit of stellar models to the photometric data unless the Schlegel et
al.~reddening estimate is increased by about 0.07 mag.  With such an adjustment
(see the top left-hand panel), the ZAHB for the observed [Fe/H] value provides
a satisfactory match to the few HB stars that are present and, in particular,
the best-fit isochrone reproduces the MS, SGB, and lower RGB stars about as
well as we have been able to achieve for most of the GCs.

\citet{igi95} discovered that M$\,$54 (NGC$\,$6715), Arp 2, and Terzan 8 are
associated with the Sagittarius dwarf galaxy, while a similar origin of Pal 12
was deduced by \citet{dmg00} from an analysis of its orbit.  In support of
previous age determinations (e.g., \citealt{van00}), Pal 12 appears to be an
unusually young system, with an age near 9 Gyr.  However, the ages derived here
for M$\,$54, Arp 2, and Terzan 8 are all quite similar to those found for other
globular clusters of similar metallicities.  As CBG09 did not determine the
mean iron abundance of stars in Ter 8, its [Fe/H] value has been taken from the
spectroscopic study by \citet{mwm08}, who obtained $-2.34$ from their analysis
of Fe I and Fe II lines.  This estimate would appear to be on nearly the same
scale as the one by CBG09 given that Mottini et al.~also determined that Arp 2
has [Fe/H] $= -1.77$ (from Fe I lines), which is in very good agreement with the
CBG09 determination of $-1.74$.  Only in the case of Pal 12 have models for
[$\alpha$/Fe] $= 0.0$ (e.g., \citealt{bwz97}) been used to fit the cluster CMD.
According to Mottini et al., the observed $\alpha$-element enhancements in
Arp 2 and Ter 8 are ``not too different" from those found in Galactic GCs at
similar [Fe/H] values, and we have assumed that this conclusion also applies to
M$\,$54.  To be specific, models for [$\alpha$/Fe] $=0.46$ have been fitted to
the CMDs of Arp 2, Ter 8, and M$\,$54. 

Interestingly, Mottini et al.~derived [O/Fe] $= 0.21 \pm 0.22$ for Arp 2 and
$0.71 \pm 0.17$ for Ter 8.  These differences have not been taken into account.
However, it is very difficult to measure the oxygen abundances in metal-poor
stars (see, e.g., \citealt{rar06}) and the assumed value, [O/Fe] $= 0.50$,
which appears to be close to the best estimate of the mean oxygen overabundance
in stars of low metallicity (see, e.g., Ram\'irez et al.~2012), is just outside
the $1\,\sigma$ uncertainties of the derived values.  It is surprising that a
ZAHB for [Fe/H] $= -1.44$ provides such a good fit to the HB population of
M$\,$54 given that \citet{cbg10a} have found a $\sim 0.2$ dex dispersion in its
metallicity.  Although an 11.75 Gyr isochrone for [Fe/H] $= -1.44$ provides a
reasonably good match to its CMD, M$\,$54 is clearly a complex system that could
well have somewhat younger or older components.  No attempt has been made here
to explore the impact of age and/or chemical abundance variations, though our
derived age is expected to be a reasonably accurate estimate for its dominant
stellar population.

\section{Globular Cluster Ages and Their Correlations with Other Properties}
\label{sec:summary}

The ages that have been derived in this investigation for the 55 GCs that
comprise our sample are listed in the fourth column of Table~\ref{tab:tab2}.
The first two columns identify the clusters, and the third column lists the
adopted [Fe/H] values from the study by CBG09 (or, in the case of Terzan 8,
from \citealt{mwm08}).  The letters ``V" or ``H" in the fifth column indicate,
in turn, whether the adopted age is based primarly on the vertical, or the
horizontal, method.  For those clusters in which the letter ``A" appears in
the fifth column, the $\delv$\ and $\delc$\ results have been averaged in order
to obtain the tabulated age.   To support these age esimates, at least one
plot is provided for each GC: the sixth column contains the reference(s) to the
relevant figure(s).  The last five columns list, in the direction from left to
right, the HB type (from \citealt{mv05}), the Galactocentric distance (in kpc),
the absolute integrated visual magnitude (both R$_{\rm G}$ and $M_V$ have been
taken from the 2010 edition of the catalog by \citealt{har96}), the central
escape velocity (in km/s), and the common logarithm of surface density of stars
at the cluster center (in $\msol/pc^2$).  (The calculation of the last two
quantities is described in \S~\ref{subsec:twogcs}.)  In this study, the HB type
is defined to be (B$-$R)/(B$+$V$+$R), where B, V, and R represent the number of
stars that lie blueward of the instability strip, the number of variable
(RR Lyrae) stars, and the number of stars that are on the red side of the
instability strip, respectively.

The $\delv$\ ages that were independently determined by the UVic participants
in this collaborative project (i.e., DAV, KB, and RL) spanned the age ranges
that are specified in the seventh column.  (If, as in a few cases, exactly the
same age was found, only a single number is reported.)  In general, these
estimates agreed to within 0.25--0.50 Gyr, which is the basis for adopting
$\pm 0.25$ Gyr uncertainties for what we consider to be the best determined
ages.  (These are {\it internal} uncertainties; i.e., they do not include the
effects of distance or chemical abundance errors.  As the Victoria models
appear to satisfy the constraints provided by standard candles reasonably well,
we believe that the derived distance moduli are accurate to within $\pm 0.10$
mag, which implies an age uncertainty of $\approx \pm 1$ Gyr.  As shown by
VBD12, turnoff luminosity versus age relations for low metallicity stars are
strong functions of primarily the helium and oxygen abundances.  Although our
adopted values of $Y$ and [O/Fe] represent current best estimates, there may
well be cluster-to-cluster variations in these quantities, in the mean, that
impact age determinations at the level of 0.5--1.0 Gyr.  For instance, a $\pm
0.15$ dex uncertainty in the assumed [O/Fe] value corresponds to an age
uncertainty of $\pm 0.5$ Gyr.)  As is widely appreciated, relative ages are
more secure than absolute ages.

Because they depend quite sensitively on the adopted metallicity, ages based
primarily on the horizontal method have generally been considered to be
uncertain by $\pm 0.50$ Gyr, though this error bar was reduced to $\pm 0.38$ Gyr
(i.e., midway between $\pm 0.25$ and $\pm 0.50$ Gyr) if the two age-dating
methods that we have employed yielded fully consistent results (as in the case
of, e.g., M$\,$70).  A larger uncertainty (i.e., $\pm 0.75$ Gyr) was
ascribed to the derived ages of only two clusters, NGC$\,$5986 and M$\,$107,
in view of the fact that their CMDs are especially problematic (broad and
asymmetric stellar distributions).  Note that, because the ZAHB-based distances
are uncertain by at least $\pm 0.01$--0.02 mag, even in the most favorable cases,
our final age estimates have been rounded to the nearest 0.25 Gyr.  Note, as
well, that the adopted age is always within the range that is listed in the
seventh column (when one is specified), though not necessarily in the middle of
that range.  After $\delv$\ ages had been independently derived by DAV, KB, and
RL, the different fits of isochrones and ZAHBs to the observed CMDs that had
been obtained by the three investigators were carefully scrutinized and the one
that was judged to be the most agreeable comparison between theory and
observations was selected.  These cases are the ones that have been reproduced
in Figs.~\ref{fig:fig11}--\ref{fig:fig14} and Fig.~\ref{fig:fig16}. 

\subsection{The Age--Metallicity and Age--Galactocentric Distance Relations}
\label{subsec:agefe}

Figures~\ref{fig:fig33} and~\ref{fig:fig34} show how the ages that we have
determined vary, in turn, with [Fe/H] and with Galactocentric distance,
R$_{\rm G}$.  The first of these plots bears more than a passing resemblance
to a similar diagram that was constructed by \citet[see his Fig.~40]{van00}.  In
fact, it should not be a surprise that the age--metallicity relation (AMR)
obtained here is qualitatively very similar to the one produced 13 years ago
because the same $\delv$\ technique was used in both studies to derive the ages
and, more importantly, the respective ZAHB models predict nearly the same
dependence of $M_V$(RR Lyrae) on [Fe/H].  It is to be expected that there will
be, and are, differences in the absolute ages given that, in particular, the
current Victoria-Regina isochrones take the diffusion of helium and the latest
nuclear reactions into account.  In addition, there are some significant
differences in the adopted chemical abundances.  However, for the GCs in common
to the two studies, the predicted ages differ by $\lta 1$ Gyr, which is well
within the error bars due to distance and [O/Fe] uncertainties.  To be sure, the
present results are more robust because they have been derived from a much
larger sample of GCs with superior photometry using improved stellar models that
allow for variations in the abundances of helium and several metals.  The
availability of models for different values of $Y$, in particular, have enabled
us to determine that the morphological variations that are seen in the CMDs of
clusters with similar metallicities cannot be explained solely in terms of age
differences.

According to Fig.~\ref{fig:fig33}, GCs more metal-poor than [Fe/H] $\sim -1.7$
have a mean age of $\approx 12.5$ Gyr with a dispersion of $\sim \pm 0.5$ Gyr.
At higher [Fe/H] values, one has the visual impression that the AMR
is bifurcated, with one branch running from approximately 12.5 Gyr at [Fe/H] $=
-1.7$ to 11 Gyr at [Fe/H] $= -1.2$, while the other is offset to higher
metallicities by $\approx 0.6$ dex at a fixed age.  Remarkably, it turns out
that most of the clusters in the latter sequence, which include M$\,$12,
NGC$\,$6362, NGC$\,$6717, and 47 Tuc, have disk-like kinematics according to
\citet{dgv99}, and it is possible that the other two [Fe/H] $< -1.0$ GCs in this
small group (NGC$\,$6717 and NGC$\,$6723) do so as well, but we have been unable
to find any orbital information for them.  In contrast, the
intermediate-metal-poor clusters that consitute the other branch (including
M$\,$2, M$\,$3, M$\,$5, M$\,$13, NGC$\,$288, and NGC$\,$362, and a few others)
seem to have mostly (exclusively?) halo-type orbits (also see \citealt{dvg99}).
Although Pal 12 does not appear in this plot (because its age is outside the
range of the ordinate), its age (9.0 Gyr) and [Fe/H] value ($-0.81$) would place
it close to a linear extension of the left-hand branch of the AMR to higher
metallicities, while the clusters with [Fe/H] $> -0.9$, which have long been
known to have a flattened spatial distribution (\citealt{z85}), lie along a
continuation of the right-hand branch.

The split AMR obviously has important implications for the formation/assembly
of the GC system belonging to the Milky Way.  Since it would be worthwhile to
perform an expanded analysis of the cluster kinematics, to compare our AMR with
those derived for stars in the solar neighborhood (e.g., \citealt{csa11}) and
for such dwarf galaxies as Sagittarius and the Large Magellanic Cloud
(e.g., \citealt{lvb13}), and to examine the consistency of the results with 
different formation scenarios (e.g., Elmegreen et al.~2012), further discussion
of Fig.~\ref{fig:fig34} has been deferred to a study by 
\citet{lvm13}.  However, one issue that should be addressed here is why
our results conflict with those of MF09 (and Dotter et al.~2010), who argued
that the majority of the Galactic GCs are nearly coeval, aside from a
relatively small subset of young, [Fe/H] $\lta -1.2$ systems that follow a
single AMR with a significant slope.  In fact, the main point of contention is
whether the GCs that have [Fe/H] $> -1.0$ are of comparable age as the most
metal-deficient clusters, as they reported, or younger by $\approx 1.5$--2 Gyr,
as derived in this investigation.  In the next section, an explanation of the
main cause of this discrepancy is provided.
 
However, before turning to that discussion, a few comments should be made 
concerning Fig.~\ref{fig:fig34}.  First, considering the entire data set, there
is virtually no dependence of the mean age or the dispersion in age on R$_{\rm
G}$.  Second, whereas the clusters in our sample with [Fe/H] $\ge -1.0$ are all
located within 8 kpc from the center of the Milky Way, those of lower metal
abundances are distributed over the entire range in Galactocentric distance
that has been plotted.  (Pal 12, which has [Fe/H] $= -0.81$ and R$_{\rm G} =
15.8$ kpc, is an ``exception to the rule", and there are likely to be others,
but this cluster is believed to have originated in the Sagittarius dwarf galaxy
and therefore {\it may} belong to a different category than the other metal-rich
GCs in our sample.  Moreover, as noted above, Pal 12 appears to be connected
with a different AMR than the other GCs in our sample that have [Fe/H] $\sim
-0.8$.)  Finally, there is quite a striking difference in the ages of the
clusters that have [Fe/H] $< -1.7$ and those that have [Fe/H] $\ge -1.0$, which
is, of course, just a reflection of the AMR shown in the previous figure.

Our finding that old, very metal-poor GCs are found at any R$_{\rm G}$, while
the majority of the metal-rich clusters are younger and located at smaller
Galactocentric distances, suggests an ``outside--in" scenario for the formation
of the Galaxy.  Given the evidence of Sagittarius, there is no denying that
mergers of dwarf galaxies are responsible for many of the GCs that reside in the
Milky Way.  However, it also seems very likely that many of them formed during
the collapse of a single, massive protogalaxy, perhaps as described by
\citet[also see \citealt{san90b}]{els62}, or by \citet{har09}, who suggested
that protogalactic filaments first collapsed in a direction perpendicular to
their lengths and then along their lengths.  To improve our understanding, it 
would be helpful to include Bulge GCs in the sample, as well as clusters that 
are located at R$_{\rm G} \gta 20$ kpc. [In fact, the AMR that is analyzed by
Leaman et al.~(2013) includes the six outer-halo clusters that were the 
subject of a paper by \citet{dsa11}.]

\subsubsection{Relative Ages Based on the rMSF Method of Mar\'in-Franch et
al.~(2009)}
\label{subsubsec:mf09}

It took some careful detective work to understand why the results obtained by
MF09 conflict with ours.  In fact, the main reasons for the difference is that
(i) the rMSF method that was devised, and used, by MF09 is especially sensitive
to the adopted metal abundances and (ii) the [$m$/H] values that they adopted
involve some inconsistencies.  This can be appreciated by considering the
results shown in Figure~\ref{fig:fig35}, which plots the age--[$m$/H] relations
that are obtained using the rMSF method on the assumption of different
metallicity scales.  The top two panels reproduce their Figs.~10 and 11 for
a subset of the sample of GCs that they considered: only those clusters
with normalized relative ages $< 0.75$ and those without [Fe/H] determinations
by CBG09 have been omitted.  Note that we have added ``R97" to the labels in
the upper left-hand corner of these two panels to emphasize that the metal
abundances used by MF09 were obtained from the \citet[hereafter R97]{rhs97}
catalog. In the case of clusters not considered by R97 (those represented by
small filled circles), MF09 adopted the metallicities given by \citet[hereafter
ZW]{zw84} and transformed them to the \citet[hereafter CG]{cg97} scale using
the equation given by \citet{ccg01}.

The results plotted in panel (a), which assume the CG scale of R97, do indicate
that there is a population of coeval GCs that spans the entire metallicity
range.  However, this is not the case if the ZW scale is adopted; see panel (b).
Despite the evidence to the contrary, MF09 (see their footnote 14) conclude
that the results derived from the two metallicity scales are equivalent, and
then proceed to use only the CG results for their subsequent analysis.  However,
as seen by comparing panels (a) and (b), there are differences between them ---
and, for the following reason, we believe that the CG scale of R97 is the least
trustworthy one.  In footnotes to their table of [Fe/H] values, R97 warn the
reader that, for several clusters at the metal-rich end, [Fe/H] values on the
CG scale have been obtained by extrapolating their Ca II triplet measurements
linearly past the most metal-rich cluster that had an [Fe/H] value determined
from high-resolution spectra.  This is of particular concern because, as
discussed by R97, there is an inconsistency between the CG and ZW scales as
derived from their calibration of Ca II triplet line strengths, likely because
the Ca II lines lose their sensitivity to [Fe/H] at the metal-rich end.  

In fact, most of that inconsistency was removed by \citet{ccg01} when they
obtained and analyzed spectra for stars in two very metal-rich clusters.  Had
MF09 employed the non-linear equation derived by Carretta et al.~to transform
{\it all} of the ZW metallicities to the CG scale, instead of just a subset of
them, they would have obtained the AMR shown in panel (c).\footnote{In
constructing this panel, we discovered that the [Fe/H] values given by MF09 (in
their Table 1) for NGC$\,$6144, NGC$\,$6584, NGC$\,$6652, and M$\,$56 are
incorrect, though the [$m$/H] values that are given for these same clusters in
their Table 4 are fine.  We also found that the transformation equation from
Carretta et al.~was not used to derive the CG/R97 [$m$/H] values of M$\,$3
and M$\,$92: the [Fe/H] values for these clusters were presumably taken
directly from \citet{cg97}.}  At the metal-rich end, this resembles the 
relationship plotted in panel (b) more so than in panel (a) and, importantly,
it also looks similar to the AMR shown in panel (d).  This is based on the more
recent metallicity scale of CBG09 (as adopted in our investigation), where a
non-linear relationship between the Ca II triplet measurements and [Fe/H] is
further supported by high-resolution spectroscopy of two additional
high-metallicity clusters.  To obtain the results of panels (c) and (d), we
used the turnoff magnitudes given by MF09 (from their Table 4) and [$m$/H]
values that were calculated from the corresponding [Fe/H] determinations using
their equation (3).  To derive the relative ages that are plotted in these
panels, we assumed that $d({\rm age})/d($[$m$/H]) $\sim -4.5$ Gyr/dex, which is
a good approximation to the predicted variation of age with [$m$/H] in their
Fig.~7, together with their relative ages and corresponding age zero-point to
obtain new absolute ages.  These were then transformed back to relative ages
using the same zero-point as MF09; namely, the mean age of clusters with [$m$/H]
$< -1.4$.  Indeed, when the same CBG09 metallicities are adopted, relative ages
based on the MF09 formalism (panel d) appear to be quite similar to our findings.

The results shown in panels (a) to (d) demonstrate that the MF09 approach is
exceedingly sensitive to the adopted chemical abundances (which are likely
to undergo further adjustments in the future).  However, as pointed out in
\S~\ref{sec:methods}, the $\delv$\ method that is the basis of most of our age
determinations is not particularly dependent on the assumed [Fe/H] scale: the
effect of metallicity on the luminosity of the HB is compensated to a
significant extent by its effect on the turnoff luminosity at a given age.  This
is a huge advantage.  Even if we had adopted the CG/R97 abundance scale, we
would not have obtained an age-metallicity relation that is similar to the one
reported by MF09.  Also worth emphasizing is the fact that, as acknowledged by
MF09, the rMSF technique is sensitive to the abundances of such elements as Mg
and Al, which have been observed to vary from cluster-to-cluster (see
\citealt{cbg09b}).  These, and several other abundant heavy elements affect the
color offset between the MSTO and the RGB at a fixed age much more so than the
luminosity of the turnoff (see VBD12 and our Fig.~\ref{fig:fig17}).  At low
metallicities, the turnoff luminosity versus age relations are strong functions
of the helium and oxygen abundances, and little else.  (It is also a clear
disadvantage of the $\delc$\ method of determining relative ages that it is a
sensitive function of [Fe/H] and the detailed heavy-element mixture.  This
method, which should be applied {\it only} to clusters that have the same
metal abundances, is clearly of limited usefulness.)

A further concern with the MF09 results is that, if the CBG09 metallicities
are correct, there is a 0.65 dex difference in the [$m$/H] values of the
reference clusters that were used for the two most metal-rich groups of GCs.
Coupled with the likelihood that the clusters with high [Fe/H] values are
more metal-rich and younger than assumed by MF09, the errors in the inferred
ages from the rMSF method will be increased.  Consider, in particular, their
Fig.~6 (or our Fig.~\ref{fig:fig4}), which shows that the lower MS slopes of
computed isochrones vary significantly with [Fe/H], especially at higher
metallicities.  Such variations make the result of the rMSF method dependent
on the exact magnitude range that is used and on the difference in the metal
abundances between that of the reference cluster for a given metallicity bin
and those of each of the target clusters in that group.  In fact, the
self-consistency test that they show in their Fig.~6 considers only relatively
small variations in [$m$/H]: the impact of varying the age was not examined.
Moreover, their test does not extend above [$m$/H] $= -0.5$, where the
difficulties will be exacerbated.

MF09 found from their consideration of a few different grids of evolutionary
models that their results were essentially independent of this choice.  This is
probably to be expected given that it is the differences between the isochrones
for different ages that matter rather than their location on the CMD in an
absolute sense.  As shown in Figure~\ref{fig:fig36}, both the DSEP and the
BASTI (\citealt{pcs06}) isochrones predict similar, though not identical,
separations between the isochrone RGBs for different ages at a fixed [Fe/H]
(left-hand panels), and for variations in [Fe/H] at a given age (right-hand
panels) when the isochrones are registered to the usual abscissa and ordinate
zero-points.  These results compare quite well with those derived from the
Victoria-Regina isochrones used in this investigation (see the right-hand panels
of Fig.~\ref{fig:fig2} and Fig.~\ref{fig:fig3}, respectively).  There are
certainly some morphological differences between the three sets of models, and
the separation of the giant branch from the MSTO is obviously much larger in
the case of the BASTI isochrones than in the DSEP or Victoria-Regina
predictions.  This is probably caused primarily by the neglect of diffusive
processes in the BASTI computations, though differences in the treatment of
convection, the atmospheric boundary condition, the adopted color--$\teff$\
relations, or the detailed heavy-element mixture may also be partially
responsible for this.  (Since the giant branches of the best-fit
Victoria-Regina isochrones tend to be on the red side of observed RGBs when the
predicted and observed turnoffs are matched, the published BASTI isochrones can
be expected to be more discrepant, if the same distance and metallicity 
scales are adopted.)

To conclude: we believe that we have provided compelling arguments that the
MF09 findings are suspect, particularly for metal-rich GCs.  In this regard,
the importance of the binary in 47 Tuc should not be overlooked, as it indicates
a clear preference for a relatively young cluster age (recall
Fig.~\ref{fig:fig15}).  (Any increase in the assumed helium abundance of this
system would only serve to reduce the derived age.)  In addition,
the rMSF method, but not our implementation of the $\delv$\ method, requires
that the TO luminosity be determined as accurately as possible.  This quantity
is very hard to define in the CMDs of metal-rich GCs, in particular, given
that they tend to be nearly vertical over a $\sim 0.2$ mag range in the
vicinity of the turnoff, and it is likely to be especially ambiguous in clusters
containing multiple stellar populations.  Regardless of this practical
difficulty, a compelling reason to avoid using the TO luminosity was provided
by \citet{mdc95}, who showed that interpolations within grids of evolutionary
tracks that differed only in the treatment of superadiabatic convection yielded
isochrones with different TO luminosity versus age relations.  This was
attributed to the effects of {\it interpolating} in tracks that were
morphologically quite different since, as their work also demonstrated, the TO
luminosities of evolutionary tracks (as opposed to isochrones) have no more than
a slight dependence on how the convective gradient is determined in surface
convection zones.  In addition, because of the strong sensitivity of the rMSF
technique to the assumed [Fe/H] determinations --- something which was properly
examined by MF09 and not studied at all by \citet{mca10} --- coupled with the
uncertainties and the errors (albeit small errors, see \S~\ref{subsec:general})
inherent to the actual fitting process, we dispute the claim made by
\citet{mca10} that their approach is ``much more suitable than other techniques
for retrieving relative GC ages".

\subsubsection{Cautionary Remarks Concerning the Fitting of Isochrones to
Observed CMDs}
\label{subsubsec:caution}

In this investigation, globular cluster ages have been determined using what we
believe is the most robust, objective, and least model-dependent method
currently in use, as it places almost no reliance on predicted temperatures and
colors and it has less of a dependence on metal abundances than other
techniques.  To derive the best estimate of the age of a given GC, for the 
assumed distance and chemical abundances, our isochrones have been adjusted in
color by whatever amount is necessary (typically by $\sim 0.02$ mag) in order to
match the observed turnoff color, thereby facilitating the identification of
that isochrone which reproduces the observed CMD {\it just} in the vicinity of
the TO.  The morphology from $\approx 1$ mag below the TO through to the
beginning of the SGB (where the subgiants have colors that differ from the
median turnoff color by $\lta 0.05$ mag) is predicted to be essentially
independent of age, helium and metal abundances, and even the value of the
mixing-length parameter (see Figs.~\ref{fig:fig2}, \ref{fig:fig3}, and
\ref{fig:fig5}) --- at least for values of these quantities relevant to GCs.
In general, our isochrones for ages that have been derived in this way provide
reasonably satisfactory fits to the entire CMDs, except that the predicted RGBs
tend to lie on the red side of the observed giant branches.  The uncertainties
in the inferred ages primarily reflect the uncertainties in the adopted
distances and chemical abundances. 

It is important to appreciate that discrepancies between predicted and observed
colors are of little concern for the ages so obtained.  Indeed, isochrones will
generally fail to provide the best possible match to an observed CMD when
well-supported estimates of the distance, reddening, and metal abundances are
assumed because, e.g., the adopted cluster parameters (which have significant
uncertainties) may not be quite right, the color-$\teff$ relations suffer from
small zero-point or systematic errors, or there are problems with the stellar
models concerning, among other things, the treatment of convection, diffusion,
and/or the atmospheric boundary conditions.  There is, in particular, no
justification for favoring that isochrone which provides the best simultaneous
fit to the observed MSTO and the RGB, or for making small adjustments to the
cluster $(m-M)_0$, $E(B-V)$, [Fe/H], $Y$, or [$\alpha$/Fe] values if they are
made solely to improve the quality of the fit to the CMD --- because both
approaches assume that the models {\it should} reproduce the photometric data.
This is not necessarily the case.

For instance, super-adiabatic convection is usually treated using the
mixing-length theory (\citealt{bo58}), which involves the free parameter $\amlt$
(and a few others).  It is unlikely that this parameter is a constant (i.e.,
independent of mass, chemical composition, and evolutionary state), even though
it is treated as such.  (Although studies of GCs appear to rule out large
variations of $\amlt$, as discussed in \S~\ref{subsec:msrg}, the uncertainties
associated with the basic properties of stars are still too large to rule out
variations at, say, the $\sim 10$--15\% level.)  In fact, \citet{ts11} have
argued, from their 3-D simulations of surface convection in solar abundance
stars (the only metallicity that they considered in their initial
investigation), in support of a particular variation of $\alpha_{\rm MLT}$ with
$\teff$\ and gravity.  (As far as we are aware, the implications of these
predictions for stellar models have yet to be determined.)  A free parameter,
$D_{\rm turb}$ also appears in formulations of the extra mixing (see
\citealt{rmr02}, VBD12) that appears to be needed below outer convection zones
in order for diffusive models to account for the low abundance of lithium in
the Sun and to successfully predict the near independence of the Li abundance
with $\teff$\ at the hot end of the ``Spite plateau" (\citealt{ss82}).  It could
turn out that, when the models incorporate a better treatment of convection or
of turbulent mixing, the resultant isochrones will provide improved fits to
observed CMDs.  If the physics in current models is lacking in some way,
some discrepancies between theoretical models and observations {\it should} be
expected.

For the same reasons, suggestions that a more objective, least-squares or 
maximum likelihood numerical method should be used to select which isochrone
best represents an observed CMD are indefensible.  A few studies over the years
have even advocated that some kind of global optimization method be employed
to derive such cluster properties as their distances, reddenings, and
metallicities, as well as their ages (e.g., \citealt{md11}).  As long as free
parameters are used in some of the physics ingredients of stellar models, one
must continue to be wary of relying on the latter (primarily the predicted 
$\teff$\ and color scales) in an absolute, or even relative, sense.  Even if
the physics were more robust than it is at present, there will continue to be
uncertainties in the opacities and nuclear reactions, as well as in the observed
properties of stellar populations.  Many things must be known to very high
accuracy in order to obtain ``perfect" fits of models to photometric data.  It
is, in fact, quite encouraging that current models reproduce observed CMDs as
well as they do.

Analyses of observations of complex stellar populations, such as those found in
dwarf galaxies or in the Galactic Bulge, or those which are so distant that only
giant branch, and possibly HB, stars can be observed clearly require
well-constrained stellar models in order to obtain the best possible
interpretations of the data.  For such studies, it is important to ``calibrate"
the isochrones (e.g., suitably adjust the color--$\teff$\ relations) to fit the
observations of nearby open and globular star clusters, so that the inferred
ages and other properties of the target system are then determined {\it
relative} to those of the calibrating objects.  This is the approach taken by,
e.g., \citet{brfs04}.  Sophisticated statistical methods that use, e.g., Monte
Carlo or Bayesian techniques to analyze observed CMDs should {\it only} be
employed after the stellar models that are used have been thoroughly tested and
corrected, as necessary, in order to satisfy empirical constraints as well as
possible.

\subsection{On the Separation of the [Fe/H] $\lta -1.5$ GCs into Two Groups}
\label{subsec:twogcs}

Our discovery that the majority of the most metal-deficient GCs can be divided
into two distinct groups, depending on the slope of the subgiant branch in their
observed CMDs (see Table~\ref{tab:tab1}), is the most intriguing result of this
investigation. It is also surprising that the total number of clusters with
[Fe/H] $\lta -1.5$ is almost evenly divided into the two groups (see below).
Such a near fifty-fifty split
brings to mind the work of \citet{vsid93}, who speculated from his analysis of
GC orbits that there may have been a merger of a massive object (or objects)
containing M$\,$3--like stellar populations with the protoGalaxy.  (Such a
merger, if it happened, could only have occurred very early in the evolution of
the Milky Way, since the Galaxy appears to have had a relatively quiescent
history since thick-disk formation; see \citealt{hpc07}.)  He noted, for
instance, that these systems predominately have ``plunging" (highly elongated)
and/or retrograde orbits (also see \citealt{rp84}), as well as below-average
luminosities, especially if they are located at large Galactocentric distances.
However, orbital information was not available at that time for many of the GCs
in our sample and, in the case of a few clusters in common to the two studies,
the orbits were classified by van den Bergh as ``indeterminate" (see his Table
6).

Fortunately, the orbital characteristics of many of the metal-poor GCs
considered here are given by Dinescu et al.~(1999a): we have extracted from
their paper the information that is given in Table~\ref{tab:tab3}.  While
there is a tendency for the M$\,$3-like clusters to have larger eccentricities
($e$) and (especially) apoGalactic distances (R$_a$), and to reach greater
heights above the plane ($z_{\rm max}$), than the clusters in the group
containing M$\,$13, there are notable exceptions no matter what orbital property
is considered.  For instance, M$\,$92 has a higher eccentricity than most of
the GCs that have relatively flat SGBs, and M$\,$13 has significantly higher
values of $e$, R$_a$, and $z_{\rm max}$ than M$\,$3.  Moreover, as shown by
Dinescu et al.~(see their Fig.~6), metal-poor clusters with red or blue HBs are
not segregated in any way on the total energy versus orbital angular momentum
plane.

Admittedly, we were beginning to question whether the division of the metal-poor
GCs into two groups was real $\ldots$ until we produced Figure~\ref{fig:fig37}.
This shows that the clusters which have similar SGB slopes as M$\,$3 (those
represented by {\it open circles}) are predominately low-luminosity
systems that have relatively large values of R$_{\rm G}$ (and R$_a$, as already
mentioned).\footnote{While this appears to confirm the results obtained by
\citet{vsid93}, it should be appreciated that his definition of an M$\,$3-like
cluster is not the same as ours.  In his investigation, GCs that either have a
relatively red HB, like that of M$\,$3, or belong to Oosterhoff class I were
considered to be M$\,$3--like.  Such systems as NGC$\,$6101 and M$\,$70, which
have blue or very blue HBs, would not have been included in that group, and yet
we have classified them as M$\,$3--like on the basis of their SGB slopes.  Thus,
while the respective cluster samples are quite similar, there are some
important differences.}  The clusters belonging to the group represented by
M$\,$13 (those plotted as {\it filled circles}) are intrinsically bright
objects and, with the exception of M$\,$13, their orbits do not extend beyond
$\sim 10$ kpc from the Galactic center (see Table~\ref{tab:tab3}).  Note that
composite symbols have been used for three GCs to indicate that they have 
intermediate SGB slopes for their metallicities and, consequently, could not
be assigned to either of the aforementioned groups. 

The upper panel of Figure~\ref{fig:fig38} plots the HB types of the same sample
of clusters as a function of [Fe/H], using the same symbols as in the previous
figure.  Not unexpectedly, the M$\,$13-like clusters have exclusively blue HBs,
but it is also apparent that there are many GCs with very blue horizontal
branches (especially at $-2.1 <$ [Fe/H] $< -1.7$) that apparently belong to
the other group.  The middle panel, reveals that, at [Fe/H] $\gta -2.1$, there
is no separation of the two groups insofar as their ages are concerned.  Hence,
differences in age are not responsible for the observed variations in the HB
types at a fixed metal abundance.  Only at the lowest [Fe/H] values is there an
indication of a correlation of HB type with age, but even here, the $1\,\sigma$
error bars on the derived ages overlap one another.  Moreover, as shown in
the classic paper by \citet{ldz94}, much larger age differences ($\approx 2$
Gyr) are needed to explain the wide variation in the HB types at similar
metallicities if age is the controlling second parameter.  
If we are correct in attributing the differences between the M$\,$3
and M$\,$13 families of GCs to helium abundance variations, these results imply
that the most important second parameter is $Y$.  (However, the possibility
should be kept in mind that cluster-to-cluster variations in the total C$+$N$+$O
abundance at a given metallicity may be present, which could impact both the
observed HB morphologies and the predicted ages.  Unfortunately, the extent of
such variations between clusters and within each GC are not presently known for
the majority of the clusters in our sample.)

A recent examination of several of the global properties of $\sim 150$ GCs was
carried out by \citet{vsid11}, who showed, among other things, that their
central concentrations ($c = \log\,r_t/r_c$, where $r_t$ and $r_c$ are,
respectively, the tidal and core radii) are independent of metallicity, that
clusters with collapsed cores tend to located close to the center of the Galaxy,
and that there is no more than a weak correlation between [Fe/H] and
R$_{\rm G}$.  He also found no unambiguous correlation of the cluster
ellipticity with other parameters.  We have not subjected our small sample of
28 metal-poor clusters to the same analysis, though we were motivated to check
whether the ellipticity, which is presumably a tracer of the orbital, and
perhaps stellar, rotational velocities, provided any discrimination between the
two groups.  It is known (see, e.g., \citealt[and references therein]{pet83})
that the horizontal-branch stars in GCs that have anomalously blue HBs for
their metallicity, such as M$\,$13, rotate significantly faster than those
found in systems with intermediate or red HB types.  Increased mass loss during
the giant-branch phase, which would promote bluer HBs, could well be an
important consequence of higher rotational velocities (\citealt{fr75}), just as
the spread in mass that is needed to explain observed HBs is likely to have some
connection to star-to-star variations in the stellar rotation rates.  The fact
that the derived [O/Fe] and O/Na abundances vary with luminosity along the RGB
of M$\,$13 (\citealt{jp12}, \citealt{kss97}), probably due to rotation-driven
deep mixing, while such correlations are not seen in M$\,$3 giants
(\citealt{skg04}, \citealt{cm05a}), may be telling us that cluster-to-cluster
variations in the rotational properties of their stellar populations are large
enough to have far-reaching ramifications.

As shown in the bottom panel of Fig.~\ref{fig:fig38}, there is perhaps a 
slight offset in the mean ellipticity of the clusters that are plotted as filled
circles, on the one hand, and those represented by open circles, on the other
--- at least at [Fe/H] $> -2.1$, if the GCs with uncertain classifications are
omitted.  (The ellipticities were taken from the table given by
\citealt{vsid11}.)   However, the evidence in support of ellipticity being a
useful discriminant is clearly rather weak.  What is worth considering in some
detail is the extent to which the two groups of clusters are able to retain the
gas that is shed by stars as they evolve.  This is the subject of the next
section.

\subsubsection{On the Retention of Mass-Loss Material by Globular Clusters}
\label{subsubsec:gas}

We have suggested that the M$\,$13--like GCs have higher helium abundances, in
the mean, than those in the group typified by M$\,$3.  In order for multiple
stellar populations to form, the matter which is lost by stars belonging to the
first generation (FG) must be able to cool and accumulate in the cluster
centers.  This cannot occur in {\it present-day} GCs that have masses $\lta
10^5 \msol$.  [The current thought is that GC masses were much higher early in
their evolutionary histories when the second generation (SG) stars formed: we
will return to this point after describing the gas retention properties of the
clusters at the present time.]  According to the gas-flow models computed by
\citet{ff77} under the most conservative of assumptions (i.e., highly centrally
concentrated structures, the neglect of photoionization energy input, etc.),
steady-state outflows (i.e., GC winds) will be very effective in removing the
mass-loss material from low-mass clusters for all gas ejection energies down to
$\sim 15$ km/s, which essentially encompasses all plausible mass-loss
mechanisms.

In Figure~\ref{fig:fig39}, the masses of several GCs of interest are plotted as
a function of their central escape velocities, $v_{e,0}$.  The masses were
derived from the $M_V$ values listed in Table~\ref{tab:tab2} assuming a
mass-to-light ratio $<$${\cal M}/L_V$$>$ $ = 1.6 ({\cal M}/L_V)_\odot$
(\citealt{ill76}; \citealt{pmf91}; \citealt{ams02}), whereas
$v_{e,0}$ was calculated using the method described by \citet[see their
\S V]{vf77}.  (Masses derived in this way are probably uncertain by at least
$\sim \pm 25$\%: more massive, centrally concentrated clusters appear to have
average mass-to-light ratios closer to 2; see, e.g., the study of M$\,$15 by
\citealt{pdp04}.)  Interestingly, none of the M$\,$13-like GCs (the filled
circles), except NGC$\,$6397, have masses $< 10^5 \msol$.  For our cluster
sample, the apparent overlap of clusters represented by both open and filled
circles at a value of $\log M/\msol$ just above 5.1 (the location of the
horizontal, dashed line) appears to mark the mean transition mass between lower
mass GCs that develop steady-state winds and higher mass systems that could
acquire growing gas reservoirs at their centers as a result of gas inflows.
(This transition mass is predicted to vary inversely with the value of the
concentration parameter, $c$; see the paper by VandenBerg \& Faulkner)

In support of this possibility, we note that NGC$\,$6752 and M$\,$30 are
collapsed-core GCs with $c = 2.50$ (according to the 2010 edition of the catalog
compiled by Harris 1996); consequently, it would not be too surprising that they
might be able to retain the gas from low-velocity winds when other clusters of
similar mass but lower central concentrations (such as NGC$\,$6584 and
NGC$\,$6934, which both have $c \approx -1.5$) develop steady-state outflows.
NGC$\,$6397 and M$\,$70 are also collapsed-core, $c= 2.50$ GCs, and their
present masses, which are less than those of NGC$\,$6752 and M$\,$30, appear to
be low enough ($\lta 10^5 \msol$) that winds should be able to dissipate the gas
which is produced by normal stellar mass-loss processes.  Although GCs less
massive than M$\,$30, including M$\,$70, apparently have relatively flat SGBs,
NGC$\,$6397 seems to be an exception to the rule.  Indeed, a double MS has been 
discovered in its CMD by \citet{mmp12}, who conclude that this system contains
two stellar populations with slightly different helium and light element
abundances.  This may indicate that NGC$\,$6397 was much more massive when it
formed and that it contains two distinct generations of stars, as Milone et
al.~have argued.  However, if that is the correct explanation and if it is
generally the case that GCs have lost $\gta 90$--95\% of their initial masses
over their lifetimes, why do nearly all of the M$\,$13-like clusters lie near
or above the horizontal dashed line in Fig.~\ref{fig:fig39}? 

Crosses mark the locations of clusters in our sample that have [Fe/H] $\gta
-1.5$, and although only a subset of them have been identified in the plot (for
the sake of clarity), the fact that NGC$\,$362 and M$\,$5 have relatively high
masses while NGC$\,$288 is a low-mass system could well be relevant to our
understanding of the second-parameter phenomenon.  In particular, the retention
of mass-loss material will be especially difficult in the case of NGC$\,$288
given both its low mass and its low central concentration ($c = 0.99$), and it
may have been more difficult than in the case of NGC$\,$362 and M$\,$5 at
earlier times as well if its initial mass was also significantly less than those
of the latter.  (Interestingly, despite having unfavorable properties for the
retention of gas {\it at the present time}, NGC$\,$288 has been found to have
discrete sequences of stars in $uv$ CMDs, which are likely caused by differences
in the light-element abundances and possibly a small variation in $Y$; see
\citealt{pmm13}.)   The other second-parameter cluster in the same set (see
\S~\ref{subsec:n288n362}) is M$\,$12.  It is located just below M$\,$56
and NGC$\,$6934 in Fig.~\ref{fig:fig39}, and since its mass is within the $\sim
0.1$--0.15 dex uncertainty associated with the dashed line (see \citealt{vf77}),
it may or may not be able to develop a steady-state outflow.  Of the GCs that
are represented by open circles, the ones that seem to have the most anomalous
locations on the mass--$v_{e,0}$ plane are M$\,$3 and M$\,$53.

However, there is another potentially important mechanism for the removal of
gas from globular clusters and that is ram-pressure sweeping by the interstellar
medium of the Galactic halo.  According to \citet{fg76}, the halo density (in
g/cm$^3$) required to sweep a cluster as a result of the dynamical ram pressure
arising from the motion of a GC through the halo medium at a velocity $v_{cl}$ 
($\approx 200$ km/s) is
          $$\rho_g(halo) \approx \alpha\,\sigma_0v_{e,0}/v_{cl}^2$$
where $\alpha$ is the proportional rate of mass loss from stars (i.e., the 
global mass loss rate divided by the GC mass; see \citealt{ff77}) and $\sigma_0$
is the surface density of stars at the cluster center (in $\msol$/pc$^2$).
(Values of $\log\,\sigma_0$ for all of the GCs in our sample are listed in
Table~\ref{tab:tab2}: these have been calculated using the relations described
by \citealt{vf77}.)  If the value of $\alpha$ is taken to be $4 \times 10^{-19}$
s$^{-1}$, as adopted by Faulkner \& Freeman, ram-pressure sweeping by a halo
medium with a density of $10^{-26}$ g/cm$^3$ would be effective in clusters
that have $\log\,\sigma_0v_{e,0} \le 5.7$.  Since $\rho_g(halo)$ is proportional
to $\alpha$, the same constraint may be obtained for lower values of the halo
density, which are probably more realistic (e.g., \citealt{si74}), simply by
assuming smaller values of the uncertain parameter $\alpha$.  In fact, a
significantly smaller value, $\alpha = 7\times 10^{-20}$ s$^{-1}$, was derived 
by \citet{prs11} in their recent investigation of GC wind models, and if this
determination is adopted, $\log\,\sigma_0v_{e,0} = 5.61$ (the location of the
vertical dotted line in Figure~\ref{fig:fig40}) is obtained if $\rho_g(halo) =
1.5\times 10^{-27}$ g/cm$^{-3}$.  Under these assumptions, the majority of the
GCs with masses $\gta 10^5 \msol$\ (those with higher values of
$\sigma_0v_{e,0}$) should be able to resist the ram-pressure purging of any
gas that has accumulated between passages through the Galactic disk.  (To avoid
ram-pressure stripping by the disk, they would need to have
$\log\,\sigma_0v_{e,0} > 8$ for any reasonable choices of $\alpha$.)

Just as there is a transition mass between GCs that will, or will not, develop
steady-state gas outflows (the horizontal dashed line), there is an apparent
separation between the filled and open circles on the mass--$\sigma_0v_{e,0}$
diagram (the vertical dotted line).  Thus, even though the density of the halo
is very low, the ram pressure associated with the passage of GCs through this
medium is sufficient to prevent the accumulation of mass-loss material in
clusters with $\log\,\sigma_0v_{e,0} \lta 5.6$.  This separation is not perfect,
as M$\,$3 lies slightly to the right, while NGC$\,$4833, M$\,$10, and M$\,$56
lie to the left, of the dotted line.  However, differences in the orbit and the
orbital velocity, among other things, could well explain these few exceptions.
(Indeed, some of the clusters may not have been allocated to the right group as
the  $1\,\sigma$ uncertainties of the SGB slope determinations are quite large;
see Table~\ref{tab:tab1} and Fig.~\ref{fig:fig29}.)  In any case, it is
interesting that, of the GCs with masses $> 10^{5.5} \msol$, M$\,$3 and
M$\,$53 (in particular) have values of $\sigma_0v_{e,0}$ that are at the low
end of the observed range.  

It is remarkable that the clusters which show the strongest evidence for
discrete multiple stellar populations are located to the right of the dotted
line and above the dashed line (reproduced from Fig.~\ref{fig:fig39}), given 
that the masses and the structural properties of the GCs will have undergone
significant changes over their evolutionary histories due to tidal interactions,
the ``evaporation" of low-mass stars, and other dynamical effects. \citet{bek11}
has argued, for instance, that initial masses of up to $10^7 \msol$ are needed
to explain the observed large fraction of second-generation (SG) stars that have
been detected in several GCs.  It is possible that most of the
clusters in the so-called M$\,$3--like group (those plotted as open circles)
were never massive enough to form significant numbers of SG stars.  This is
suggested by the fact that the CMDs of most of these clusters tend to be very
tight and well-defined (see, e.g., those for NGC$\,$5053 and NGC$\,$5466 in
Fig.~\ref{fig:fig11} or of NGC$\,$4147 in Fig.~\ref{fig:fig12}).
The star-to-star variations in age or in the abundances of helium or C$+$N$+$O
in these clusters must be rather small given that the luminosities of their SGB
stars at a given color vary so little.

As O--Na anticorrelations have been found in most GCs, including those with
masses $10^4 \lta {\cal M}/\msol \lta 10^5$ (\citealt{cbg10b}), they may simply
be a property of the gas out of which the FG stars formed.  If the clusters
that have been plotted as open circles had been able to generate large numbers
of SG stars like their more massive cousins, but lost a bigger fraction of their
initial stellar populations over time, there should still be enough FG stars in
their cores to cause significant spreads in the colors of the MS, SGB, and RGB
stars in their observed CMDs.  The lack of such spreads suggests that these
clusters contain just a single stellar generation, albeit one that shows
primordial variations in the abundances of the light elements.  A noteworthy
cluster in this regard is NGC$\,$7492, which shows the O--Na anticorrelation
(\citealt{cm05b}) despite having $\log\,\sigma_0v_{e,0} \approx 3.0$.  This
outer-halo (R$_{\rm G} \approx 26$ kpc) cluster is located so far to the left
of the dotted line in Fig.~\ref{fig:fig40} that it seems highly improbable that
it was ever able to retain any of the mass-loss material from the first (and
only?) generation of stars that formed, even if its initial mass were 20 times
its current mass (which is $\log\,{\cal M}/\msol \approx 4.46$).   Another
cluster with a very low value of $\log\,\sigma_0v_{e,0}$ ($\approx 3.1$), but a
sufficiently high mass (just under $10^5 \msol$) that it is expected show an
O--Na anticorrelation, is NGC$\,$5053.  However, as far as we are aware, the
necessary spectroscopic studies to determine whether this and associated
variations of the light elements are found in this GC have not yet been
undertaken.

It may turn out that only those GCs in the upper right-hand corner of
Fig.~\ref{fig:fig40} (i.e., those that currently have $\log\,{\cal M}/\msol
\gta 5.1$ and $\log\,\sigma_0v_{e,0} \gta 5.6$) show appreciable enhancements
in helium, and possibly CNO (or other metals), though C$+$N$+$O generally
appears to be constant in them to within the uncertainties; e.g., see the
studies of M$\,$13 by \citet{cm05a} and of NGC$\,$6752 by \citet{cbg07}.  The
light element variations that appear to be common to all clusters more massive
than $\log\,{\cal M}/\msol \approx 4.5$, and in particular, the large
cluster-to-cluster variations in such characteristics as the ratio of CN-weak to
CN-strong stars and the degree to which O and Na or Al and Mg are anticorrelated 
may not indicative of multiple stellar generations, but are possibly due instead
to differences in the star formation history and the chemical evolution of the
individual protoclusters prior to the formation of the stars that currently
reside in them.\footnote{Support for this possibility is provided by
\citet{dh13}, whose preprint appeared while our paper was being refereed.
They have proposed that the observed chemical abundance anomalies in GCs were
produced by a very early generation of super-massive stars which polluted the
primordial cluster gas.  Unlike all other scenarios proposed to date, their
model provides an excellent fit to {\it all} of the observed light-element
correlations and anticorrelations, and it naturally explains why GC, but not
the field halo, stars have, e.g., high sodium and low oxygen abundances.
Moreover, if it is only in compact, highly centrally concentrated systems like
GCs where such super-massive stars have contributed to the very early chemical
evolution of the gas, the absence of such anomalies in dwarf galaxies can also
be explained.  Another attractive feature of this model is that it does not
require GCs to have been much more massive initially than they are today.}
The presence of multiple stellar populations is certainly much
less conspicuous in bona fide GCs (e.g., \citealt{pma12}) than in objects that
were likely to have been the nucleated cores of dwarf galaxies ($\omega\,$Cen,
M$\,$54, and possibly NGC$\,$2808 and NGC$\,$1851; see, e.g., \citealt{by12}).
(A thoughtful discussion of the different manifestations of the multiple
stellar populations phenomenon in terms of the initial mass and the progenitor
structure is provided by \citealt{vc11}.)

\section{Concluding Remarks}
\label{sec:final}

Using an improved version of the venerable $\delv$\ method, ages have been
determined for the majority of the 55 globular clusters considered in this
investigation.  For the most part, the ZAHBs that are the basis of the derived
distances reproduce the morphologies of the observed HB distributions very
well, which, together with the fact that they satisfy empirical constraints on
the luminosities of RR Lyrae stars to within their $1\,\sigma$ uncertainties,
gives us added confidence in them.  An important advantage of our implementation
of the $\delv$\ method is that isochrones are fitted to just the turnoff portion
of an observed CMD, where the morphology is predicted to be nearly independent
of age and metallicity, and that the inferred ages are based on the location of
the beginning of the SGB instead of the turnoff luminosity, which is a poorly
defined quantity in observed CMDs.  Indeed, in high-quality CMDs such as those
analyzed here, the uncertainties in the derived ages arising from the fitting
procedure are typically at the level of $\lta \pm 0.25$ Gyr, which is a small
fraction of the uncertainties associated with the adopted distances and chemical
abundances ($\sim \pm 1.5$--2 Gyr).  To be sure, the ages of GCs that have
extremely blue HBs, or broad photometric sequences due to the effects of
differential reddening and/or the presence of multiple stellar populations are
less precise.

There are a few observations/concerns that should be mentioned with regard to
the results of this study, which are reported and quite thoroughly discussed in
\S~\ref{sec:summary}.  For one thing, the unusual HB morphology of M$\,$13
(see Figs.~\ref{fig:fig21} and~\ref{fig:fig22}) --- namely, the dense
concentration of stars that begins just at the knee of the HB and extends to
bluer colors together with a sparser population that is offset to brighter
magnitudes and extending to the reddest colors --- is common to many of the GCs
that have exclusively blue horizontal branches (e.g., NGC$\,$6101, M$\,$12,
NGC$\,$6541, NGC$\,$6752, M$\,$56, and probably NGC$\,$288, NGC$\,$6397, and
M$\,$10, if not a few others).  Some are low-mass systems, some are not,
and while they tend to have ages $\gta 12.5$ Gyr, this is not always the case.
It is not at all clear whether the faintest HB stars in all of these clusters
have close to the primordial helium abundance or a higher value of $Y$, and yet
our adopted distance moduli are consistent with $Y \approx 0.25$.  The main
difficulty with the assumption of a higher helium abundance is that the
luminosity of the HB is {\it very} sensitive to $Y$; consequently, the adoption
of an appreciably higher $Y$ would imply a significantly increased distance
modulus and a younger age.  However, this would lead to problems with the
predicted and observed MSTO-to-RGB color differences.

Indeed, it was necessary to postulate that M$\,$13 stars have helium abundances
ranging from $Y \approx 0.25$ to possibly 0.33, with a mean value of $\approx
0.29$, in order to obtain a sufficiently short distance modulus and a high
enough age to reconcile its $\delc$ parameter with that of M$\,$3 {\it and} to
explain the observed differences in their SGB slopes (Fig.~\ref{fig:fig20}).  If
all of the M$\,$13 stars had initial helium contents close to $Y = 0.29$, the
ZAHB-based distance modulus would have been $(m-M)_{APP} \approx 14.60$,
implying an age near 10.25 Gyr, and the overall fit to the observed CMD would
have been considerably less agreeable.  Thus, we were lead to a scenario in
which most (all?) of the stars in some GCs (M$\,$3 and lower-mass systems that
have similar SGB morphologies) have ``normal" helium abundances, while others
(M$\,$13 and GCs that share its CMD characteristics)  have a higher $Y$ {\it in
the mean}, but which still contain a significant population of stars with $Y
\approx 0.25$.   On the other hand, if the faintest HB stars in M$\,$13 stars
have $Y \approx 0.25$, why are they so much bluer than their counterparts in
M$\,$3?  Clearly something is missing in our understanding.  (If both clusters
have nearly the same age, as we and MF09 have concluded, and very similar CNO
abundances, differences in the stellar rotation rates would appear to be the
next most plausible explanation for their different HB types.) 

\begin{figure}[t]
\includegraphics[width=1.0\textwidth]{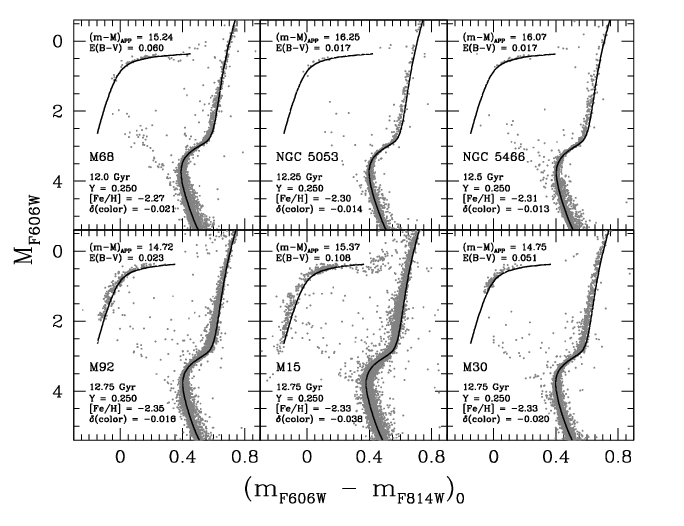}
\caption{The CMDs of six globular clusters that have [Fe/H] $< -2.2$ (according
to CBG09) are plotted on the assumption of the reddenings and the apparent 
distance moduli, as derived from the fit of a ZAHB to the lower bound of the
distribution of cluster HB stars, that are given in the top left-hand corner of
each panel.  The derived ages, the assumed helium and [Fe/H] abundances, and the
adjustments in color that were needed in order for the selected isochrones to
match the observed turnoff colors are specified below the cluster names.  The
main purpose of this and subsequent figures is to show that the ZAHB models
provide good fits to the observed HB stars, especially to those with $0.0 \lta
(m_{F606W} - m_{F814W})_0 \lta 0.3$ in this case, and that the best-fit 
isochrones do, indeed, reproduce the cluster photometry in the vicinity of the
turnoff very well.}
\label{fig:fig11}
\end{figure}

\begin{figure}[t]
\centering
\includegraphics[width=1.0\textwidth]{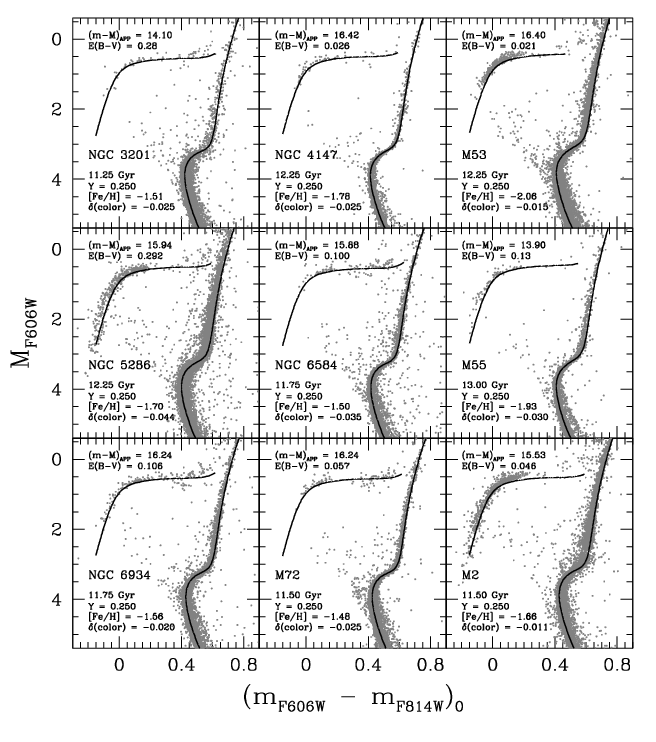}
\caption{Similar to the previous figure; in this case, the CMDs of globular
clusters that have $-1.50 >$ [Fe/H] $\ge -2.2$ have been fitted by ZAHB loci and
isochrones for the appropriate metallicities.  The panels have been organized
such that the cluster NGC numbers increase in the direction from left to right,
beginning in the top row and ending in the bottom left-hand corner --- though
the Messier number is used if a given cluster has one.}
\label{fig:fig12}
\end{figure}

\begin{figure}[t]
\centering
\includegraphics[width=1.0\textwidth]{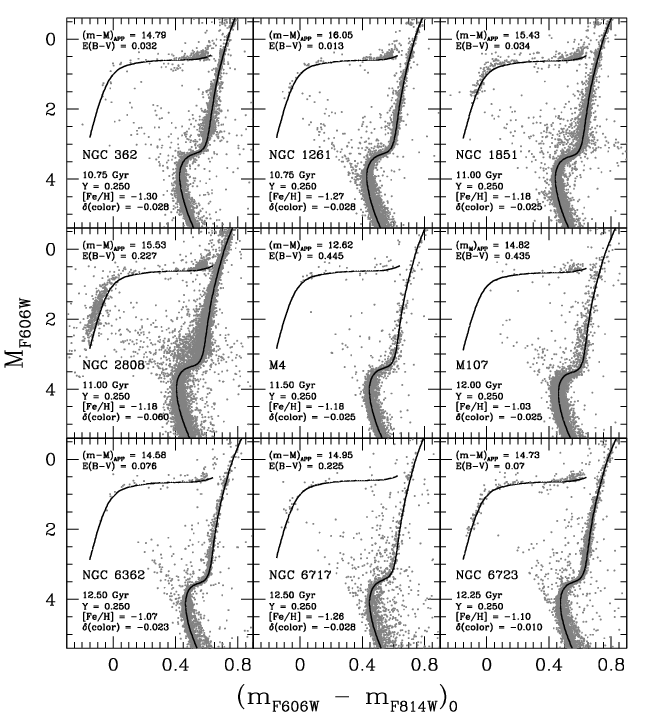}
\caption{Similar to the previous figure; in this case, the CMDs of globular
clusters that have $-1.00 >$ [Fe/H] $\ge -1.50$ have been fitted by ZAHB loci
and isochrones for the appropriate metallicities.}
\label{fig:fig13}
\end{figure}

\begin{figure}[t]
\centering
\includegraphics[width=0.8\textwidth]{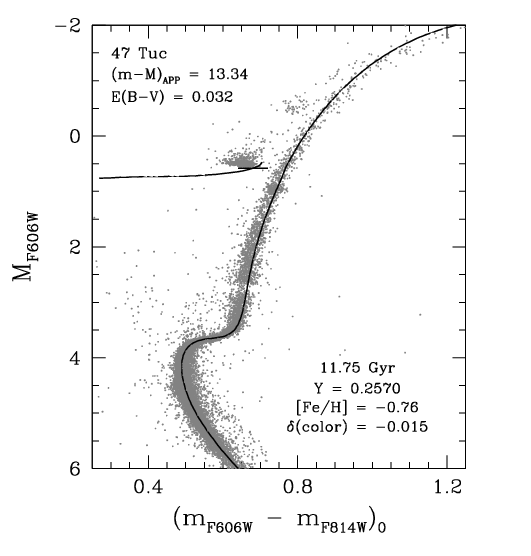}
\caption{As in previous comparisons of isochrones and ZAHB loci with
photometric data; in this case, for 47 Tucanae.  Note that the red end of the
ZAHB gives the location of star that has not undergone any mass loss prior to
reaching the HB.  Its mass is the same as that of the model at the RGB tip of
the best-fit isochrone for the indicated age.  The point at the intersection of
the short horizontal line with the ZAHB indicates the location of a model that
has a lower mass by $0.20 {{\cal M}_\odot}$.  (The apparent gap at $M_{F606W}
\approx 2.8$ is an artifact arising from the selection of stars that have
photometric errors $< 0.015$ mag, given that short and long integrations were
employed to obtain the photometry for the cluster giants and main-sequence
stars, respectively.)}
\label{fig:fig14}
\end{figure}

\begin{figure}[t]
\includegraphics[width=1.0\textwidth]{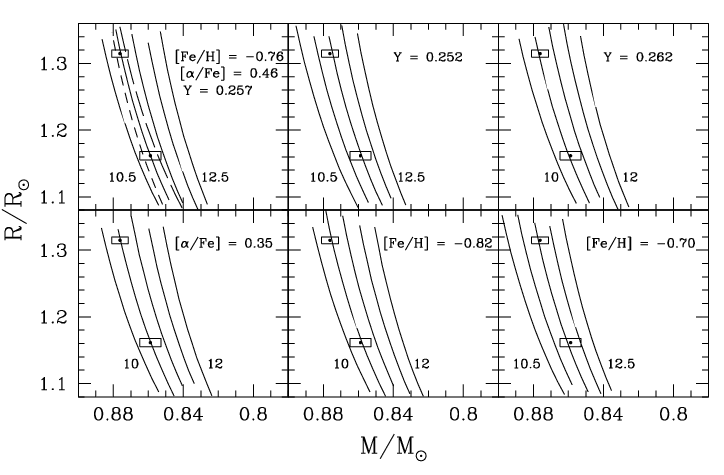}
\caption{The filled circles and error boxes plot the masses and radii of the
components of the binary, V69, as determined by \citet{tkr10}, along with their
uncertainties.  The variations of radius with mass that are predicted by the
same isochrones which have been used in the previous figure, but for ages of
10.5 to 12.5 Gyr in 0.5 Gyr intervals, have been plotted as solid curves in the
upper {\it left-hand panel}.  The adopted chemical properties for these models
are listed in the top right-hand corner of this panel.  If the temperatures
along the 11.0 Gyr isochrone are increased by 75 K, the result is the
short-dashed curve, while the long-dashed curve illustrates the effect of
reducing the predicted temperatures by 75 K.  The other panels show how the
inferred age is affected if the values of the various abundance parameters are
varied, in turn, by small amounts (as noted in each panel).}
\label{fig:fig15}
\end{figure}

\begin{figure}[t]
\centering
\includegraphics[width=1.0\textwidth]{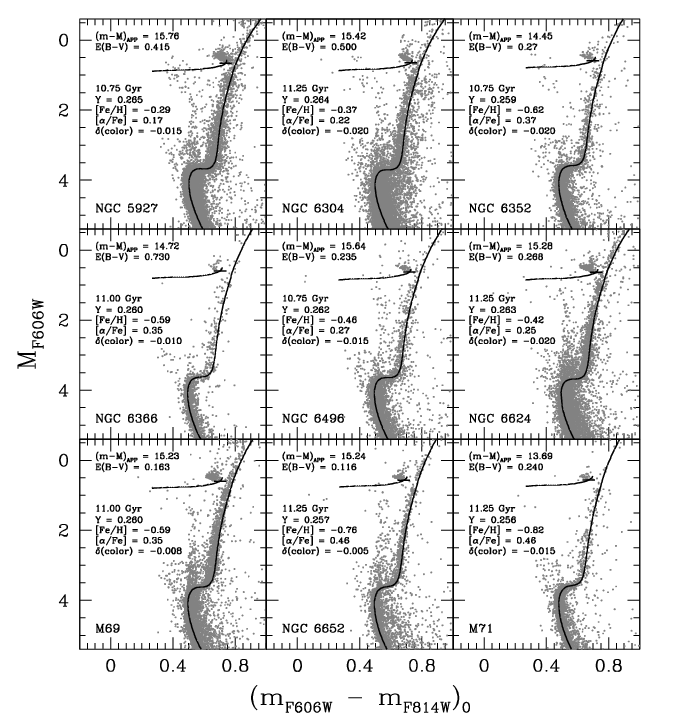}
\caption{Similar to Fig.~\ref{fig:fig14}; in this case, the CMDs of globular
clusters that have [Fe/H] $\ge -1.0$ have been fitted by ZAHB loci and
isochrones for the appropriate metallicities.}
\label{fig:fig16}
\end{figure}

\begin{figure}[t]
\centering
\includegraphics[width=0.8\textwidth]{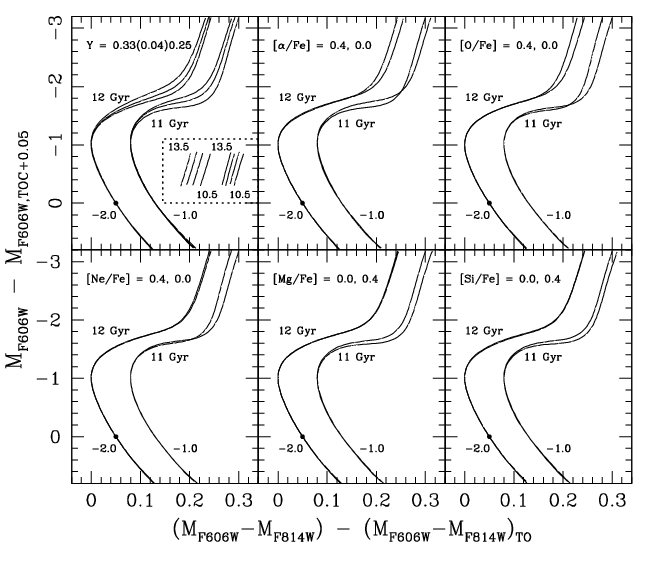}
\caption{Isochrones for different choices of the chemical abundance parameters,
as indicated in the top left-hand corner of each panel, have been superimposed
such that they have the same turnoff colors and the same magnitudes at the point
along the upper MS that is 0.05 mag redder than the TO.  (These fiducial points
define the abscissae and ordinate zero-points.)  Isochrones for [Fe/H] $=
-2.0$ and the different helium and heavy-element mixtures have been generated
for an age of 12 Gyr, whereas those for [Fe/H] $= -1.0$ assume an age of 11 Gyr.
Note that the grids of evolutionary tracks that are the basis of these
isochrones were computed by VandenBerg et al.~(2012) on the assumption of the
\citet{ags05} solar metals mix, with enhanced abundances of a number of metals.
(These are the only computations that are available to us at the present time
in which the abundances of individual metals are enhanced, in turn, by 0.4 dex.)
Note as well that, for the sake of clarity, the isochrones for [Fe/H] $= -1.0$
have been offset to the red by 0.1 mag.  The loci contained within the dotted
rectangle represent the RGB segments of 10.5 to 13.5 Gyr isochrones, in 1.0 Gyr
intervals, for the reference mixture at ordinate values of $\approx -2.4$ to
$-3.2$.  By comparing these results with the predicted horizontal offsets of
the RGB loci that are produced by the various abundance choices, an equivalent
age difference can be estimated.}  
\label{fig:fig17}
\end{figure}

\begin{figure}[t]
\includegraphics[width=1.0\textwidth]{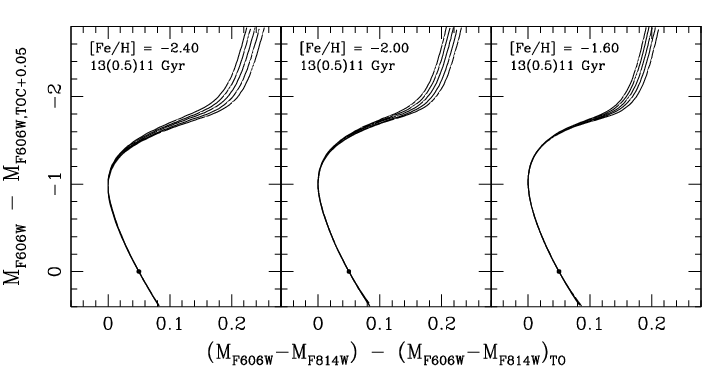}
\caption{Similar to the previous figure, except that 11--13 isochrones for the
GSCX mix and three values of [Fe/H], as indicated, have been registered to
one another.}
\label{fig:fig18}
\end{figure}

\begin{figure}[t]
\includegraphics[width=1.0\textwidth]{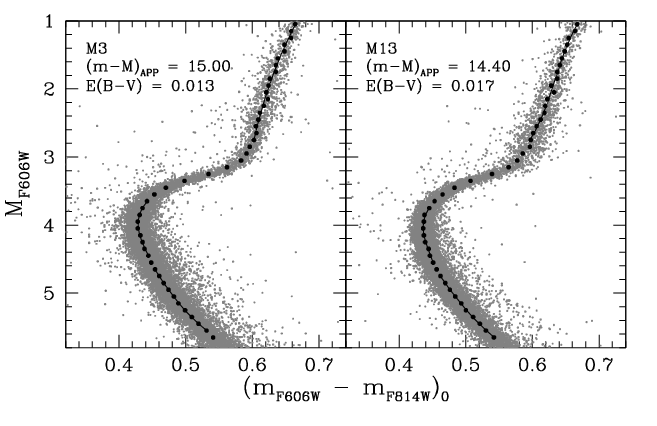}
\caption{The CMDs of M$\,$3 (NGC$\,$5272) and M$\,$13 (NGC$\,$6205) showing the
median points ({\it filled circles}) that have been derived in 0.1 mag bins
along the photometric sequences.  {\it Solid curves} illustrate the
least-squares fits to the median points in three different parts of the CMD:
the RGB, the vicinity of the turnoff, and the region enclosing that point
on the upper MS which is 0.05 mag redder than the TO; see the text.}
\label{fig:fig19}
\end{figure}

\begin{figure}[t]
\centering
\includegraphics[width=1.0\textwidth]{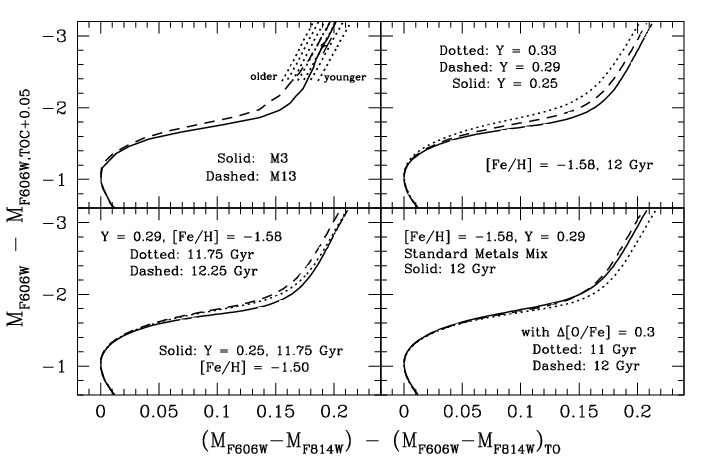}
\caption{{\it Upper left:} The CMDs of M$\,$3 and M$\,$13 have been registered
to the usual VBS90 abscissa and ordinate zero-points.  The dotted lines 
represent the RGB segments of isochrones for [Fe/H] $= -1.50$ that differ in
age, in turn, by 0.5 Gyr in the direction indicated.  The offset of the short
sloped line from the small filled circle at an ordinate value of $-2.8$
indicates the horizontal correction that should be applied to the M$\,$3 RGB to
account for the difference in its [Fe/H] value ($-1.50$) and that of M$\,$13
($-1.58$).  {\it Upper right:} 12 Gyr isochrones for [Fe/H] $= -1.58$ and
$Y = 0.33$, 0.29, and 0.25 (in the direction from left to right) have been
similarly registered.  {\it Bottom right:} 12 Gyr isochrones for [Fe/H] $=
-1.58$ and $Y = 0.29$ with, or without, 0.3 dex enhancements in [O/Fe] (in the 
direction from left to right) have been similarly registered.  {\it Bottom
left}: Registration of isochrones for those ages and chemical abundances that 
appear to be close to the values needed to explain the observations in the
upper, left-hand panel.}
\label{fig:fig20}
\end{figure}

\begin{figure}[t]
\centering
\includegraphics[width=0.8\textwidth]{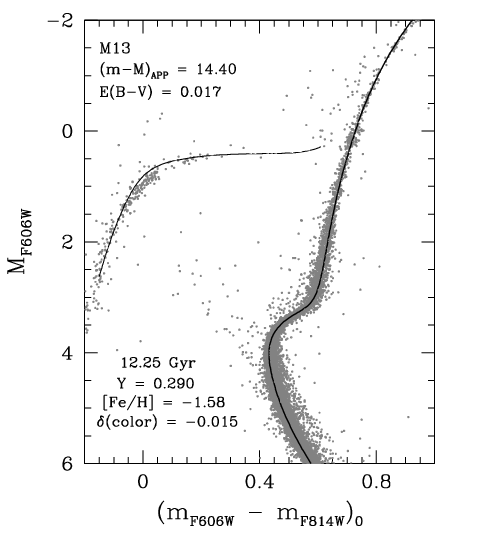}
\caption{A 12.25 Gyr isochrone for $Y = 0.29$ and [Fe/H] $= -1.58$ has the
same turnoff luminosity as M$\,$13 (as shown), if the cluster has $E(B-V) =
0.017$ (Schlegel et al.~1998) and an apparent distance modulus of 14.40 mag.
Although the isochrones match the morphologies of the MS, SGB, and RGB quite
well, a large fraction of the cluster HB stars are fainter than the
corresponding ZAHB, which argues against this particular interpretation of
the observed CMD.}
\label{fig:fig21}
\end{figure}

\begin{figure}[t]
\centering
\includegraphics[width=1.0\textwidth]{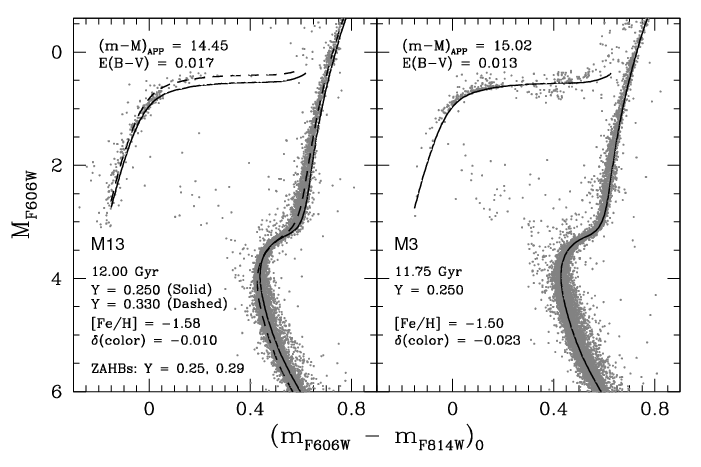}
\caption{Fits of isochrones and ZAHB loci for the indicated parameters to the
CMDs of M$\,$3 and M$\,$13 ({\it right-} and {\it left-hand panels},
respectively).  In both cases, the adopted distance moduli are such that ZAHBs
for $Y = 0.25$ provide satisfactory fits to the lower bounds of the observed HB
populations.  Isochrones for the same age, but different $Y$, are shown in the
{\it left-hand panel}: morphological differences in the SGBs of M$\,$3 and
M$\,$13 may be explained if the latter has a higher helium abundance, in the
mean, than the former (see the text).  If this inference is correct and the
stars with the highest helium abundances in M$\,$13 evolve to ZAHB locations
well to the blue of the instability strip, its observed HB may be satisfactorily
reproduced by models that encompass a significant range in $Y$.} 
\label{fig:fig22}
\end{figure}

\begin{figure}[t]
\centering
\includegraphics[width=1.0\textwidth]{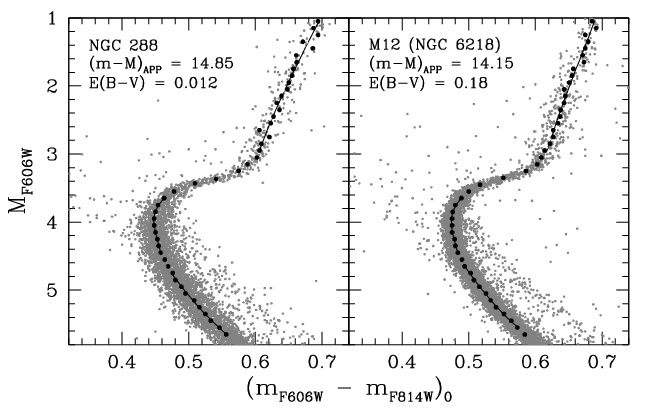}
\caption{Similar to Fig.~\ref{fig:fig19}; in this case, the CMDs and the derived
fiducal sequences are shown for NGC$\,$288 and M$\,$12 (NGC$\,$6218).
Interestingly, the different MS widths of the two clusters suggests that
star-to-star chemical abundance variations are significantly larger in
NGC$\,$288 than in M$\,$12.}
\label{fig:fig23}
\end{figure}

\begin{figure}[t]
\centering
\includegraphics[width=1.0\textwidth]{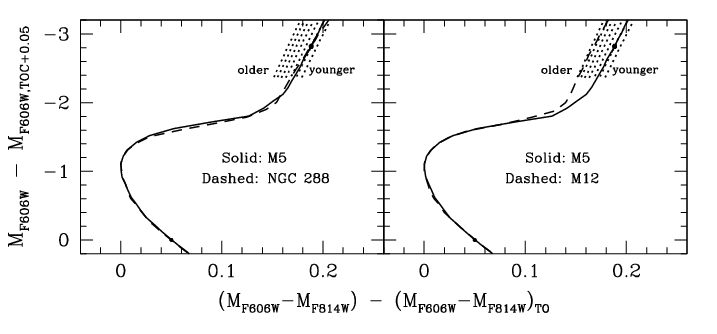}
\caption{Similar to the upper left-hand panel in Fig.~\ref{fig:fig20}; in this
case, the fiducial sequence of M$\,$5 (NGC$\,$5904) has been registered to that
of NGC$\,$288 ({\it left-hand panel}) and to that of M$\,$12 (NGC$\,$6218) ({\it
right-hand panel}).  This suggests that NGC$\,$288 and M$\,$5 are nearly coeval,
while M$\,$12 is significantly older than M$\,$5 (but see the discussion in
the text).}
\label{fig:fig24}
\end{figure}

\begin{figure}[t]
\centering
\includegraphics[width=0.8\textwidth]{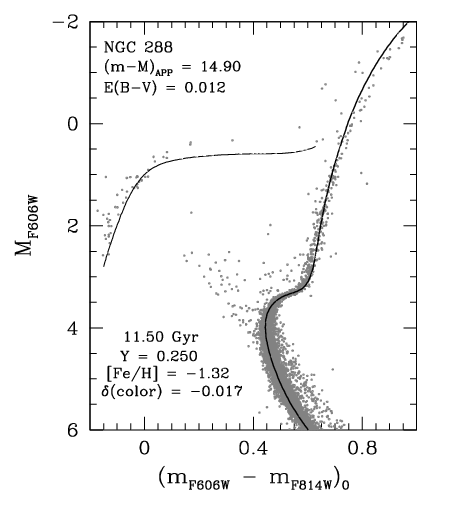}
\caption{Fit of an isochrone for the indicated age and chemical abundances and
a fully consistent ZAHB to the CMD of NGC$\,$288 on the assumption of the
apparent distance modulus and reddening that are specified in the upper
left-hand corner.}
\label{fig:fig25}
\end{figure}

\begin{figure}[t]
\centering
\includegraphics[width=1.0\textwidth]{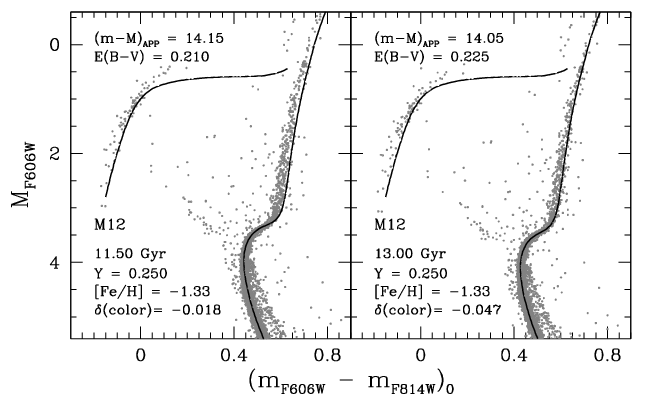}
\caption{Similar to the previous figure; in this case, isochrones for the same
chemical abundances (as indicated) and ages of 11.5 and 13.0 Gyr have been 
fitted to the photometry of M$\,$12 in the {\it left-} and {\it right-hand
panels}, respectively.  Note that there is no obvious difference in the quality
of the fit to the cluster HB stars if the reddening is adjusted to accommodate
a different distance modulus.}
\label{fig:fig26}
\end{figure}

\begin{figure}[t]
\centering
\includegraphics[width=0.9\textwidth]{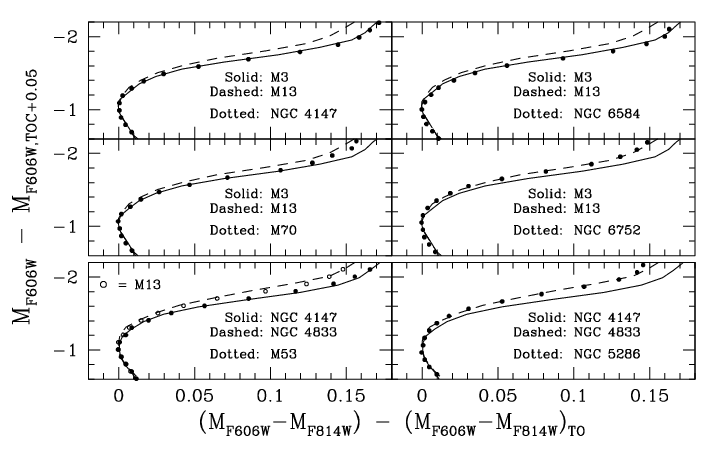}
\caption{{\it Uppermost four panels:} the photometric sequences for the turnoff
and subgiant stars of four target GCs ({\it dotted curves}) are compared with
those of M$\,$3 and M$\,$13 ({\it solid} and {\it dashed} loci, respectively).
With the exception of NGC$\,$4147 (see the text), all of the clusters have the
same [Fe/H] values to within 0.10 dex.  Note that NGC$\,$6752 is the only 
target cluster which is ``M$\,$13-like" insofar as the slope of its SGB is
concerned.  {\it Bottom two panels:} the fiducial sequences for two
lower-metallicity clusters, M$\,$53 and NGC$\,$5286, are compared with those
of NGC$\,$4147, which has a relatively flat SGB (like M$\,$3) and NGC$\,$4833,
which has a steeper SGB (comparable to that of M$\,$13).  All of the cluster
fiducials have been adjusted in the horizontal and vertical directions in order 
that they have the same turnoff color (the abscissa zero-point) and the same
magnitude on the upper main sequence that is 0.05 mag redder than the TO (the
ordinate zero-point).}
\label{fig:fig27}
\end{figure}

\begin{figure}[t]
\centering
\includegraphics[width=0.9\textwidth]{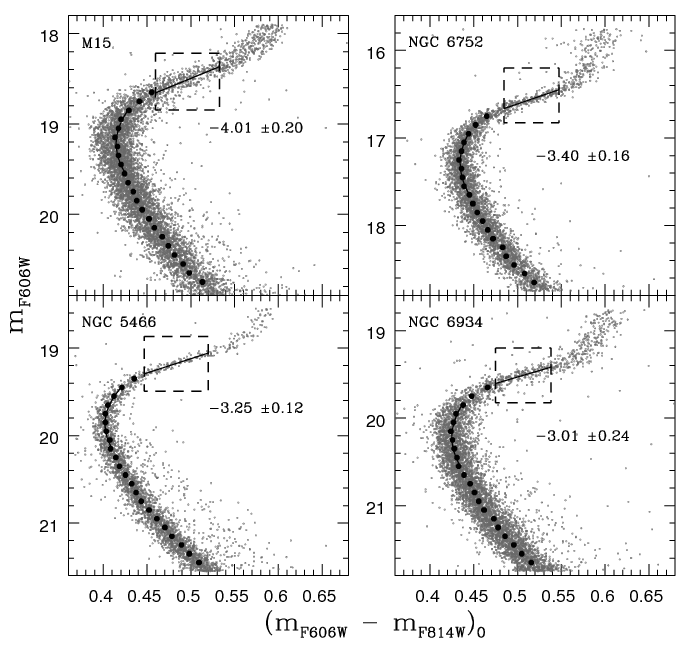}
\caption{Representative examples of the linear least-squares fits that have
been performed to the SGBs of all of the clusters that have [Fe/H] $\lta -1.5$
in order to determine the subgiant slopes and their uncertainties.  Only those
stars inside the {\it dashed rectangles} were fitted (see the text), resulting
in the straight lines within them.  Numerical values for the slopes of these
lines, and their uncertainties, are given below and to the right of the
rectangles.  As in Fig.~\ref{fig:fig26}, the large filled circles define the
fiducial cluster sequences for the upper MS and TO stars. }
\label{fig:fig28}
\end{figure}

\begin{figure}[t]
\centering
\includegraphics[width=1.0\textwidth]{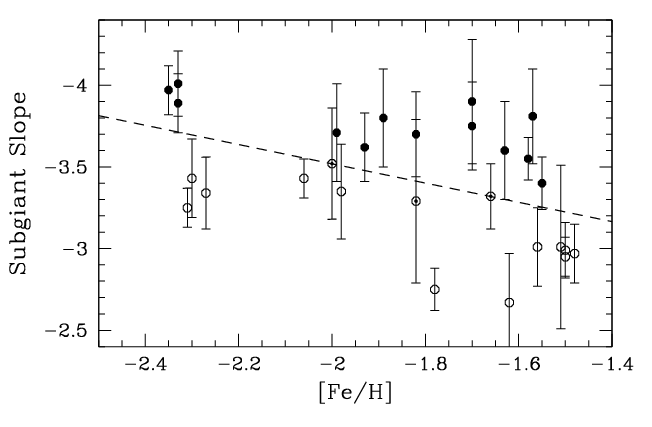}
\caption{Plot of the subgiant slopes and their uncertainties as a function of
[Fe/H] for all of the clusters listed in Table~\ref{tab:tab1}.  {\it Filled
circles} identify clusters that have relatively steep SGB slopes
(``M$\,$13-like"), while those with shallower slopes (``M$\,$3-like") are
respresented by {\it open circles}.  Composite symbols have been used for 
three clusters that have intermediate SGB slopes.  A linear least-squares fit
to all of the points resulted in the dashed line.  Note that the interpretation
of these results is complicated by the predictions that (i) at a fixed age,
isochrones at lower metallicities have slightly steeper SGBs (see
Fig.~\ref{fig:fig3}), and (ii), at a fixed [Fe/H] value, there is some
dependence of the SGB slope on age (especially at [Fe/H] $\lta -2$; see
Fig.~\ref{fig:fig18}).}
\label{fig:fig29}
\end{figure}

\begin{figure}[t]
\centering
\includegraphics[width=0.85\textwidth]{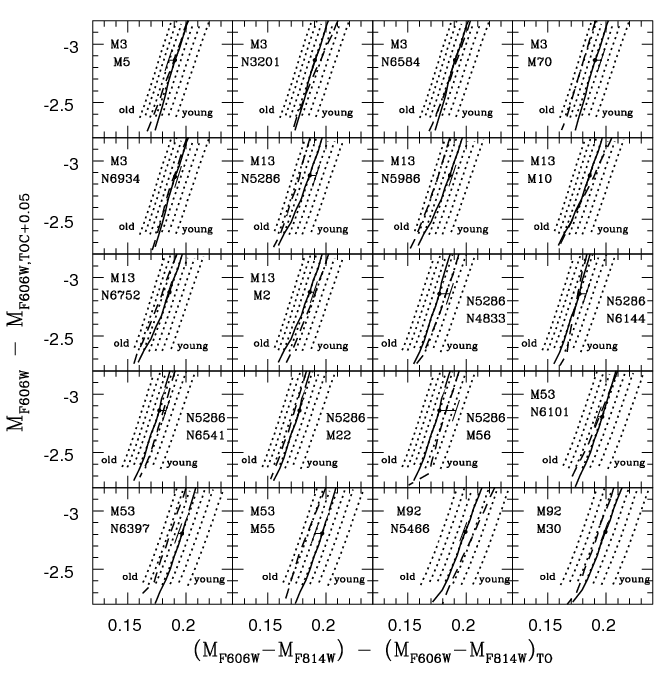}
\caption{Differences in the ages of globular clusters as determined from the
$\delc$\ technique.  The clusters considered in each panel are identified in the
upper left-hand corner (or to the right of center in the case of comparisons
involving NGC$\,$5286),  where the upper and lower names correspond to the
reference and target clusters, respectively: their lower-RGB fiducials are
represented, in turn, by the solid and dashed curves.  Giant-branch segments of
isochrones, in 0.5 Gyr increments, for the [Fe/H] value of the target cluster
are shown as dotted lines.  The correction, if any, that should be applied to
the location of the fiducial sequence of the reference cluster to account for
the difference in [Fe/H] between it and the target cluster is given by the
short horizontal line at an ordinate value of $\approx -2.8$.}
\label{fig:fig30}
\end{figure}

\begin{figure}[t]
\centering
\includegraphics[width=1.0\textwidth]{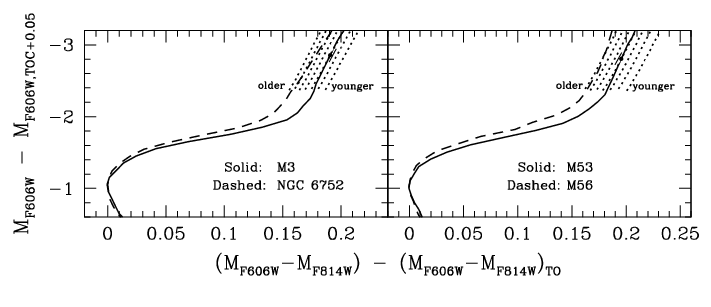}
\caption{Similar to Fig.~\ref{fig:fig24}; in this case, the principal
photometric sequences for the upper-MS to lower-RGB stars in M$\,$3 and 
NGC$\,$6752 ({\it left-hand panel}) and in M$\,$53 and M$\,$56 ({\it
right-hand panel}) are compared after they have been registered to the usual
abscissa and ordinate zero-points (see the text).}
\label{fig:fig31}
\end{figure}

\begin{figure}[t]
\centering
\includegraphics[width=1.0\textwidth]{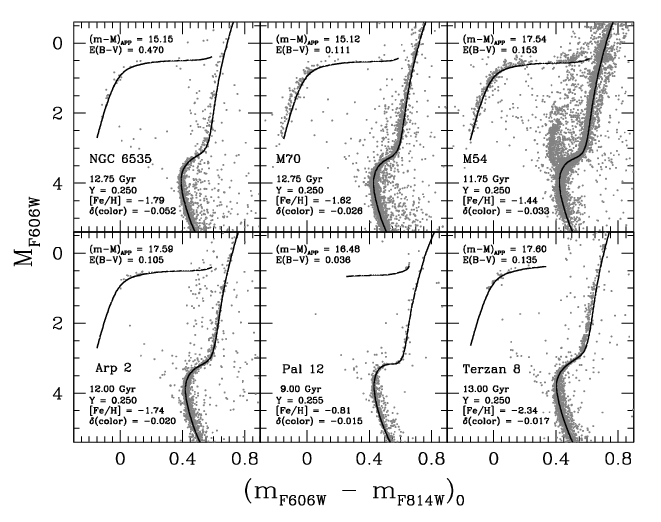}
\caption{Similar to Fig.~\ref{fig:fig11}; in this case, isochrones and ZAHB
loci have been fitted to the CMDs of NGC$\,$6535 and M$\,$70, as well
as to those of four GCs that are associated with the Sagittarius dwarf galaxy.
In the case of Pal 12, [$\alpha$/Fe] $= 0.0$ has been adopted: the other
systems have been assumed to have normal $\alpha$-element abundances for 
their metallicities.}
\label{fig:fig32}
\end{figure}

\begin{figure}[t]
\centering
\includegraphics[width=1.0\textwidth]{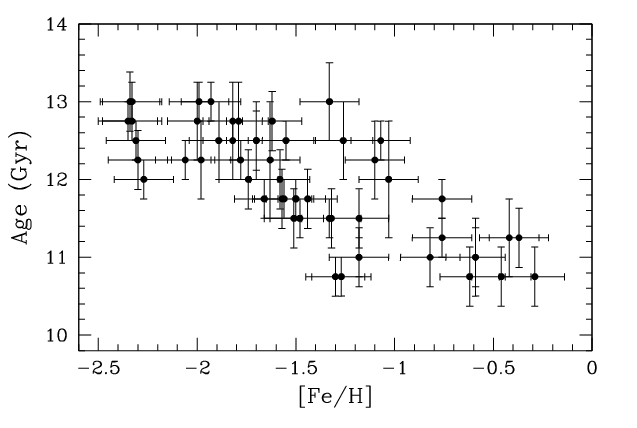}
\caption{The age--[Fe/H] relationship that has been derived in this
investigation; see Table~\ref{tab:tab2} for numerical values of the data that
have been plotted.}
\label{fig:fig33}
\end{figure}

\begin{figure}[t]
\centering
\includegraphics[width=1.0\textwidth]{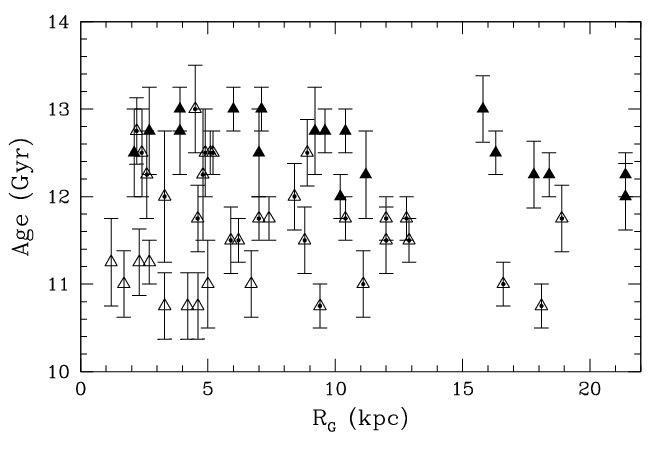}
\caption{The age--R$_{\rm G}$ relationship that has been derived in this study.
The data that have been plotted are given in Table~\ref{tab:tab2}.  {\it Filled
triangles, open triangles,} and {\it composite symbols} have been used to
represent GCs that have [Fe/H] $< -1.7$, [Fe/H] $\ge -1.0$ and $-1.7 \le$ [Fe/H]
$< -1.0$, respectively.}
\label{fig:fig34}
\end{figure}

\begin{figure}[t]
\centering
\includegraphics[width=0.85\textwidth]{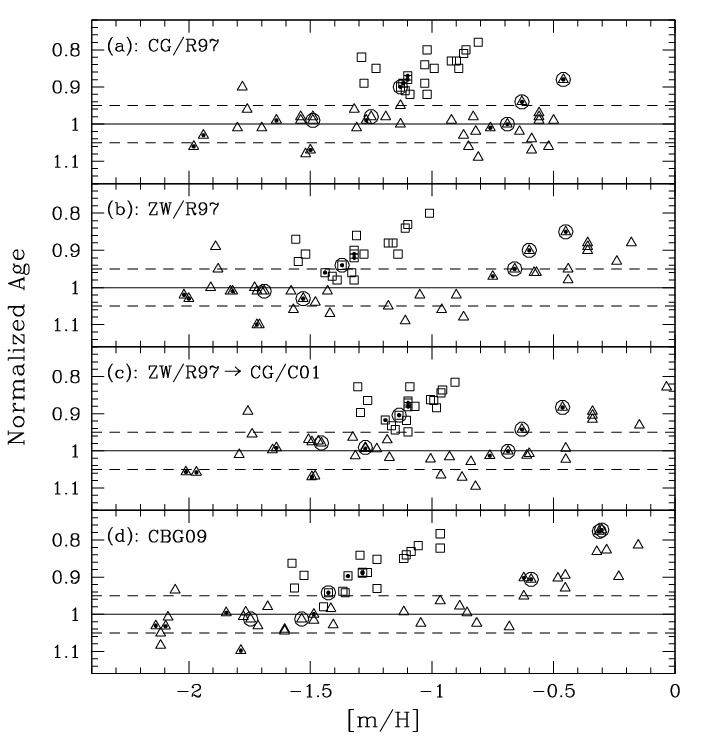}
\caption{Normalized ages of globular clusters as determined using the rMSF
method of MF09.  Squares and triangles represent, in turn, GCs that belong to a
young and an old group according to MF09.  Large open circles denote clusters
that are not included in our sample.  Panels (a) and (b) reproduce the results
reported by MF09 (their Figs.~10 and 11, respectively).  Small filled circles
denote those clusters which were not considered in the R97 study.  Panels (c)
and (d) portray, in turn, the results that MF09 would have obtained if they had 
transformed all of their ZW/R97 metal abundances to the CG scale using the
equation provided by Carretta et al.~(2001), or if they had been able to use the
metallicities from CBG09, which were not published at that time.}
\label{fig:fig35}
\end{figure}

\begin{figure}[t]
\centering
\includegraphics[width=0.85\textwidth]{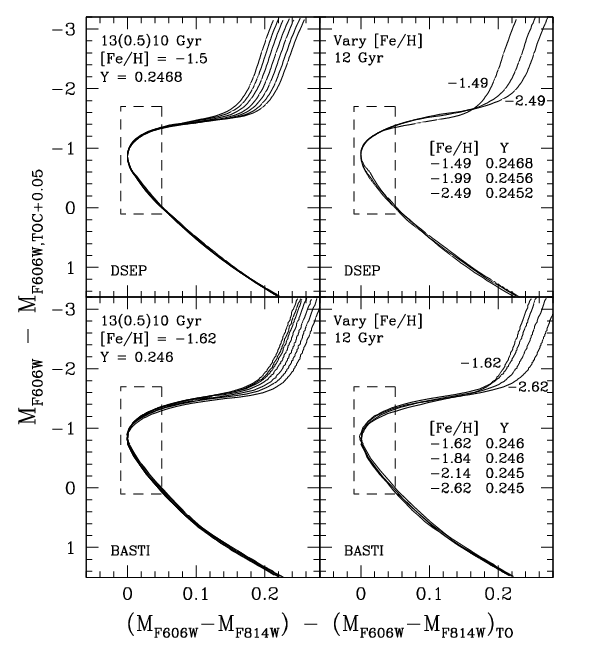}
\caption{The registration of selected DESP and BASTI isochrones (from their
respective web sites) to the usual zero-points to illustrate how their
MSTO-to-RGB separations vary with age at a given metallicity (the {\it
left-hand panels}) and with [Fe/H] at a fixed age (the {\it
right-hand panels}).}
\label{fig:fig36}
\end{figure}

\begin{figure}[t]
\centering
\includegraphics[width=1.0\textwidth]{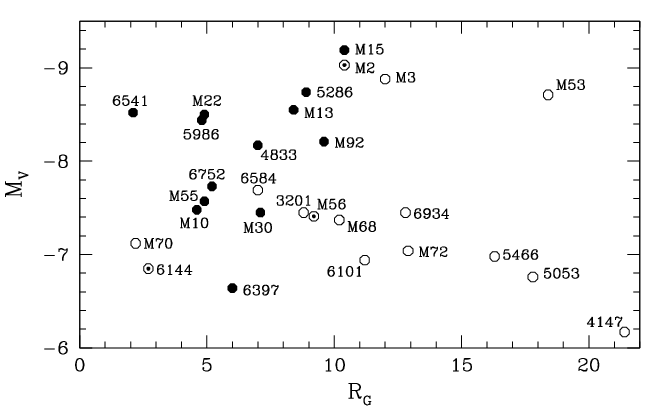}
\caption{The variation of $M_V$ (absolute integrated visual magnitude) with
Galactocentric distance, R$_{\rm G}$, for the clusters in our sample that have
[Fe/H] $\lta -1.5$.  The clusters, which are identified by their Messier or NGC 
numbers, have been classified as M$\,$3--like ({\it open circles}) or
M$\,$13--like ({\it filled circles}) based on the slope of the subgiant branch
in the observed CMDs.  A black dot at the center of an open circle indicates
that the classification is uncertain; i.e., the slope of the cluster SGB is
intermediate to those in the aforementioned groups.}
\label{fig:fig37}
\end{figure}

\begin{figure}[t]
\centering
\includegraphics[width=0.85\textwidth]{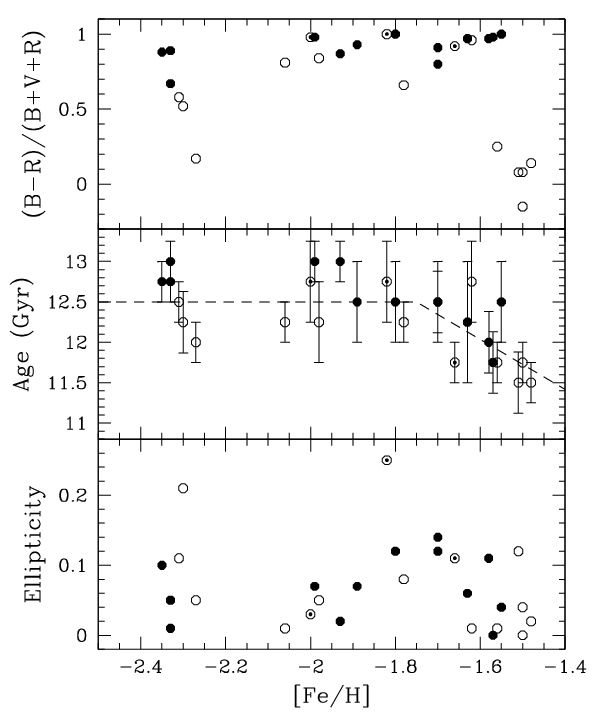}
\caption{The HB types ({\it upper panel}), ages ({\it middle panel}), and
cluster ellipticities ({\it lower panel}) of the same GCs that were considered
in the previous figure are plotted as a function of their [Fe/H] values.  The
same symbols have been used to represent the M$\,$3--like and M$\,$13--like 
clusters, as well as those with uncertain classifications.  The sloped and
dashed parts of the dashed line in the middle panel represent, respectively, the
metal-poor branch of the bifurcated age-metallicity relation shown in
Fig.~\ref{fig:fig33} and its extension to the lowest [Fe/H] values.} 
\label{fig:fig38}
\end{figure}

\begin{figure}[t]
\centering
\includegraphics[width=1.0\textwidth]{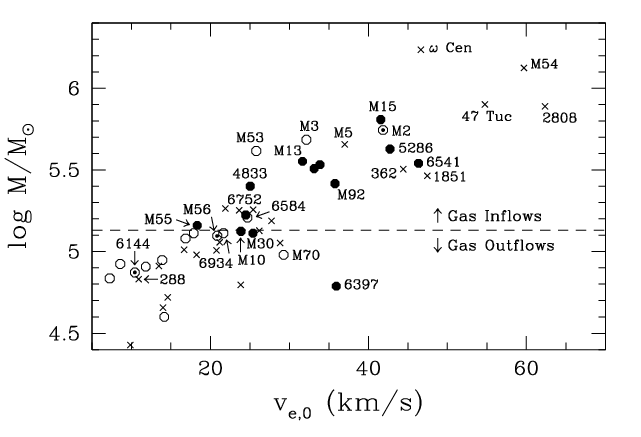}
\caption{The masses of the GCs, in solar units, are plotted as a function of
their central escape velocities, $v_{e,0}$.  Open and filled circles have been
used to represent the M$\,$3--like and M$\,$13--like clusters, respectively, 
while composite symbols indicate those clusters with uncertain classifications
(as in the previous two figures).  Crosses indicate GCs that have [Fe/H] $\gta
-1.5$ or those that show variations in [Fe/H], like $\omega\,$Cen.  Clusters
that are expected to develop GC winds, according to the models by \citet{ff77},
lie below the horizontal dashed line, whereas those which may build up a central
reservoir of the gas that is shed by mass-losing stars lie above that line.
The location of this line is approximate as it depends on several factors (see
the text).  Of the clusters that have been explicitly identified, 47 Tuc,
M$\,$15, M$\,$30, M$\,$53, NGC$\,$6397, M$\,$70, and NGC$\,$6752 all have
central concentrations $c > 2.0$ (\citealt{har96}.}
\label{fig:fig39}
\end{figure}

\begin{figure}[t]
\centering
\includegraphics[width=1.0\textwidth]{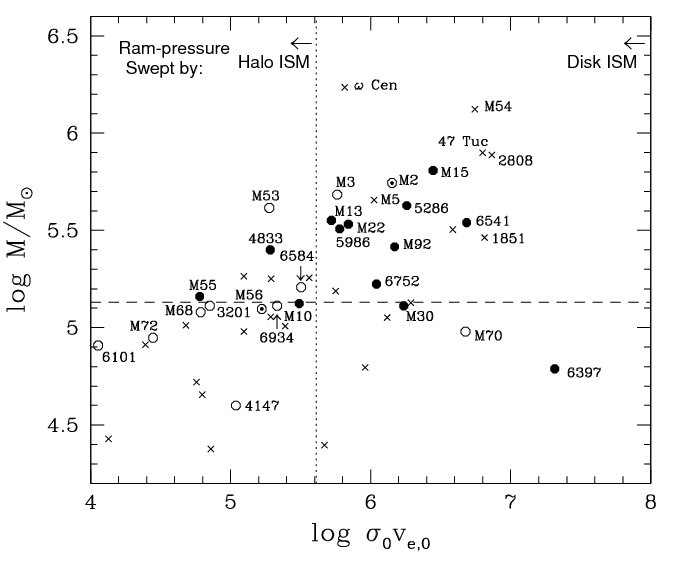}
\caption{The masses of the GCs, in solar units, are plotted as a function of the
logarithm of the product of the surface density of the stars at the cluster
center, $\sigma_0$, and the central escape velocity, $v_{e,0}$.  The latter 
is correlated with the ability of a given GC to resist ram-pressure stripping
of any gas that it contains as a result of its passage through the Galactic halo
or disk.  The vertical dotted line separates clusters in which ram-pressure
sweeping would, or would not, be effective in removing gas from them by the
halo interstellar medium.  The location of this line is uncertain, but it is
expected to be approximately where it has been drawn if the density of the halo
medium is $1.5 \times 10^{-27}$ g/cm$^3$ and the proportional rate of mass loss
from the cluster stars is $\alpha = 7 \times 10^{-20}$ s$^{-1}$ (see the text).
Clusters that lie to the right of the dotted line would be ram-pressure swept
of any gas that they are able to accumulate only when they pass through the
Galactic disk.  The symbols and the horizontal dashed line have the same
definitions as in the previous figure.}
\label{fig:fig40}
\end{figure}

Another important issue is the extent to which the C$+$N$+$O abundances vary,
both from cluster-to-cluster and from star-to-star within each GC.  With
relatively few exceptions (e.g., NGC$\,$1851, see \citealt{ygd09}), 
spectroscopic studies have found little or no variation in the total CNO
abundance within GCs, though it is expected that any
significant enhancement in $Y$ will be accompanied by increased CNO abundances
if mass loss from intermediate-mass AGB stars is the origin of the
helium-enriched gas (e.g., \citealt{fck04}, \citealt{kfs06}).  No allowance has
been made in this study for such a correlation, even though turnoff luminosity
versus age relations for metal-poor stars depend almost entirely on the absolute
C$+$N$+$O abundance (see VBD12), because it is not known at the present time
whether the observed luminosity widths of the SGBs in clusters that have thick
subgiant branches is caused by variations in CNO or age (or both).   The most
obvious way of explaining why, for instance, M$\,$15 has a much redder HB than
M$\,$92 or M$\,$30 (see Fig.~\ref{fig:fig11}), despite all three clusters having
nearly identical [Fe/H] values, is that the former has higher [CNO/Fe] than the
latter since increased CNO abundances have the effect of driving HB models to
lower effective temperatures (and hence redder colors), with minimal effects on
their luminosities (\citealt{ct77}).   If this is indeed the case, then M$\,$15
should be somewhat younger than M$\,$92 and M$\,$30, assuming that all of the
other parameters which affect age determinations are left unchanged.

Fortunately, the predicted mean luminosity (but not the slope) of the SGB at a
fixed age is essentially independent of $Y$, but to disentangle the age and
abundance effects, large spectroscopic surveys of the cluster subgiant
populations will be required.  However, perhaps the best way of getting a
handle on the chemistry of GCs is to search for and identify eclipsing binary
members because they can provide tight contraints on the masses and radii of
stars at their locations in observed CMDs, as well as on their distances if
accurate and precise estimates of their temperatures can be derived.  Such work
should be given very strong support: this avenue of research is likely to lead
to the biggest improvement in our understanding of GCs during the next few
years.  (A good example of the value and importance of such studies is provided
by the binary V69 in 47 Tuc, which provides a compelling case against the
possibility that, as found by MF09, this [Fe/H] $\sim -0.8$ cluster is coeval
with the most metal-deficient GCs.)

The most unanticipated result of this study is the discovery that the most
metal-poor GCs (those with [Fe/H] $\sim -1.5$) can be divided into two groups
on the basis of differences in their SGB morphologies.  That nearly the same
separation is obtained between clusters that, {\it at the present time}, are
able, or unable, to retain the gas which is lost via normal mass-loss processes
through the development of GC winds or which are subject to ram-pressure
stripping by the halo interstellar medium is especially surprising.  How is it
possible that the gas-retention properties of present-day GCs discriminate so
well between those systems that show the strongest evidence for multiple stellar
populations (specifically a significantly higher $Y$) and those for which the
tightness of the CMD appears to imply little or no variations in $Y$, [CNO/Fe],
or age?   This may be telling us that the masses of GCs have not changed by a
large factor during their lifetimes.  If all clusters formed with $\sim 10$--25
times their present masses (as predicted by, e.g., \citealt{dvd08},
\citealt{con12}), surely the cluster-to-cluster variations in mass loss that 
occured since their formation due to 2-body and tidal interactions, which depend
on the orbit and various cluster properties, would have been large enough that
the separation of the M$\,$3-like and M$\,$13-like clusters would not have been
so clear-cut.

Some support for a much reduced mass loss over the lifetimes of GCs to date is
provided by the recent work of \citet{lsb12}, who found that the globular
clusters in Fornax, which show very similar variations of light-element
abundances as those seen in Galactic GCs, contain 20--25\% of all of the stars
in this dwarf galaxy that have [Fe/H] $< -2.0$.  Even if all of the field stars
that satisfy this inequality came from the clusters, the latter could not have
been more massive than 4--5 times their present masses.  Although this result
may be complicated by the issue of whether or not Fornax is the result of a
merger of two dwarf galaxies (e.g., \citealt{cdb05}, \citealt{yb12}),
initial-to-final mass ratios of $\lta 5$ are also suggested by the results of
$n$-body simulations that attempt to model how an assumed initial mass function
(IMF) would have been altered over the evolutionary history of a cluster through
the evaporation of low-mass stars and the effects of tidal interactions with the
Galaxy (e.g., \citealt{bdk08}; \citealt{zkb11}).  Interestingly, it seems that
the present-day MFs can be reproduced quite well if all clusters have the same
IMF, which is suggested to be the case by the work of \citet{lus12}.   In
addition, the great difficulty of finding halo field stars with low oxygen and
high sodium abundances is hard to understand if GCs have lost at least 90--95\%
of their masses over the past 11--13 Gyr.  While Ram\'irez et al.~(2012) have
reported the discovery of two such dwarfs, those stars have low $\alpha$-element
abundances (see their Fig.~1), which are not typical of GC stars.  Where are
the field halo stars with high $\alpha$/Fe ratios and low oxygen abundances?
This is clearly less of a problem if GCs were initially much less massive than
many studies have proposed.

Inferences concerning the existence of significant populations of
second-generation stars in GCs from the morphology of the horizontal branch ---
in particular, from the length of the blue HB or the existence of gaps along
this feature (e.g., \citealt{dc08}, \citealt{gcb10}) --- are quite speculative
as well, though it is possible to differentiate between sub-populations on the
HB using $uv$ photometry (\citealt{dsf13}, \citealt{gls13}, and references
therein).  This is demonstrated by the work of \citet{cd11} who suggested that
M$\,$53 consists primarily of first-generation stars because (i)
it has quite a stubby blue HB for its metallicity, and (ii) \citet{msb08}
proposed that one way of understanding the small range in the observed C and
N abundances in this cluster is that the polluting gas had not been processed
through the full CNO-cycle, in which case the Ne-Na cycle would not have been
operative either.  As a consequence, Caloi \& D'Antona predicted that stars in
this sytem would not show the O--Na anticorrelation.  

We do not know if the necessary observations have been carried out to test this
prediction, but we suspect that an O--Na anticorrelation will be found because
its presence has been reported in stars belonging to M$\,$55
(\citealt{cbg09c}), which have nearly the same [Fe/H] as their counterparts in
M$\,$53 and, what is especially noteworthy, even more uniform CN band strengths
(\citealt{bsh93}).   Both Martell et al.~(2008) and Briley et al.~suggest that
this can be explained if deep mixing has produced low C/Fe ratios in bright
giants of both clusters.  This could well be the explanation given that a
decline in the carbon abundance with increasing luminosity along the upper RGB
has been observed in a number of moderate to very metal-deficient clusters,
including M$\,$92 (\citealt{clb82}), M$\,$15 (\citealt{tcl83}), and M$\,$4 and
NGC$\,$6752 (\citealt{ss91}), among others.  In the case of M$\,$13, even the
oxygen abundance appears to follow a similar trend (\citealt{jp12}).  This
raises the possibility that deep mixing also plays a role in the observed O--Na
anticorrelations, which are seen in, e.g., the lowest metallicity ([Fe/H] $<
-1.6$) stars in $\omega\,$Cen (\citealt[see their Fig.~19]{jp10}).

It should also be kept in mind that, as discussed in the extensive review by
\citet[also see \citealt{ren08}]{gcb12}, current models for the AGB phase and
for rapidly rotating massive stars (\citealt[and references therein]{dmc07})
are not able to reproduce the observed abundance patterns in a fully
satisfactory and consistent way.  Although it may turn out that better agreement
will be obtained when some of the many parameters that affect the
models are fine-tuned, these difficulties (coupled with those mentioned above)
may instead be telling us that the scenario which has been developed over the
past few years, requiring very high initial GC masses, among other things, is
incorrect.  In particular, it is still within the realm of possibility that
most of the light-element variations that are observed were present in the gas
out of which the observed generation(s) of cluster stars formed (see
\citealt{dh13}); i.e., that they arose at earlier times in the evolution of
protoclusters.  As is widely appreciated, there are problems and inconsistencies
with every proposal that has been made to date --- which is, of course, the
reason why a satisfactory solution has not yet been found despite a tremendous
effort by many researchers.

\acknowledgements
We thank Ata Sarajedini for providing us with the ACS photometry that we have
used as well as the web site address where helpful information concerning
photometric zero-points is given.  We also thank Santi Cassisi for sending us
his color transformations to the F606W and F814W bandpasses, based on
Castelli-Kurucz model atmospheres, as they were used to extend the MARCS
transformations to effective temperatures $> 8000$ K.  For many helpful
discussions, as well as useful references, we are grateful to Pavel Denisenkov,
David Hartwick, and Sidney van den Bergh.  Christian Johnson and Caty
Pilachowski pointed out some pertinent results in their recent papers in 
response to our inquiries, while Bruce Elemgreen provided some important
clarifications concerning the formation of globular clusters: these
contributions to our understanding are very much appreciated as well.  This
paper has also benefitted from suggestions provided by Tom Brown, Aldo Valcarce,
and, in particular, two helpful referees.  KB acknowledges the support received
from the Carlsberg Foundation and from the Villum Foundation, while RL
acknowledges NSERC Discovery Grant support to Kim Venn and the financial
support to the DAGAL network from the People Programme (Marie Curie Actions) of
the European Union's Seventh Framework Programme FP7/2007-2013 under REA grant
agreement number PITN-GA-2011-289313.  DAV is grateful for the support of a
Discovery Grant from the Natural Sciences and Engineering Research Council of
Canada.

\clearpage

\clearpage

\clearpage
 
\clearpage

\clearpage

\clearpage

\clearpage

\clearpage

\clearpage

\clearpage
 
\clearpage

\clearpage

\clearpage

\clearpage

\clearpage

\clearpage

\clearpage

\clearpage

\clearpage

\clearpage

\clearpage

\clearpage

\clearpage

\clearpage

\clearpage

\clearpage

\clearpage

\clearpage

\clearpage

\clearpage

\clearpage

\clearpage

\clearpage

\clearpage

\clearpage

\clearpage

\clearpage

\clearpage

\clearpage

\clearpage

%
\clearpage
\begin{deluxetable}{cccc}
\tabletypesize{\footnotesize}
\tablewidth{360pt}
\tablecaption{Subgiant Slopes of [Fe/H] $\lta -1.5$ Globular Clusters
 \label{tab:tab1}}
\tablewidth{0pt}
\tablehead{\colhead{NGC} & \colhead{Name} & \colhead{[Fe/H]} &
 \colhead{SGB slope}}
\startdata
\multispan4 {\bf M3-like} \\
\noalign{\vskip 3pt}
 3201 &           & $-1.51$ & $-3.01\pm 0.50$ \\
 4147 &           & $-1.78$ & $-2.75\pm 0.13$ \\
 4590 & M$\,$68   & $-2.27$ & $-3.34\pm 0.22$ \\
 5024 & M$\,$53   & $-2.06$ & $-3.43\pm 0.12$ \\
 5053 &           & $-2.30$ & $-3.43\pm 0.24$ \\
 5272 & M$\,$3    & $-1.50$ & $-2.95\pm 0.12$ \\
 5466 &           & $-2.31$ & $-3.25\pm 0.12$ \\
 6101 &           & $-1.98$ & $-3.35\pm 0.29$ \\
 6584 &           & $-1.50$ & $-2.99\pm 0.17$ \\
 6681 & M$\,$70   & $-1.62$ & $-2.67\pm 0.30$ \\
 6934 &           & $-1.56$ & $-3.01\pm 0.24$ \\
 6981 & M$\,$72   & $-1.48$ & $-2.97\pm 0.18$ \\
\noalign{\vskip 3pt}
\multispan4 {\bf M13-like} \\
\noalign{\vskip 3pt}
 4833 &           & $-1.89$ & $-3.80\pm 0.30$ \\
 5286 &           & $-1.70$ & $-3.75\pm 0.27$ \\
 5986 &           & $-1.63$ & $-3.60\pm 0.30$ \\
 6205 & M$\,$13   & $-1.58$ & $-3.55\pm 0.13$ \\
 6254 & M$\,$10   & $-1.57$ & $-3.81\pm 0.29$ \\
 6341 & M$\,$92   & $-2.35$ & $-3.97\pm 0.15$ \\
 6397 &           & $-1.99$ & $-3.71\pm 0.30$ \\
 6541 &           & $-1.82$ & $-3.70\pm 0.26$ \\
 6656 & M$\,$22   & $-1.70$ & $-3.90\pm 0.38$ \\
 6752 &           & $-1.55$ & $-3.40\pm 0.16$ \\
 6809 & M$\,$55   & $-1.93$ & $-3.62\pm 0.21$ \\
 7078 & M$\,$15   & $-2.33$ & $-4.01\pm 0.20$ \\
 7099 & M$\,$30   & $-2.33$ & $-3.89\pm 0.18$ \\
\noalign{\vskip 3pt}
\multispan4 {\bf Intermediate/Uncertain} \\
\noalign{\vskip 3pt}
 6144 &           & $-1.82$ & $-3.29\pm 0.50$ \\  
 6779 & M$\,$56   & $-2.00$ & $-3.52\pm 0.34$ \\
 7089 & M$\,$2    & $-1.66$ & $-3.32\pm 0.20$ \\
\enddata
\end{deluxetable}

\clearpage
\begin{deluxetable}{cccccccccccc}
\tabletypesize{\scriptsize}
\tablewidth{460pt}
\tablecaption{Ages and Other Properties of the Globular Cluster Sample
 \label{tab:tab2}}
\tablewidth{0pt}
\tablehead{\colhead{NGC} & \colhead{Name} & \colhead{[Fe/H]} & \colhead{Age} &
 \colhead{Method\tablenotemark{a}} & \colhead{Fig(s).} &
 \colhead{Range\tablenotemark{b}} &
 \colhead{HB type} & \colhead{R$_{\rm G}$} & \colhead{$M_V$} &
 \colhead{$v_{e,0}$} & \colhead{$\log_{10}\,\sigma_0$} }
\startdata
 \phantom{0}104 & 47 Tuc & $-0.76$ & $11.75\pm 0.25$ & V & 14 & 11.50--11.75 & 
   $-0.99$ & \phantom{1}7.4 & $-9.42$ & 54.8 & 5.061 \\
 \phantom{0}288 & & $-1.32$ & $11.50\pm 0.38$ & H & 24 &    & $+0.98$ &
   12.0 & $-6.75$ & 10.9 & 2.953 \\
 \phantom{0}362 & & $-1.30$ & $10.75\pm 0.25$ & V & 13 & 10.75--11.00 & 
   $-0.87$ & \phantom{1}9.4 & $-8.43$ & 44.4 & 4.938 \\
 1261 &   & $-1.27$ & $10.75\pm 0.25$ & V & 13 & 10.75--11.25 & $-0.71$ &
   18.1 & $-7.80$ & 23.6 & 3.913 \\
 1851 &   & $-1.18$ & $11.00\pm 0.25$ & V & 13 & 10.75--11.25 & $-0.32$ &
   16.6 & $-8.33$ & 47.6 & 5.136 \\
 2808 &   & $-1.18$ & $11.00\pm 0.38$ & V & 13 & 11.00--11.25 & $-0.49$ &
   11.1 & $-9.39$ & 62.4 & 5.070 \\
 3201 &   & $-1.51$ & $11.50\pm 0.38$ & A & 12,30 & 11.25--11.75 & $+0.08$ &
   \phantom{1}8.8 & $-7.45$ & 17.9 & 3.599 \\
 4147 &   & $-1.78$ & $12.25\pm 0.25$ & V & 11 & 12.25--12.50 & $+0.66$ & 
   21.4 & $-6.17$ & 14.2 & 3.886 \\
 4590 & M$\,$68 & $-2.27$ & $12.00\pm 0.25$ & V & 11 & 12.00 & $+0.17$ &
   10.2 & $-7.37$ & 16.9 & 3.559 \\
 4833 &   & $-1.89$ & $12.50\pm 0.50$ & A & 30 &  & $+0.93$ &
   \phantom{1}7.0 & $-8.17$ & 25.0 & 3.885 \\
 5024 & M$\,$53 & $-2.06$ & $12.25\pm 0.25$ & V & 12 & 12.25--12.50 & $+0.81$ &
   18.4 & $-8.71$ & 25.8 & 3.866 \\
 5053 &   & $-2.30$ & $12.25\pm 0.38$ & A & 12 & 12.25--12.50 & $+0.52$ &
   17.8 & $-6.76$ & \phantom{1}7.2 & 2.196 \\
 5272 & M$\,$3 & $-1.50$ & $11.75\pm 0.25$ & V & 22 &  & $+0.08$ &
   12.0 & $-8.88$ & 32.1 & 4.254 \\
 5286 &    & $-1.70$ & $12.50\pm 0.38$ & A & 12,30 & 11.75--12.25 & $+0.80$ &
   \phantom{1}8.9 & $-8.74$ & 42.7 & 4.628 \\
 5466 &   & $-2.31$ & $12.50\pm 0.25$ & V & 11,30 & 12.25--12.50 & $+0.58$ &
   16.3 & $-6.98$ & \phantom{1}8.6 & 2.453 \\
 5904 & M$\,$5 & $-1.33$ & $11.50\pm 0.25$ & V & 8 & 11.50--11.75 & $+0.31$ &
   \phantom{1}6.2 & $-8.81$ & 37.0 & 4.457 \\
 5927 &   & $-0.29$ & $10.75\pm 0.38$ & V & 16 & 10.50--10.75 & $-1.00$ &
   \phantom{1}4.6 & $-7.81$ & 25.4 & 4.156 \\
 5986 &   & $-1.63$ & $12.25\pm 0.75$ & A & 30 &  & $+0.97$ &
   \phantom{1}4.8 & $-8.44$ & 33.1 & 4.259 \\
 6101 &   & $-1.98$ & $12.25\pm 0.50$ & H & 30 &  & $+0.84$ & 
   11.2 & $-6.94$ & 11.8 & 2.980 \\  
 6121 & M$\,$4 & $-1.18$ & $11.50\pm 0.38$ & V & 13 & 11.25--11.50 & $-0.06$ &
   \phantom{1}5.9 & $-7.19$ & 20.8 & 4.070 \\
 6144 &   & $-1.82$ & $12.75\pm 0.50$ & H & 30 &   & $+1.00$ &
   \phantom{1}2.7 & $-6.85$ & 10.4 & 2.975 \\
 6171 & M$\,$107 & $-1.03$ & $12.00\pm 0.75$ & V & 13 &   & $-0.73$ &
   \phantom{1}3.3 & $-7.12$ & 18.2 & 3.832 \\
 6205 & M$\,$13 & $-1.58$ & $12.00\pm 0.38$ & A & 20,22 &   & $+0.97$ &
   \phantom{1}8.4 & $-8.56$ & 31.7 & 4.219 \\
 6218 & M$\,$12 & $-1.33$ & $13.00\pm 0.50$ & A & 24,26 &   & $+0.98$ &
   \phantom{1}4.5 & $-7.31$ & 21.2 & 3.961 \\
 6254 & M$\,$10 & $-1.57$ & $11.75\pm 0.38$ & H & 30 &   & $+0.98$ &
   \phantom{1}4.6 & $-7.48$ & 23.9 & 4.112 \\
 6304 &   & $-0.37$ & $11.25\pm 0.38$ & V & 16 & 11.00--11.25 & $-1.00$ &
   \phantom{1}2.3 & $-7.30$ & 28.8 & 4.656 \\
 6341 & M$\,$92 & $-2.35$ & $12.75\pm 0.25$ & V & 11 & 12.75--13.25 & $+0.91$ &
   \phantom{1}9.6 & $-8.21$ & 35.8 & 4.618 \\
 6352 &   & $-0.62$ & $10.75\pm 0.38$ & V & 16 & 10.50--11.00 & $-1.00$ &
   \phantom{1}3.3 & $-6.47$ & 14.6 & 3.592 \\
 6362 &   & $-1.07$ & $12.50\pm 0.25$ & V & 13 & 12.25--12.75 & $-0.58$ &
   \phantom{1}5.1 & $-6.95$ & 13.5 & 3.261 \\
 6366 &   & $-0.59$ & $11.00\pm 0.50$ & V & 16 & 11.00 & $-0.97$ &
   \phantom{1}5.0 & $-5.74$ & \phantom{1}9.8 & 3.135 \\
 6397 &   & $-1.99$ & $13.00\pm 0.25$ & A & 30 & 13.00 & $+0.98$ &
   \phantom{1}6.0 & $-6.64$ & 35.9 & 5.759 \\
 6496 &   & $-0.46$ & $10.75\pm 0.38$ & V & 16 & 10.50--10.75 & $-1.00$ &
   \phantom{1}4.2 & $-7.20$ & 16.7 & 3.459 \\
 6535 &   & $-1.79$ & $12.75\pm 0.50$ & V & 32 &   & $+1.00$ &
   \phantom{1}3.9 & $-4.75$ & \phantom{1}8.2 & 3.327 \\
 6541 &   & $-1.82$ & $12.50\pm 0.50$ & H & 30 &  & $+1.00$ &
   \phantom{1}2.1 & $-8.52$ & 46.3 & 5.019 \\
 6584 &   & $-1.50$ & $11.75\pm 0.25$ & A & 12,30 &   & $-0.15$ &
   \phantom{1}7.0 & $-7.69$ & 24.7 & 4.112 \\
 6624 &   & $-0.42$ & $11.25\pm 0.50$ & V & 16 & 11.00--11.25 & $-1.00$ &
   \phantom{1}1.2 & $-7.49$ & 26.2 & 4.869 \\
 6637 & M$\,$69 & $-0.59$ & $11.00\pm 0.38$ & V & 16 & 11.00--11.25 & $-1.00$ &
   \phantom{1}1.7 & $-7.64$ & 27.7 & 4.308 \\
 6652 &   & $-0.76$ & $11.25\pm 0.25$ & V & 16 & 11.00--11.25 & $-1.00$ &
   \phantom{1}2.7 & $-6.66$ & 23.9 & 4.583 \\
 6656 & M$\,$22 & $-1.70$ & $12.50\pm 0.50$ & H & 30 &   & $+0.91$ &
   \phantom{1}4.9 & $-8.50$ & 33.9 & 4.311 \\
 6681 & M$\,$70  & $-1.62$ & $12.75\pm 0.38$ & A & 30,32 &   & $+0.96$ &
   \phantom{1}2.2 & $-7.12$ & 29.3 & 5.210 \\
 6715 & M$\,$54 & $-1.44$ & $11.75\pm 0.50$ & V & 32 &   & $+0.54$ &
   18.9 & $-9.98$ & 59.7 & 4.968 \\
 6717 &   & $-1.26$ & $12.50\pm 0.50$ & V & 13 &   & $+0.98$ &
   \phantom{1}2.4 & $-5.66$ & 16.3 & 4.458 \\
 6723 &   & $-1.10$ & $12.50\pm 0.25$ & V & 13 & 12.25--12.75 & $-0.08$ &
   \phantom{1}2.6 & $-7.83$ & 21.9 & 3.755 \\
 6752 &   & $-1.55$ & $12.50\pm 0.25$ & A & 30 &   & $+1.00$ &
   \phantom{1}5.2 & $-7.73$ & 24.5 & 4.655 \\
 6779 & M$\,$56 & $-2.00$ & $12.75\pm 0.50$ & H & 30 &   & $+0.98$ &
   \phantom{1}9.2 & $-7.41$ & 20.8 & 3.905 \\
 6809 & M$\,$55 & $-1.93$ & $13.00\pm 0.25$ & A & 12,30 & 12.75--13.25 &
   $+0.87$ & \phantom{1}3.9 & $-7.57$ & 18.3 & 3.515 \\
 6838 & M$\,$71 & $-0.82$ & $11.00\pm 0.38$ & V & 16 & 11.00 & $-1.00$ &
   \phantom{1}6.7 & $-5.61$ & 13.0 & 3.746 \\
 6934 &   & $-1.56$ & $11.75\pm 0.25$ & V & 12,30 & 11.50--12.00 & $+0.25$ &
   12.8 & $-7.45$ & 21.6 & 3.998 \\
 6981 & M$\,$72 & $-1.48$ & $11.50\pm 0.25$ & V & 12,30 & 11.25--11.75 & 
   $+0.14$ & 12.9 & $-7.04$ & 13.9 & 3.304 \\
 7078 & M$\,$15 & $-2.33$ & $12.75\pm 0.25$ & V & 11 & 12.50--12.75 &
   $+0.67$ & 10.4 & $-9.19$ & 41.6 & 4.827 \\
 7089 & M$\,$2 & $-1.66$ & $11.75\pm 0.25$ & A & 12,30 & 11.50--11.75 &
   $+0.92$ & 10.4 & $-9.03$ & 41.8 & 4.531 \\
 7099 & M$\,$30 & $-2.33$ & $13.00\pm 0.25$ & A & 11,30 & 12.75--13.00 &
   $+0.89$ & \phantom{1}7.1 & $-7.45$ & 25.4 & 4.831 \\
    & Arp 2 & $-1.74$ & $12.00\pm 0.38$ & V & 32 &   & $+0.53$ & 21.4 &
   $-5.29$ & \phantom{1}3.5 & 1.545 \\
    & Pal 12 & $-0.81$ & \phantom{1}$9.0\pm 0.38$ & V & 32 & & $-1.00$ &
   15.8 & $-4.47$ & \phantom{1}4.3 & 3.297 \\
    & Ter 8 & $-2.34$ & $13.00\pm 0.38$ & V & 32 & & $+1.00$ & 19.4 &
   $-5.07$ & \phantom{1}4.3 & 1.938 \\
\enddata
\tablenotetext{a}{Ages are based on the vertical (V) or horizontal (H) methods, or
 an average (A) of the former.}
\tablenotetext{b}{This range encompasses the ages independently derived by DAV,
 KB, and RL.}
\end{deluxetable}

\clearpage
\begin{deluxetable}{cccccc}
\tabletypesize{\footnotesize}
\tablewidth{400pt}
\tablecaption{Orbital Parameters of [Fe/H] $\lta -1.5$ Globular Clusters
 \label{tab:tab3}}
\tablewidth{0pt}
\tablehead{\colhead{NGC} & \colhead{Name} & \colhead{$e$\tablenotemark{a}} & 
 \colhead{R$_a$\tablenotemark{a,b}} & \colhead{R$_p$\tablenotemark{a,b}} &
 \colhead{$z_{\rm max}$\tablenotemark{a,b}}}
\startdata
\multispan6 \hfil {\bf M3-like SGB} \hfil \\
\noalign{\vskip 3pt}
 4147 &         & $0.72\pm 0.10$ & $25.3\pm \phantom{1}2.6$  
                & $\phantom{1}4.1\pm 2.2$  & $13.1\pm \phantom{1}1.7$ \\
 4590 & M$\,$68 & $0.48\pm 0.03$ & $24.4\pm \phantom{1}3.1$  
                & $\phantom{1}8.6\pm 0.3$  & $\phantom{1}9.1\pm \phantom{1}1.3$ \\
 5024 & M$\,$53 & $0.40\pm 0.12$ & $36.0\pm 16.8$           
                & $15.5\pm 1.9$  & $24.2\pm \phantom{1}8.1$ \\
 5272 & M$\,$3  & $0.42\pm 0.07$ & $13.4\pm \phantom{1}0.8$  
                & $\phantom{1}5.5\pm 0.8$ & $\phantom{1}8.7\pm \phantom{1}0.5$ \\
 5466 &         & $0.79\pm 0.03$ & $57.1\pm 24.6$           
                & $\phantom{1}6.6\pm 1.5$ & $34.1\pm 14.4$ \\
 6584 &         & $0.87\pm 0.05$ & $12.6\pm \phantom{1}2.4$  
                & $\phantom{1}0.9\pm 0.7$ & $\phantom{1}3.1\pm \phantom{1}2.3$ \\
 6934 &         & $0.72\pm 0.07$ & $37.5\pm 15.2$            
                & $\phantom{1}6.0\pm 1.6$ & $21.2\pm \phantom{1}9.5 $\\
\noalign{\vskip 3pt}
\multispan6 \hfil {\bf M13-like SGB}\hfil \\
\noalign{\vskip 3pt}
 6205 & M$\,$13 & $0.62\pm 0.06$ & $21.5\pm 4.7$           
                & $5.0\pm 0.5$ & $13.2\pm 2.1$ \\
 6254 & M$\,$10 & $0.19\pm 0.05$ & $\phantom{1}4.9\pm 0.2$ 
                & $3.4\pm 0.4$ & $\phantom{1}2.4\pm 0.2$ \\
 6341 & M$\,$92 & $0.76\pm 0.03$ & $\phantom{1}9.9\pm 0.4$ 
                & $1.4\pm 0.2$ & $\phantom{1}3.8\pm 0.5$ \\
 6397 &         & $0.34\pm 0.02$ & $\phantom{1}6.3\pm 0.1$ 
                & $3.1\pm 0.2$  & $\phantom{1}1.5 \pm 0.1$ \\
 6656 & M$\,$22 & $0.53\pm 0.01$ & $\phantom{1}9.3\pm 0.7$ 
                & $2.9\pm 0.2$  & $\phantom{1}1.9\pm 0.1$ \\
 6752 &         & $0.08\pm 0.02$ & $\phantom{1}5.6\pm 0.2$ 
                & $4.8\pm 0.3$ & $\phantom{1}1.6\pm 0.1$ \\
 7078 & M$\,$15 & $0.32\pm 0.05$ & $10.3\pm 0.7$           
                & $5.4\pm 1.1$ & $\phantom{1}4.9\pm 0.8$ \\
 7099 & M$\,$30 & $0.39\pm 0.06$ & $\phantom{1}6.9\pm 0.3$ 
                & $3.0\pm 0.4$   & $\phantom{1}4.4\pm 0.3$ \\
\noalign{\vskip 3pt}
\multispan6 \hfil {\bf Intermediate SGB Slope} \hfil \\
\noalign{\vskip 3pt}
 6144 &         & $0.25\pm 0.15$ & $\phantom{1}3.0\pm \phantom{1}0.7$ 
                & $1.8\pm 0.2$ & $\phantom{1}2.4\pm 0.2$ \\  
 6779 & M$\,$56 & $0.86\pm 0.03$ & $12.4\pm \phantom{1}1.5$          
                & $0.9\pm 0.3$ & $\phantom{1}1.1\pm 0.7$ \\
 7089 & M$\,$2  & $0.68\pm 0.06$ & $33.6\pm 12.5$                     
                & $6.4\pm 1.1$ & $20.0\pm 6.5$ \\
\enddata
\tablenotetext{a}{From Dinescu et al.~(1999).}
\tablenotetext{b}{In kpc.}
\end{deluxetable}

\end{document}